\documentclass{aa}

\usepackage{color}
\definecolor{red}{rgb}{1,0,0}
\definecolor{green}{rgb}{0,1,0}
\definecolor{blue}{rgb}{0,0,1}

\usepackage[T1]{fontenc}

\usepackage{graphicx}	
\usepackage{gensymb}

\usepackage{mathtools}

\usepackage{amssymb}	

\usepackage{amsmath}	
\usepackage{comment}
\usepackage{longtable}
\usepackage{xtab}
\usepackage[skip=0.5ex]{subcaption}

\usepackage{txfonts}
\usepackage{hyperref}%
\hypersetup{colorlinks=true,linkcolor=[rgb]{1.,0.2,0.2},citecolor=[rgb]{0.1,0.4,1.},filecolor=[rgb]{0.7,0.2,0.2},urlcolor=[rgb]{0.7,0.2,0.2}}
\usepackage{silence}
\begin{document}

   \title{The YMCA (Yes, Magellanic Clouds Again) survey: probing the outer regions of the Magellanic system with VST\thanks{Based on the European Southern Observatory programs 098.D-0587; 099.D-0662; 0100.D-0565; 0101.D-0349; 0102.D-0574; 0103.D-0629; 0104.D-0427; 105.20DM}}


   \author{M. Gatto
          \inst{1}\fnmsep\thanks{Email: massimiliano.gatto@inaf.it}
          \and
          V. Ripepi\inst{1}
          \and
          M. Bellazzini\inst{2}
          \and
          M. Tosi\inst{2}
          \and 
          M. Cignoni\inst{2,3,4}
          \and
          C. Tortora\inst{1}
          \and
          M. Marconi\inst{1}
          \and
          M. Dall'Ora\inst{1}
          \and
          M.-R. L. Cioni\inst{5}          
          \and
          I. Musella\inst{1}
          \and
          P. Schipani\inst{1}
          \and
          M. Spavone\inst{1}
          }

   \institute{INAF - Osservatorio Astronomico di Capodimonte, Via Moiariello 16, I-80131 Naples, Italy.
         \and
             INAF - Osservatorio di Astrofisica e Scienza dello Spazio, Via Gobetti 93/3, I-40129 Bologna, Italy.
             \and
             Physics Department, University of Pisa, Largo Bruno Pontecorvo 3, 56127 Pisa, Italy.
             \and
             INFN - Largo Bruno Pontecorvo 3, 56127 Pisa, Italy.
             \and
             Leibniz-Institut f\"ur Astrophysik Potsdam, An der Sternwarte 16, D-14482 Potsdam, Germany.
             }

   \date{}

 
  \abstract
  {The Magellanic Clouds (MCs) are the Milky Way's most massive dwarf satellites. As they also represent the closest pair of galaxies in an ongoing tidal interaction, while simultaneously infalling into the Milky Way halo, they provide a unique opportunity to study in detail an ongoing three-body encounter.}
   {We present the ``YMCA (Yes, Magellanic Clouds Again) survey: probing the outer regions of the Magellanic system with VST'' based on deep optical photometry carried out with the VLT Survey Telescope (VST).}
   {YMCA targeted 110 square degrees, in the $g$ and $i$ filters, in the periphery of both the MCs, including a long strip in between the Large Magellanic Cloud (LMC) and the Small Magellanic Cloud (SMC). The photometry of YMCA is sufficiently deep (50\% complete down to $g \simeq 23.5-24.0$~mag) to allow for a detailed analysis of main-sequence stars in regions of the MCs remained relatively unexplored at these faint magnitudes.}
   {The resulting colour-magnitude diagrams reveal that the outskirts of the MCs are predominantly characterized by intermediate-age and old stellar populations, with limited or negligible evidence of recent star formation. 
   The analysis of the age distribution of star clusters (SCs) within the surveyed area, both already known and newly discovered candidates, hints at a close fly-by between the LMC and SMC that occurred $\simeq 2.5-3.0$~Gyr ago, in agreement with previous results.
   We also report the discovery of candidate SCs with ages within the so-called ``age-gap'', questioning its real existence.}
   {}

               
\titlerunning{The YMCA survey}
   \maketitle
%

\section{Introduction}

One of the key questions in astrophysics is how galaxies form and evolve. The widely accepted $\Lambda$-CDM paradigm predicts that most galaxies, including the smaller ones, should have an extended old, metal-poor, stellar halo, populated by the debris of past merging events \citep[e.g.][]{Bullock&Johnston2005,Cooper-2010,Annibali-2016}. 
Halos and their characteristic underlying old stellar populations thus provide a fossil record of past mergers and interactions in the form of extended tidal debris \citep[e.g.][]{Amorisco-2014}. 
The distribution and extent of such debris are both sensitive to the total mass and size of the galaxy, while their star formation history (SFH) is sensitive to the occurred accretion and outflow events.
Hence, probing the low surface brightness edges of galaxies can constrain their formation history \citep[e.g.][]{Deason-2014}. 
The debris originated by the past merger events are expected to have a very low surface brightness (i.e. 28-30 mag/arcsec$^2$) and to extend for many kpc \citep[e.g.][]{Cooper-2010,Cooper-2015}, hence only a combination of very deep photometry and wide sky coverage can unveil them in nearby galaxies.\par
The Magellanic Clouds (MCs) are the most massive dwarf satellites of the Milky Way (MW). Historically this system has been pivotal for many different astrophysical branches, spanning from improving the stellar evolutionary and pulsation models \citep{Keller&Wood2002,Keller&Wood2006,Brocato-2003,Marconi&Clementini2005,Girardi&Marigo2007,Lebzelter&Wood2007,Marconi-2017} to a finer evaluation of the Hubble constant \citep[e.g.,][and references therein]{Riess-2021,Ripepi-2021}.
As they also represent the closest pair of galaxies in an ongoing tidal interaction \citep[$D\sim$~50-60 kpc,][]{deGrijs&Bono2015,Pietrzynski-2019}, while simultaneously infalling into the MW halo \citep[see, e.g.,][and references therein]{D'Onghia&Fox2016}, they offer a unique benchmark to study in detail a gas-rich interacting system whose outskirts can be investigated down to the intrinsically faintest and oldest constituents.\par
Amongst the most evident signatures of their mutual interaction and of their interaction with the MW, it is worth mentioning the Magellanic Stream, an extended stream of HI gas that spans for more than 200\degree~around the Galactic South pole of the MW \citep[][]{Mathewson-1974,Putman-2003,Bruns-2005,Nidever-2010,D'Onghia&Fox2016}, and the Magellanic Stream counterpart, dubbed the Leading Arm, which is located ahead of the LMC \citep{Putman-1998}. Recent works discovered a population of stars that appear to be the stellar counterpart of both the trailing and leading arms of the Magellanic Stream \citep{Price-Whelan-2019,Zaritsky-2020,Petersen-2022,Chandra-2023}.
A further noticeable interacting feature is the Magellanic Bridge (MB), a stream made of gas and stars that connects the two galaxies from the east side of the SMC to the south-west of the LMC \citep{Hindman-1963,Irwin-1985,Demers&Battinelli1998,Harris2007,Bagheri-2013,Nidever-2013,Noel-2013,Dobbie-2014,Skowron-2014,Noel-2015,Carrera-2017,Mackey-2017}.\par
The advent of recent powerful instrumentation at different wavelengths has changed our view of the MC system.
It is now well accepted, thanks to very precise tangential velocities achieved with the Hubble Space Telescope (HST) combined with dynamical models, that the two galaxies may have not always orbited around the MW, but entered into the Galaxy halo some Gyr ago \citep[e.g.][]{Kallivayalil-2006b,Kallivayalil-2013,Besla-2012,Vasiliev2024}. 
Moreover, the recent discovery of a large number of MW satellites in a small area around the MCs \citep[][]{Koposov-2015,Martin-2015,Koposov-2018,Torrealba-2018,Cerny-2021}, in combination with proper motion measurements obtained from the Gaia space mission \citep[][]{Gaia-Brown-2018}, suggested that at least a fraction of the new objects might have once been part of the Magellanic group \citep[][]{Kallivayalil-2018,Pace&Li2019,Pardy-2020,Patel-2020}, thus reinforcing earlier hints that some of the satellite galaxies of the MW may be associated with MCs \citep[e.g.][]{Lynden-Bell1976}.
Additionally, in the last decade, many studies reported the presence of new substructures in the form of elongated, ring-like, or arc-like stellar streams, discovered in the periphery of the MCs, likely originated by their mutual gravitational interaction, and with the MW as well \citep[][]{Belokurov&Koposov2016,Pieres-2017,Choi-2018b,Mackey-2016,Mackey-2018,Belokurov&Erkal2019,Massana-2020,ElYoussoufi-2021,Gatto-2022b,Petersen-2022}.
It is then crucial to investigate the outskirts of the MCs and collect new insights about their interaction history and their recent connection with the MW, and also about their early evolution, as it is poorly constrained yet.\par
In this context, we present the first results from the VST (VLT Survey Telescope) survey ``Yes, Magellanic Clouds Again'' (YMCA; PI: V. Ripepi). 
We describe the YMCA survey in Sect.~2, whereas the observation strategy is explained in Sect.~3. The data reduction, including the absolute photometric calibration of the sources and the construction of the final catalogue, is presented in Sect.~4, as well as the artificial star test to derive the completeness of the YMCA survey. Throughout the Sect.~5, we describe the analysis of the stellar populations of the MCs within the entire YMCA footprint. Sect.~6 presents a detailed investigation of MCs star clusters (SCs), that comprises SCs already reported in literature as well as unknown candidate SCs discovered in this work. A summary closes the article.

\section{The YMCA survey}

The YMCA survey probes 110 square degrees in the periphery of the MCs, as well as a strip above the MB connecting them, through deep and homogeneous photometry in the optical $g$ and $i$ filters.
The survey aims at reconstructing the evolutionary history of the MCs but also at better constraining their interaction history.
We plan to tackle this task by deriving the SFH over the whole Hubble time for a large region around the MCs not covered by other surveys of equivalent photometric depth. The main tool for obtaining the SFH is the colour-magnitude diagram (CMD) of the MC resolved stars, and to reach this purpose, we obtained photometry 1-2 magnitudes fainter than the main sequence turn-off (MSTO) of the oldest population. 
The photometric depth of the survey also allows us to search for faint and old SCs that may still lie hidden and undetected in the outskirts of the system.
Further goals of the survey include the search of low-surface brightness structures in the periphery of the galaxies, as proxies of tidal induced disturbances.\par
Besides the YMCA, a few ongoing or proposed surveys plan to cover the MCs and their immediate surroundings.
{\it ``The SMC in Time: Evolution of a Prototype interacting late-type dwarf galaxy''} \citep[STEP: PI V. Ripepi,][]{Ripepi-2014} is a completed optical photometric survey complementary to the YMCA. It covers 53 square degrees of the SMC main body and the MB with the $g,r, i,$ and $H_\alpha$ filters. Some of its main objectives are to: i) derive the SFH of the SMC; ii) unveil how the stellar component of the Bridge was formed; iii) make a homogeneous statistical analysis of the SMC star clusters.
{\it ``The Survey of the MAgellanic Stellar History''} \citep[SMASH,][]{nidever-2017} maps 480 square degrees around the MCs with the Dark Energy Camera (DECam) in the $u,g,r,i,z$ filters, with the primary goal of resolving low surface brightness structures related to the MCs tidal interactions. 
The {\it ``VIsible Soar photometry of star Clusters in tApii and Coxi HuguA''} survey \citep[VISCACHA,][]{Maia-2019} aims at performing deep photometric observations of MC star clusters through the SOuthern Astrophysical Research (SOAR) telescope.
The fourth phase of the {\it ``Optical Gravitational Lensing
Experiment''} \citep[OGLE-IV,][]{Udalski-2015} probes the MCs in $V$ and $I$ bands with repeated observations to characterize the variable stars of the Magellanic system.
The {\it Magellanic Satellites Survey} \citep[MagLiteS,][]{Drlica-Wagner-2016} observed 1200 square degrees in the MCs periphery with the DECam to search for MC satellites.
Finally, the VISTA survey of the Magellanic Clouds \citep[VMC, ][]{Cioni-2011} obtained $YJK_s$ photometry for 184 square degrees in the LMC and SMC main bodies, the Bridge and the Stream as well.\par
The YMCA survey targeted with unprecedented depth and photometric accuracy periphery regions of the MCs not covered by the above-mentioned complementary surveys, making it the only dataset which allows us to resolve and analyze the very old MC stellar population in those fields.
The YMCA survey already demonstrated its capability in spotting and subsequently analyzing faint old stellar systems in the MC peripheries thanks to its deep and accurate photometry. For example, in a pilot work made by our group we tested a new cluster-finder algorithm on some preliminary observed fields. In this framework, we reported the discovery of a list of candidate LMC SCs with ages falling in the so-called ``age gap'' \citep[][]{Gatto-2020}, a period ranging from $\sim$4 up to 10 Gyr ago almost devoid of SCs \citep[e.g.][]{Jensen-1988,DaCosta1991}. 
An analysis made on the CMD of the catalogued SC KMHK~1762 by using new YMCA data, allowed us to assert that it is the third age gap SC ever discovered in the LMC \citep[][]{Gatto-2022c}.
The algorithm also led to the detection of a new stellar system, dubbed YMCA-1 \citep[][]{Gatto-2021}, which is old, likely associated with the LMC, and has structural properties in between the known LMC globular clusters (GCs) and the ultra-faint dwarf galaxies (UFDg) of the Local Group \citep[][]{Gatto-2022a}. 

\begin{table}
\centering
 
 \caption{Overview of the surveys.}
 \label{tab:surveys}
 \footnotesize\setlength{\tabcolsep}{5.3pt}
 \begin{tabular}{lccccc}
  \hline\hline
  Survey & Area & Bands & Depth & Seeing & Pxl scale\\
    & [Sq. Deg.] &  & [mag] & [\arcsec] & [\arcsec]\\
  \hline
    OGLE-IV$^a$ & 3000 & {\it VI}  & $I \leq 20$ & 1.25--1.35  & 0.26\\
    MagLiteS$^b$ & 1200 & {\it gr} & $g > 23.0$ &  $\leq 1.5$ & 0.263\\
    SMASH$^c$ & 480 & {\it ugriz} & $g \simeq 24.8$ & 1.13 & 0.263\\
    STEP & 53 & {\it griH$_\alpha$} & $g \simeq 25.0$ & 1.19 & 0.21\\
    VMC$^d$ & 184 & {\it YJK$_s$} & $Y \simeq 21.1$ & 1.03 & 0.339\\
    YMCA & 110 & {\it gi}  & $g \simeq 24.8$ & 1.13 & 0.21\\
    \hline
 \end{tabular}
 \tablefoot{The different columns show the name of the survey, the total portion of the sky observed in square degrees, the filters adopted, the $5\sigma$ limiting magnitude, the median seeing and the resolution of the pixel. To preserve the readability of the table, we indicated the depth and the seeing only for the deepest filter.\\
 References are: (a) \citet{Udalski-2015} (b) \citet{Drlica-Wagner-2016}; (c) https://datalab.noirlab.edu/smash/smash.php; (d) \citet{Cioni-2011}.}
\end{table}

\section{YMCA Observations}

\begin{figure*}[h!]
    \centering
    \includegraphics[width=.95\textwidth]{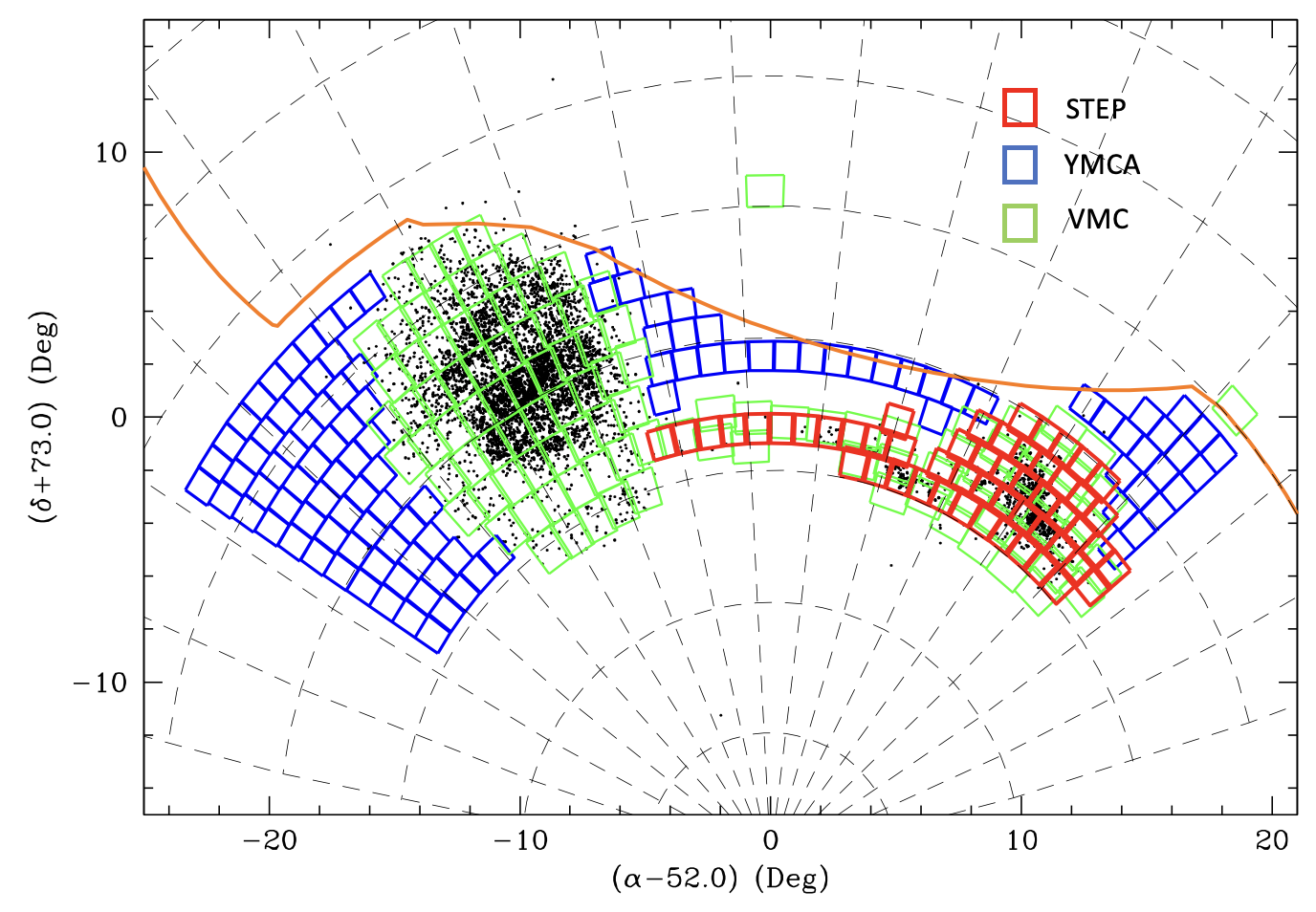}
    \caption{Footprint of the OmegaCam pointings of the YMCA survey (blue boxes) in a zenithal equidistant projection. For comparison, the footprints of the STEP (red boxes) and the VMC survey (green boxes) are also displayed in the figure. The orange solid line indicates the Southermost border of the Dark Energy Survey \citep[DES,][]{DES-Abbott-2016}, which observed 5000 square degrees of the sky. Black points mark the position of all objects presented by \citet{Bica-2008}.}
    \label{fig:ymca_footprint}
\end{figure*}

The YMCA survey has been carried out with the VLT Survey Telescope \citep[VST,][]{Capaccioli&Schipani2011}, a 2.6-meter telescope located at Cerro Paranal (Chile).
OmegaCam is a 32-CCD, 16k x 16k detector mosaic with a wide field of view \citep[FoV, 1 deg$^2$,][]{Kuijken2011} mounted on VST, it has a pixel scale of 0.214$\arcsec$, with thin gaps between the CCDs and an overall geometric filling factor of 91.4\%. 
The combination of a wide field of view (FOV) and good optical quality makes OmegaCam@VST the ideal instrument to homogeneously sample the stellar components of the MCs down to the faint magnitudes required to achieve the main goals of the survey.\par
Figure~\ref{fig:ymca_footprint} shows a footprint of YMCA in a zenithal equidistant projection along with some complementary optical and infrared surveys. As evident from the figure, most YMCA fields are centred on the LMC outer regions, while the remaining observed areas are in the north-western periphery of the SMC and in a strip in the sky linking the north-east side of the SMC with the westernmost fields of the LMC.
The observations were carried out between October 2016 and December 2020, except the field 4\_37 whose images were obtained in 2021 (see Table~\ref{tab:log_observation}). 
We adopted the {\it 5-point diag} dithering pattern, which shifts the telescope by $\pm 25\arcsec$ and $\pm 85\arcsec$ in the X and Y direction, respectively, between two consecutive exposures\footnote{The observation of tiles 3\_34, 3\_35, 4\_35, 4\_36, and 8\_44 occurred during a period characterized by the malfunction of two central CCDs. Our dithering procedure proved insufficient in compensating for these malfunctioning CCDs, resulting in a gap in the central regions of these tiles. To avoid the central gap, the field 4\_37 was observed with a different pattern.}.
Each {\it tile} (e.g. a single OmegaCam pointing) has been observed in a ``deep'' mode which consists of 10 long exposures of 180~s (200~s for the fields observed in 2016-2017) and 140~s (in $g$ and $i$ filters, respectively) and in a ``shallow'' mode, namely 5 exposures of 25~s (both in $g$ and $i$), to avoid saturation of the brightest MC stars. 
When the images are stacked, we achieve an average 50\% completeness level at $g \simeq 23.5 - 24.0$~mag (see Sect.~\ref{sec:completeness}), which is sufficient for the main goals of the survey.\par
\begin{figure}
    \centering
    \includegraphics[width=.5\textwidth]{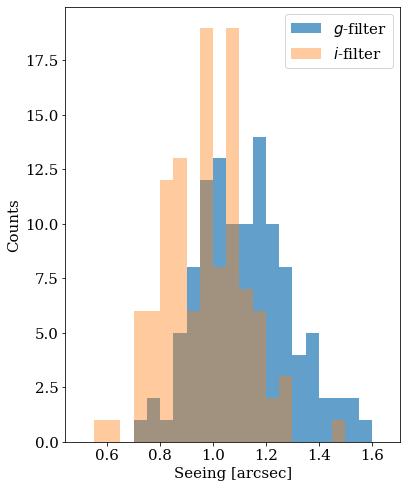}
    \caption{Histograms of the seeing, as reported in the last columns of Table~\ref{tab:log_observation}, for observations carried out in $g$ (blue) and $i$ (red) bands.}
    \label{fig:seeing_hist}
\end{figure}

Table~\ref{tab:log_observation} reports the essential observation log for all the 110 tiles.
Figure~\ref{fig:seeing_hist} illustrates the distribution of the seeing in both filters, showing that the seeing is in the range $\sim$0.7\arcsec-1.6\arcsec and $\sim$0.7\arcsec-$\sim$1.3\arcsec, while the median is 1.13\arcsec and 0.98\arcsec, for the observations made with $g$ and $i-$filters, respectively.

\begin{table*}[!ht]
   \caption{Log of Observations.}
   \label{tab:log_observation}
   \footnotesize\setlength{\tabcolsep}{5.5pt}
\begin{tabular}{lccccc|lccccc}
  \hline\hline   
  Tile & R.A. & Dec & Date & S$_g$ & S$_i$ & Tile & R.A. & Dec & Date & S$_g$ & S$_i$\\
      &  (hms) & (dms) &      & (\arcsec) & (\arcsec) & &  (hms) & (dms) &      & (\arcsec) & (\arcsec)\\
  \hline
      1\_27 & 06:23:46.392 & -73:59:14.640 & 2016-10-11 & 1.24 & 1.01 & 
    1\_28 & 06:38:14.976 & -73:59:14.640 & 2016-10-12 & 1.34 & 1.00\\
    1\_29 & 06:52:43.584 & -73:59:14.640 & 2016-10-12 & 1.51 & 0.96 &
    1\_30 & 07:07:12.192 & -73:59:14.640 & 2016-10-23 & 1.30 & 0.87\\
    1\_31 & 07:21:40.776 & -73:59:14.640 & 2016-11-19 & 1.34 & 0.82 &
    2\_30 & 06:41:13.272 & -72:53:03.480 & 2017-10-12 & 1.30 & 1.06\\
    2\_31 & 06:54:49.056 & -72:53:03.480 & 2017-10-13 & 1.17 & 1.17 &
    2\_32 & 07:08:24.864 & -72:53:03.480 & 2017-10-12 & 1.30 & 1.01\\
    2\_33 & 07:22:00.672 & -72:53:03.480 & 2016-12-18 & 1.05 & 0.73 &
    3\_1 & 00:06:30.648 & -71:46:52.680 & 2018-12-12 & 1.23 & 0.85\\ 
    3\_10 & 02:01:54.576 & -71:46:52.680 &  2018-12-15 & 1.15 & 0.88 &
    3\_21 & 04:22:57.144 & -71:46:52.680 & 2017-10-12 & 1.09 & 1.09\\
    3\_32 & 06:43:59.688 & -71:46:52.680 & 2020-01-04 & 1.16 & 1.07 &
    3\_33 & 06:56:49.032 & -71:46:52.680 & 2020-01-04 & 1.03 & 0.80\\
    3\_34 & 07:09:38.352 & -71:46:52.680 & 2020-12-13 & 0.92 & 0.86 &
    3\_35 & 07:22:27.672 & -71:46:52.680 & 2020-12-14 & 1.03 & 0.81\\ 
    4\_1 & 00:06:09.144 & -70:40:41.520 & 2018-06-07 & 1.24 & 1.09 &
    4\_2 & 00:18:17.256 & -70:40:41.520 & 2018-06-08 & 1.28 & 1.08\\
    4\_9 & 01:43:14.040 & -70:40:41.520 & 2018-12-14 & 1.24 & 0.83 &
    4\_10 & 01:55:22.152 & -70:40:41.520 & 2018-12-15 & 1.09 & 0.98\\
    4\_11 & 02:07:30.264 & -70:40:41.520 & 2018-12-15 & 1.01 & 0.88 & 
    4\_12 & 02:19:38.376 & -70:40:41.520 & 2018-12-17 & 1.17 & 1.09\\
    4\_13 & 02:31:46.488 & -70:40:41.520 & 2018-12-17 & 1.06 & 0.80 &
    4\_14 & 02:43:54.600 & -70:40:41.520 & 2019-08-21 & 0.98 & 1.06\\ 
    4\_15 & 02:56:02.712 & -70:40:41.520 & 2019-08-22 & 0.89 & 0.90 &
    4\_16 & 03:08:10.824 & -70:40:41.520 & 2019-08-23 & 0.90 & 1.08\\
    4\_17 & 03:20:18.936 & -70:40:41.520 & 2019-08-22 & 1.07 & 0.98 &
    4\_18 & 03:32:27.048 & -70:40:41.520 & 2019-08-22 & 1.04 & 0.97\\
    4\_19 & 03:44:35.160 & -70:40:41.520 & 2017-10-12 & 1.19 & 0.98 &
    4\_20 & 03:56:43.272 & -70:40:41.520 & 2017-10-13 & 1.13 & 0.99\\
    4\_21 & 04:08:51.384 & -70:40:41.520 & 2017-12-23 & 1.08 & 0.97 &
    4\_22 & 04:20:59.496 & -70:40:41.520 & 2017-12-25 & 1.12 & 0.99\\
    4\_33 & 06:34:28.728 & -70:40:41.520 & 2020-02-22 & 1.23 & 1.09  &
    4\_34 & 06:46:36.840 & -70:40:41.520 & 2020-02-12 & 1.02 & 1.08 \\
    4\_35 & 06:58:44.952 & -70:40:41.520 & 2020-12-15 & 0.96 & 0.77  &
    4\_36 & 07:10:53.064 & -70:40:41.520 & 2020-12-16 & 0.92 & 0.82 \\
    4\_37 & 07:23:01.176 & -70:40:41.520 & 2021-11-26 & 1.6 & 0.74 &
    5\_1 & 00:05:49.992 & -69:34:30.360 & 2018-09-16 & 1.48 & 0.93\\
    5\_2 & 00:17:21.312 & -69:34:30.360 & 2018-12-13 & 1.54 & 0.87 &
    5\_3 &00:28:52.656 & -69:34:30.360 & 2018-12-13 & 1.37 & 1.05\\
    5\_21 & 03:56:16.680 & -69:34:30.360 & 2018-12-12 & 1.38 & 1.07  &
    5\_22 & 04:07:48.024 & -69:34:30.360 & 2017-12-26 & 1.13 & 1.08\\
    5\_23 & 04:19:19.368 & -69:34:30.360 & 2018-01-11 & 1.25 & 1.09 &
    5\_36 & 06:49:06.720 & -69:34:30.360 & 2020-02-27 & 1.05 & 1.17 \\
    5\_37 & 07:00:38.040 & -69:34:30.360 & 2020-02-27 & 0.93 & 1.04  &
    5\_38 & 07:12:09.384 & -69:34:30.360 & 2020-02-29 & 1.46 & 1.13 \\
    5\_39 & 07:23:40.728 & -69:34:30.360 & 2020-03-02 & 1.44 & 1.14  &
    6\_1 & 00:05:32.832 & -68:28:19.560  & 2017-11-20 & 1.13 & 0.87\\  
    6\_2 & 00:16:31.176 & -68:28:19.560 & 2017-11-20 & 1.12 & 0.90 &
    6\_3 & 00:27:29.496 & -68:28:19.560 & 2018-12-14 & 1.26 & 1.11\\  
    6\_4 & 00:38:27.816 & -68:28:19.560 & 2018-12-14 & 1.10 & 1.06 & 
    6\_5 & 00:49:26.136 & -68:28:19.560 & 2018-12-14 & 1.16 & 1.07\\ 
    6\_6 & 01:00:24.480 & -68:28:19.560 & 2018-12-15 & 1.09 & 0.93 &
    6\_23 & 04:06:56.040 & -68:28:19.560 & 2018-12-12 & 1.19 & 0.90 \\
    6\_24 & 04:17:54.360 & -68:28:19.560 & 2018-12-13 & 0.99 & 0.81  &
    6\_37 & 06:40:32.616 & -68:28:19.560 & 2020-01-04 & 1.22 & 1.27 \\
    6\_38 & 06:51:30.960 & -68:28:19.560 & 2020-01-03 & 1.31 & 0.97  &
    6\_39 & 07:02:29.280 & -68:28:19.560 & 2020-01-03 & 1.00 & 0.95 \\
    6\_40 & 07:13:27.600 & -68:28:19.560 & 2020-01-03& 0.87 & 1.16  &
    6\_41 & 07:24:25.944 & -68:28:19.560 & 2020-01-02 & 1.12 & 0.86 \\
    7\_1 & 00:05:17.400 & -67:22:08.400& 2017-11-23 & 1.27 & 1.02 &
    7\_2 & 00:15:45.960 & -67:22:08.400 & 2017-11-20 & 1.33 & 0.96\\
    7\_3 & 00:26:14.496 & -67:22:08.400 & 2017-11-20 & 1.27 & 1.25 &
    7\_4 & 00:36:43.056 & -67:22:08.400 & 2017-11-23 & 1.37 & 1.13\\
    7\_26 & 04:27:11.160 & -67:22:08.400 & 2018-12-12 & 1.14 & 1.10  &
    7\_27 & 04:37:39.720 & -67:22:08.400 & 2018-12-14 & 1.45 & 1.11  \\
    7\_38 & 06:32:53.784 & -67:22:08.400 & 2020-01-03 & 1.22 & 1.08  &
    7\_39 & 06:43:22.320 & -67:22:08.400 & 2020-01-01 & 1.23 & 0.98 \\
    7\_40 & 06:53:50.880 & -67:22:08.400 & 2020-01-01 & 1.14 & 0.97  &
    7\_41 & 07:04:19.416 & -67:22:08.400 & 2020-01-01 & 1.04 & 1.50 \\
    7\_42 & 07:14:47.976 & -67:22:08.400 & 2020-01-02 & 1.38 & 1.06  &
    7\_43 & 07:25:16.536 & -67:22:08.400 & 2020-01-02 & 1.03 & 0.95 \\
    8\_1 & 00:05:03.456 & -66:15:57.600 & 2017-11-22 & 1.30 & 1.18 &
    8\_2 & 00:15:05.016 & -66:15:57.600 & 2017-11-22 & 1.40 & 1.02\\
    8\_3 & 00:25:06.576 & -66:15:57.600 & 2017-11-28 & 1.20 & 1.21 &
    8\_28 & 04 35:45.696 & -66:15:57.600 & 2018-12-15 & 1.20 & 0.89 \\
    8\_39 & 06:26:02.880 & -66:15:57.600 & 2020-01-01 & 1.22 & 1.03  &
    8\_40 & 06:36:04.464 & -66:15:57.600 & 2020-01-03 & 0.95 & 1.11 \\
    8\_41 & 06:46:06.024 & -66:15:57.600 & 2019-12-31 & 0.96 & 0.76  &
    8\_42 & 06:56:07.584 & -66:15:57.600 & 2019-12-31 & 0.92 & 0.71 \\
    8\_43 & 07:06:09.144 & -66:15:57.600 & 2019-12-31 & 0.86 & 0.88  &
    8\_44 & 07:16:10.704 & -66:15:57.600 & 2019-12-31 & 1.02 & 0.96 \\
    8\_45 & 07:26:12.288 & -66:15:57.600 & 2020-01-01 & 0.94 & 0.93  &
    9\_40 & 06:19:53.976 & -65:09:46.440 & 2019-12-29 & 1.17 & 0.99 \\
    9\_41 & 06:29:30.984 & -65:09:46.440 & 2019-12-30 & 1.00 & 0.93  &
    9\_42 & 06:39:07.992 & -65:09:46.440 & 2019-12-31 & 0.76 & 1.08 \\
    9\_43 & 06:48:45.000 & -65:09:46.440 & 2019-12-29 & 0.97 & 1.00  &
    9\_44 & 06:58:21.984 & -65:09:46.440 & 2019-12-30 & 0.92 & 0.79 \\
    9\_45 & 07:07:58.992 & -65:09:46.440 & 2019-12-30 & 0.84 & 0.82  &
    9\_46 & 07:17:36.000 & -65:09:46.440 & 2019-12-30 & 0.98 & 0.83 \\
    9\_47 & 07:27:13.008 & -65:09:46.440 & 2019-12-30 & 0.90 & 1.00  &
    10\_42 & 06:23:36.600 & -64:03:35.640 & 2018-12-17 & 0.75 & 0.62 \\
    10\_43 & 06:32:51.168 & -64:03:35.640 & 2018-12-17 & 1.01 & 1.16  & 
    10\_44 & 06:42:05.736 & -64:03:35.640 & 2019-04-10 & 1.06 & 1.18 \\
    10\_45 & 06:51:20.304 & -64:03:35.640 & 2019-04-10 & 0.97 & 1.03  &
    10\_46 & 07:00:34.872 & -64:03:35.640 & 2019-04-11 & 1.20 & 1.14 \\
    10\_47 & 07:09:49.440 & -64:03:35.640 & 2019-12-29 & 1.16 & 0.85  &
    10\_48 & 07:19:04.008 & -64:03:35.640 & 2019-12-29 & 1.08 & 1.00 \\
    10\_49 & 07:28:18.576 & -64:03:35.640 & 2019-12-29 & 1.08 & 0.81  &
    11\_41 & 06:00:28.848 & -62:57:24.480 & 2017-12-09 & 0.98 & 0.96\\
    11\_42 & 06:09:22.848 & -62:57:24.480 & 2018-01-08 & 1.15 & 0.77 &
    11\_43 & 06:18:16.848 & -62:57:24.480 & 2018-01-14 & 1.05 & 0.72\\
    11\_44 & 06:27:10.848 & -62:57:24.480 & 2018-01-19 & 0.93 & 0.85 &
    11\_45 & 06:36:04.872 & -62:57:24.480 & 2018-01-19 & 0.97 & 0.82\\
    11\_46 & 06:44:58.872 & -62:57:24.480 & 2018-12-12 & 1.02 & 0.87  &
    11\_47 & 06:53:52.872 & -62:57:24.480 & 2018-12-13 & 1.20 & 1.26 \\
    11\_48 & 07:02:46.872 & -62:57:24.480 & 2018-12-14 & 1.20 & 1.27  &
    11\_49 & 07:11:40.896 & -62:57:24.480 & 2018-12-15 & 1.16 & 0.74 \\
    11\_50 & 07:20:34.896 & -62:57:24.480 & 2018-12-16& 0.96 & 0.71  &
    11\_51 & 07:29:28.896 & -62:57:24.480 & 2018-12-16 & 0.80 & 0.59\\
    \hline
 \end{tabular}
 \tablefoot{The different columns show the name of the tile, its centre, date of observation, and average FWHM over the images (S$_g$ and S$_i$).}
\end{table*}

\section{Data reduction}
\label{sec:psf_photometry}

In this section we describe the procedures adopted to i) remove the instrumental signature from the raw images; ii) obtain the astrometric calibration; iii) carry out the Point Spread Functions (PSF) photometry; iv) calibrate our photometry to the absolute system.

\subsection{Pre-reduction, astrometry and PSF photometry}
\label{sec:pre-reduction}

The pre-reduction, astrometry and stacking of all dithered images to create a single frame were performed by the {\tt VST-TUBE} imaging pipeline \citep[][but see also the Appendix in \citealt{Capaccioli-2015}]{Grado-2012} for all fields observed until the ESO Period 102, namely the first 21 tiles observed between October 2016 and January 2019.
The {\tt VST-TUBE} pipeline removes bias, performs flat-field correction, applies CCD gain equalization and applies an illumination correction. Once the pre-reduction procedure was completed, it returned two images in both the $g$ and $i$ filters, one corresponding to the stack of the ``shallow'' and the other to the ``deep'' observations. More details can be found in \citet{Grado-2012} and \citet{Ripepi-2014}.
For the remaining tiles, collected from the ESO Period 103 to 105 (April 2019-March 2020), these tasks have been performed with {\tt ASTROWISE} pipeline \citep[][]{McFarland-2013}, since the {\tt VST-TUBE} package was not available anymore. \par
Then, for all 110 YMCA tiles, we made use of the standard DAOPHOT IV/ALLSTAR \citep[][]{Stetson1987,Stetson1992} packages to carry out the PSF photometry for each stacked shallow and deep image in both filters. To derive the PSF, we adopted an automated pipeline based on DAOPHOT IV/ALLSTAR, SExtractor \citep{SExtractor}, and other routines written in Fortran.
Once we obtained the photometry in each tile and each filter, we adopted the {\tt STILTS} package \citep{STILTS} for merging the catalogues resulting from the shallow and deep images to obtain a unique catalogue of stars in the two separate filters. 
Finally, we cross-matched the $g$ and $i$ catalogues to get the final photometric catalogue of each tile.

\subsection{Absolute calibration}

The absolute photometric calibration has been carried out employing the local standard stars provided by the ``AAVSO Photometric All-Sky Survey'' (APASS\footnote{https://www.aavso.org/apass}).
For each tile, we carried out the following steps:
\begin{itemize}
    \item We cross-matched the photometric catalogue obtained as described in Sect.~\ref{sec:pre-reduction} with the APASS data release 10 by adopting a search radius of 0.5\arcsec, found empirically \citep[see][]{Ripepi-2014,Gatto-2020}, to mitigate the number of wrong matches (the APASS instrument's pixel-size is 2.57\arcsec). We also retained only APASS observations with signal-to-noise (S/N) ratio larger than 10, namely with a photometric uncertainty lower than 0.1 mag.
    \item We searched for and corrected any residual spatial variation of the photometric zero points. To this aim, we inspected the photometric zero point difference between our photometry and the APASS one as a function of the RA and Dec. 
    \item We corrected the colour dependence of the zero points in $g$ and $i$ filters. 
\end{itemize}
  At the end of this procedure, we achieved an average precision (1-sigma errors) of the order of 0.02 and 0.03 mag in $g$ and $i$ bands, respectively \citep[see also][]{Gatto-2020}.
To validate our photometry we retrieved standardised synthetic photometry in the SDSS system from the Gaia Synthetic Photometry Catalogue \citep[GSPC][]{Gaia_synthphot} for about 300,000 stars in common with YMCA. The median ($\pm$ standard deviation) differences in the two passbands are $g_{\rm YMCA} - g_{\rm Gaia} = -0.011 \pm 0.025$~mag e $i_{\rm YMCA} - i_{\rm Gaia} = -0.013 \pm 0.024$~mag, and no perceivable colour trend, confirming that YMCA photometric zero points are very accurate.

\subsection{Creation of the photometric catalogue}
\label{sec:final_catalog}

The final step to obtain a ready-to-use catalogue is concatenating all sub-catalogues obtained, as described in the previous sections, for each single tile. Since the dithering pattern adopted makes adjacent YMCA tiles slightly overlap at their borders, we made use of the {\tt STILTS} package \citep{STILTS} to find common stars between adjacent tiles adopting a tolerance radius of 0.25\arcsec, and get rid of duplicates. We opted for using a weighted mean of the photometry for sources in common between two or more tiles, and a simple mean for the {\it SHARPNESS} and {\it CHI} output parameters  of the DAOPHOT IV/ALLSTAR \citep[][]{Stetson1987,Stetson1992} packages. We kept in the catalogue also stars located within the overlapping regions and detected only in one tile. 
At the end of this procedure we obtained the final YMCA catalogue, made by 9,838,499 sources spanning 110 square degrees. 
Table~\ref{tab:example_cat} displays the first rows of the final YMCA catalogue, that provides, for each source, the coordinates, the calibrated magnitudes in both \emph{g} and \emph{i} filters, and finally the {\it CHI} and {\it SHARPNESS} parameters.\par 
This catalogue still includes spurious detections due to bad pixels and extended sources. 
As these sources would produce undesired features in the CMD, we applied a cut using the {\it SHARPNESS} parameter. In particular, to remove also extended objects, and spurious sources near the edge of the tiles, we filtered out all sources beyond the interval -1 $\leq$ {\it SHARPNESS} $\leq$ 1\footnote{The tiles 4\_17 ad 5\_39 represent an exception to this cut.}. Throughout this work, any plot and/or scientific exploitation of YMCA data will refer to this point-source sub-sample made by a total of 6,467,819 stars. We also provide the scientific community with this cleaned catalogue.\footnote{Readers interested in the whole catalogue with no cuts can contact the first author.}

\begin{table}
\footnotesize\setlength{\tabcolsep}{3pt}
 \caption{Main photometric catalogue of the YMCA survey.}
 \label{tab:example_cat}
 \begin{tabular}{cccccccc}
  \hline\hline
  RA & Dec & $g$ & $\Delta_g$ & $i$ & $\Delta_i$ & {\it CHI} & {\it SHARP} \\
   (J2000) & (J2000) & (mag) & (mag) & (mag) & (mag) & & \\
  \hline
359.92552 & -72.29845 & 24.03 & 0.32 & 23.08 & 0.33 & 0.98 & -0.34 \\
359.92622 & -72.30697 & 22.93 & 0.12 & 22.14 & 0.17 & 0.78 & 0.91 \\
359.92649 & -72.28756 & 23.77 & 0.19 & 22.85 & 0.33 & 0.75 & 0.11 \\
359.92701 & -72.28275 & 24.48 & 0.38 & 23.21 & 0.31 & 0.76 & 0.73 \\
359.92882 & -72.29437 & 23.88 & 0.19 & 23.31 & 0.38 & 0.70 & -0.46 \\
359.92890 & -72.26670 & 21.98 & 0.08 & 20.30 & 0.04 & 0.98 & -0.88 \\
359.92893 & -72.27306 & 23.90 & 0.29 & 23.04 & 0.28 & 1.03 & -0.21 \\
359.92894 & -72.28692 & 22.94 & 0.10 & 20.46 & 0.05 & 0.78 & 0.20 \\
... & ... & ... & ... & ... & ... & ... & ... \\
    \hline
 \end{tabular}
 \tablefoot{The different columns show the coordinates of the source, the calibrated magnitudes in both \emph{g} and \emph{i} filters and their uncertainties, the {\it CHI} and {\it SHARPNESS} parameters. \\
A portion is shown here for guidance regarding its form and content. The machine-readable version of the full table will be published at the CDS (Centre de Données astronomiques de Strasbourg, https://cds.u-strasbg.fr/).}
\end{table}



\subsection{Completeness}
\label{sec:completeness}

To derive the completeness of the survey, we followed the standard procedure by adding artificial star to the images and measuring the recovery fraction of stars as a function of magnitude.
We performed this task over the entire strip defined by the tiles 7\_26, 7\_27, 7\_38, 7\_39, 7\_40, 7\_41, 7\_42, 7\_43, which surround the LMC, to probe our photometric depth at different radial distances from the LMC centre (different crowding) and/or observing conditions (e.g. seeing).\par
We applied the same procedure on the 8 investigated tiles. Main steps are the following:
(i) Random selection of stars evenly distributed in a grid CMD with a resolution of 0.01 mag on both axes and whose intervals are: $-1.0 \leq g-i \leq 3.0$ mag and $17.5 \leq g \leq 26.0$ mag;
(ii) Sub-division of the whole tile into boxes of 10\arcsec~per side, adding only one star per box to avoid self-crowding;
(iii) PSF pipeline was carried out on the images containing the artificial stars, exactly as described in Sect.~\ref{sec:psf_photometry};
(iv) Points (i), (ii), and (iii) were repeated over the same tile until the number of artificial stars was tenfold the actual number of stars detected in the tile; (v) A star was considered recovered if it is detected in both $g$ and $i$ bands, its position and magnitude estimates are within 0.25\arcsec~and 0.75 mag, respectively, from their input values, and its {\it SHARPNESS} parameter is within the same range indicated in Sect.~\ref{sec:final_catalog}.\par
\begin{figure*}
    \centering
    \includegraphics[width=\textwidth]{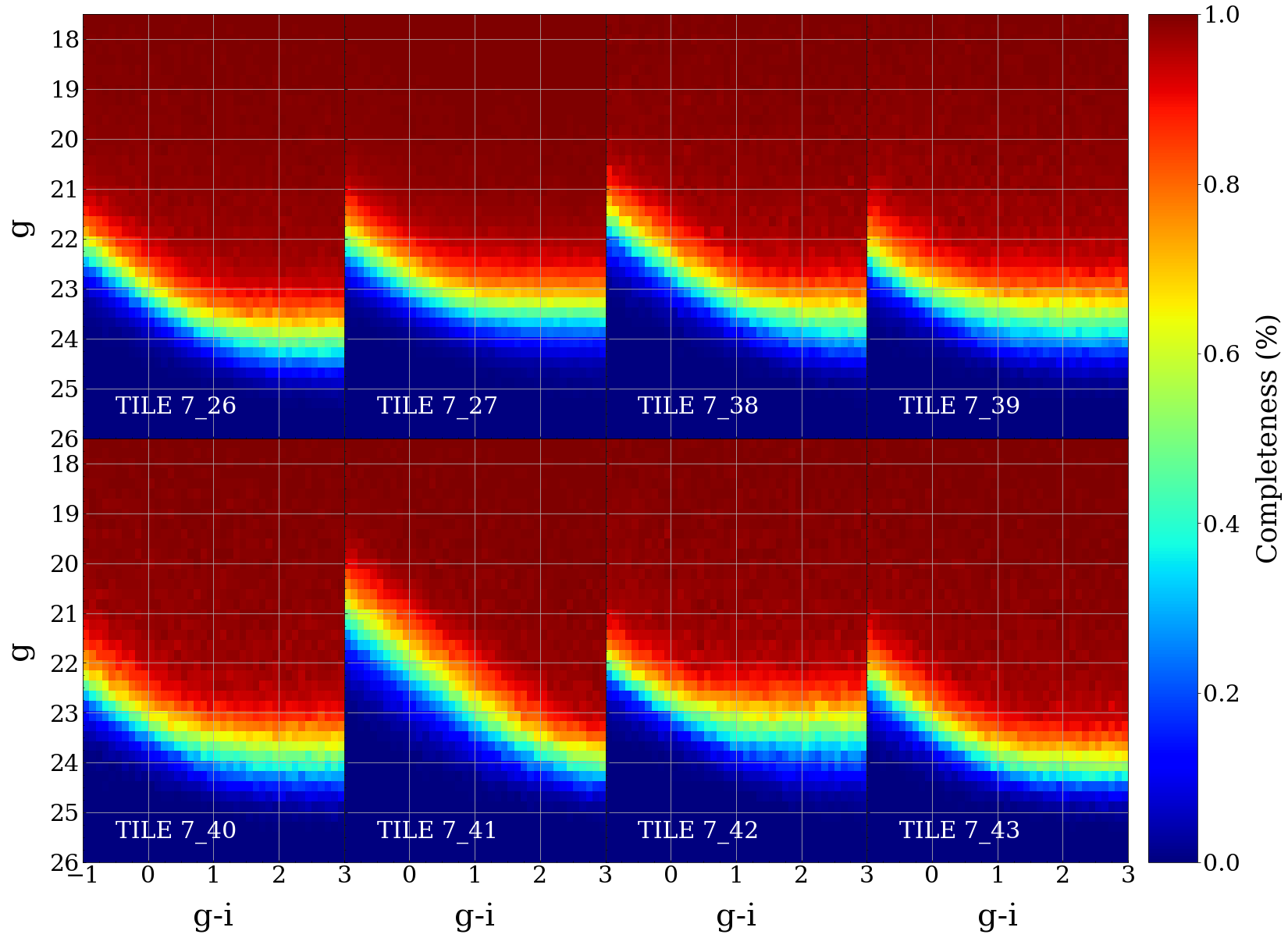}
    \caption{Completeness values in the CMD estimated in bins of 0.01 mag $\times$ 0.02 mag. Each subplot shows the measured completeness in different tiles, as indicated in the bottom left corner of each panel.}
    \label{fig:artificials}
\end{figure*}
Figure~\ref{fig:artificials} displays the recovery fraction of artificial stars in each bin of the CMD for the eight probed tiles. 
The figure shows that, if we consider the colour interval of the faint MS (e.g. $0 \leq g - i \leq 1$~mag), the 50\% completeness level is achieved on average between $g \simeq 23 - 23.5$~mag in the tiles 7\_27 and 7\_38 and $g \sim 23.5 - 24$~mag in the tiles 7\_40 and 7\_43. The only exception is the tile 7\_41, whose 50\% completeness level is reached, in the same colour intervals, at $g \simeq 22 - 23$~mag. We notice a strong correlation between the recovery fraction of artificial stars and the seeing (see also Table~\ref{tab:log_observation}) rather than tiles crowding, as the tiles showing the deepest completeness levels are those with the best seeing, regardless of their distance from the LMC centre. 
Therefore, we assume that completeness levels for the remaining YMCA tiles might, in principle, be similar to those tiles observed in similar atmospheric conditions.\par
\begin{figure}
    \centering
    \includegraphics[width=0.48\textwidth]{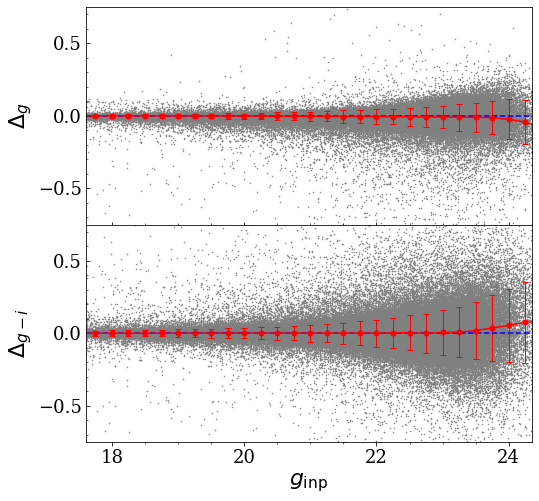}
    \caption{Magnitude difference (\emph{Top}) and colour difference (\emph{Bottom}), as a function of the input $g$-magnitude, between the injected and recovered artificial stars in the tile 7\_26. We displayed only stars in the colour interval $0 \leq g_{\rm inp} - i_{\rm inp} \leq 1$, which is typical for MC stars. Grey points show the position in these panels of each recovered artificial star, selected in the above-mentioned colour interval. Red dots and error bars indicate the mean magnitude/colour difference and their standard deviations in bins of $g_{\rm inp}$=0.25 mag.}
    \label{fig:accuracy}
\end{figure}
Figure~\ref{fig:accuracy} shows the magnitude and colour difference between each inserted and recovered artificial star that passed the selection criteria described above in the tile 7\_26, to estimate the photometric precision of the survey. Moreover, in this plot, we only included stars that had a colour input in the range $0 \leq g_{\rm inp} - i_{\rm inp} \leq 1$, that embraces all MC stars in the MS, red clump (RC), and most of the RGB evolutionary phases, adopting the typical metallicities of the LMC and SMC old ($t \geq 10$ Gyr) stellar populations.
The average magnitude offset for this tile measured at $g_{\rm inp} = 22.5$~mag, which is the magnitude of the MSTO of the oldest stellar population of the LMC, is $\Delta_g = -0.01 \pm 0.06$~mag, while the colour difference at the same input magnitude is $\Delta_{g - i} = 0.00 \pm 0.12$~mag. 
In particular, the standard dispersion of the magnitude and colour offsets can be used as a proxy for the statistical errors of the photometric data, and they appear to be quite small even at the faint magnitudes of the MSTO of the oldest LMC stellar population.\par
We examined the photometric precision also for the other seven tiles for which we performed the artificial star tests, and listed in Table~\ref{tab:phot_errors} the average $\Delta_g$, and $\Delta_{g - i}$ values and their standard deviations at the $g-$magnitude of the MSTO. 
Most of the tiles exhibit a similar level of photometric precision, with two main exceptions, namely a higher $\delta_g$ within the tile 7\_27 and a higher $\delta_{g - i}$ within the tile 7\_41. Looking at Table~\ref{tab:log_observation}, and at Figure~\ref{fig:artificials}, we speculate that the high seeing (1.45\arcsec~and 1.5\arcsec~in $g$ and $i$ for the tiles 7\_27 and 7\_41, respectively) may have affected the average accuracy and precision within these two tiles. Based on these considerations, we conclude that the YMCA survey achieves a precision of $\delta_g \sim$ 0.06 mag and $\delta_{g - i} \sim$ 0.12 mag at the MSTO magnitudes of the oldest LMC stellar population with the exceptions of tiles observed in very bad atmospheric conditions, namely with seeing higher than 1.4-1.5\arcsec.
Table~\ref{tab:phot_errors} also contains the average magnitude offset at the $g-$magnitude of the MSTO of oldest SMC stellar population, namely $g \sim$ 23 mag for each of the eight investigated tiles. The standard deviations of $\Delta_g \sim$ and $\Delta_{g - i}$ are only slightly larger than $g_{\rm inp} = $ 22.5 mag, despite the lower magnitude. In particular, the YMCA survey achieves a precision of $\delta_g \sim$ 0.08 mag and $\delta_{g - i} \sim$ 0.14 mag at the MSTO magnitudes of the oldest SMC stellar population.



\begin{table}
\centering
 
 \caption{Artificial star tests.}
 \label{tab:phot_errors}
 \begin{tabular}{lccc}
  \hline\hline
  Tile & $g_{\rm inp} $ & $\Delta_g$ & $\Delta_{g - i}$\\
  \hline
    7\_26 & 22.5 & -0.01 $\pm$ 0.06 & 0.00 $\pm$ 0.12 \\
    7\_26 & 23.0 & -0.01 $\pm$ 0.08 & 0.01 $\pm$ 0.15\\
    7\_27 & 22.5 & 0.09 $\pm$ 0.15 & 0.02 $\pm$ 0.21\\
    7\_27 & 23.0 & 0.12 $\pm$ 0.19 & 0.03 $\pm$ 0.26\\
    7\_38 & 22.5 & 0.00 $\pm$ 0.07 & 0.01 $\pm$ 0.16\\
    7\_38 & 23.0 & -0.01 $\pm$ 0.08 & 0.04 $\pm$ 0.20\\
    7\_39 & 22.5 & 0.00 $\pm$ 0.06 & 0.00 $\pm$ 0.11\\
    7\_39 & 23.0 & -0.01 $\pm$ 0.08 & 0.01 $\pm$ 0.14\\
    7\_40 & 22.5 & 0.00 $\pm$ 0.05 & 0.01 $\pm$ 0.10\\
    7\_40 & 23.0 & 0.00 $\pm$ 0.07 & 0.01 $\pm$ 0.13\\
    7\_41 & 22.5 & 0.00 $\pm$ 0.05 & 0.04 $\pm$ 0.17\\
    7\_41 & 23.0 & 0.00 $\pm$ 0.06 & 0.09 $\pm$ 0.24\\
    7\_42 & 22.5 & 0.00 $\pm$ 0.05 & 0.00 $\pm$ 0.11\\
    7\_42 & 23.0 & 0.00 $\pm$ 0.08 & 0.01 $\pm$ 0.15\\
    7\_43 & 22.5 & 0.00 $\pm$ 0.05 & 0.00 $\pm$ 0.10\\
    7\_43 & 23.0 & 0.00 $\pm$ 0.06 & 0.01 $\pm$ 0.14\\
    
    \hline
 \end{tabular}
 \tablefoot{$\Delta_g$ and $\Delta_{g - i}$ are the average magnitude and colour differences, at a fixed magnitude, between the recovered and injected artificial stars, measured in eight different representative tiles. The two fixed investigated input magnitudes (i.e. $g_{\rm inp} = 22.5 - 23.0$ mag) represent the luminosities of the MSTO of the LMC and SMC oldest stellar populations, respectively.}
\end{table}


\section{Stellar populations analysis through the CMD}

In this section, we describe the CMDs of the different regions surveyed by YMCA. In particular, we split the entire survey into three main sub-regions, namely the east side of the LMC, the LMC west side along the bridge that connects the two galaxies and, finally, the northwest of the SMC (see Fig.~\ref{fig:ymca_footprint}).
All CMDs presented in this section are reddening-corrected through the reddening maps by \citet{Schlegel-1998}, re-calibrated by \citet{Schlafly&Finkbeiner2011}, and retrieved using the python package {\tt DUSTMAPS} \citep[][]{Dustmaps}\footnote{We adopted the extinction coefficients $R_g$ = 3.303 and $R_i$ = 1.698, as provided in Table~6 in \citet{Schlafly&Finkbeiner2011}.}. For the subsequent CMD analysis, throughout this work, we adopt the {\tt PARSEC} suite of isochrones\footnote{http://stev.oapd.inaf.it/cgi-bin/cmd} \citep[][]{Bressan-2012} at varying ages and metallicities.

\subsection{The east side of the LMC}

\begin{figure}
    \centering
    \includegraphics[width=.49\textwidth]{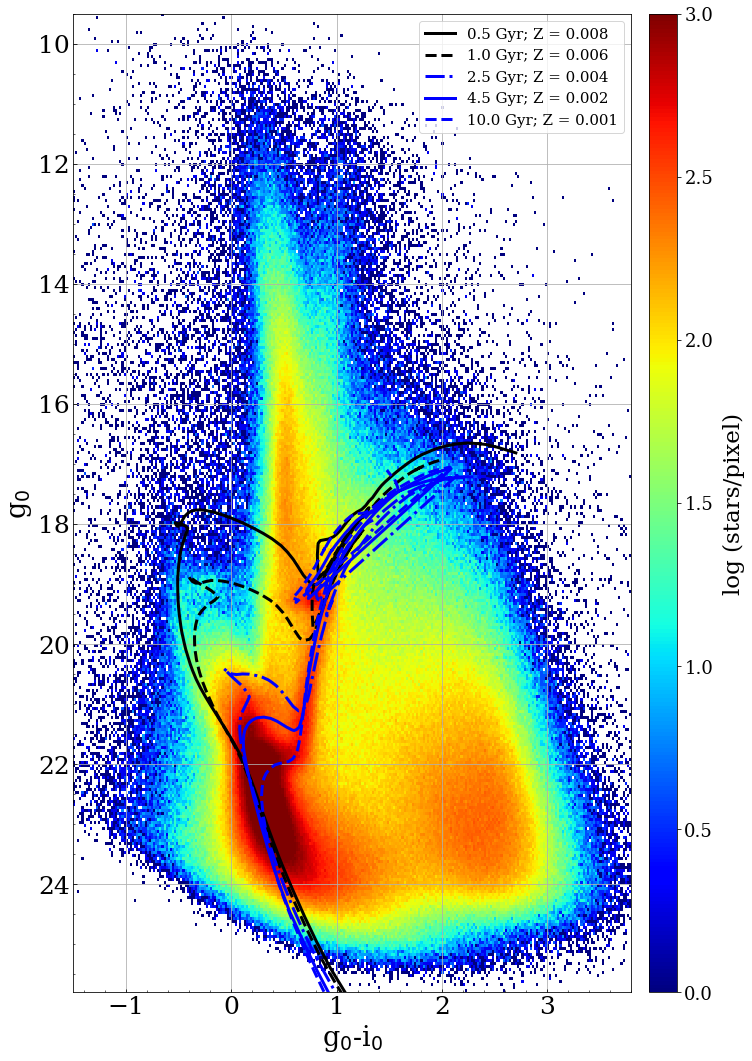}
    \caption{Hess diagram towards the eastern direction of the LMC. Each pixel is $0.02 \times 0.05$ mag in size. Isochrones with different ages and metallicities, listed in the top right corner of the plot, are also superposed on the Hess diagram.}
    \label{fig:cmd_LMC_east_total}
\end{figure}

Figure~\ref{fig:cmd_LMC_east_total} depicts the Hess diagram of the LMC towards its eastern direction, including about 4.2 million stars located within a sky region of $\sim$67 deg$^{2}$. In the same figure, we also drew different isochrones spanning a range of ages from 500 Myr up to 10 Gyr, corrected for the average distance modulus of the LMC \citep[i.e. $\rm{DM} = 18.48$~mag,][]{Pietrzynski-2019}.
\begin{figure*}[h!]
    \centering
    \includegraphics[scale=0.48]{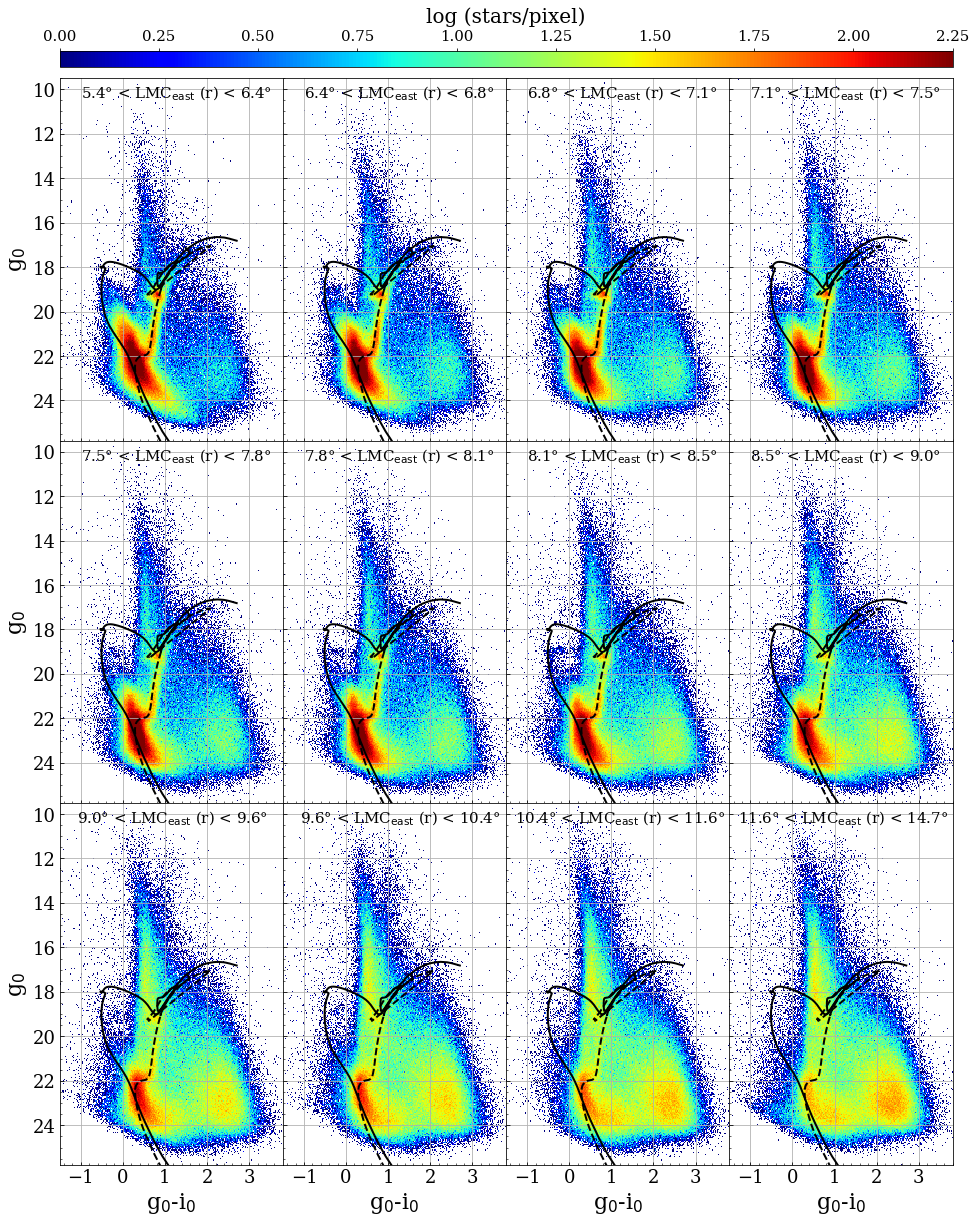}
    \caption{Hess diagrams of the LMC east side built at increasing distances from the LMC centre. Each subpanel shows about 300,000 stars. The isochrones are representative of a young and metal-richer (t = 500 Myr; Z = 0.006 dex, solid line) and old, metal-poorer (t = 10 Gyr, Z = 0.001 dex, dashed line) stellar population, corrected for the distance modulus of the LMC. At the top of each subplot, the range of angular distances from the LMC centre is indicated.}
    \label{fig:cmd_LMC_east_subplots}
\end{figure*}
\begin{figure*}[h!]
    \centering
    \includegraphics[width=.48\textwidth]{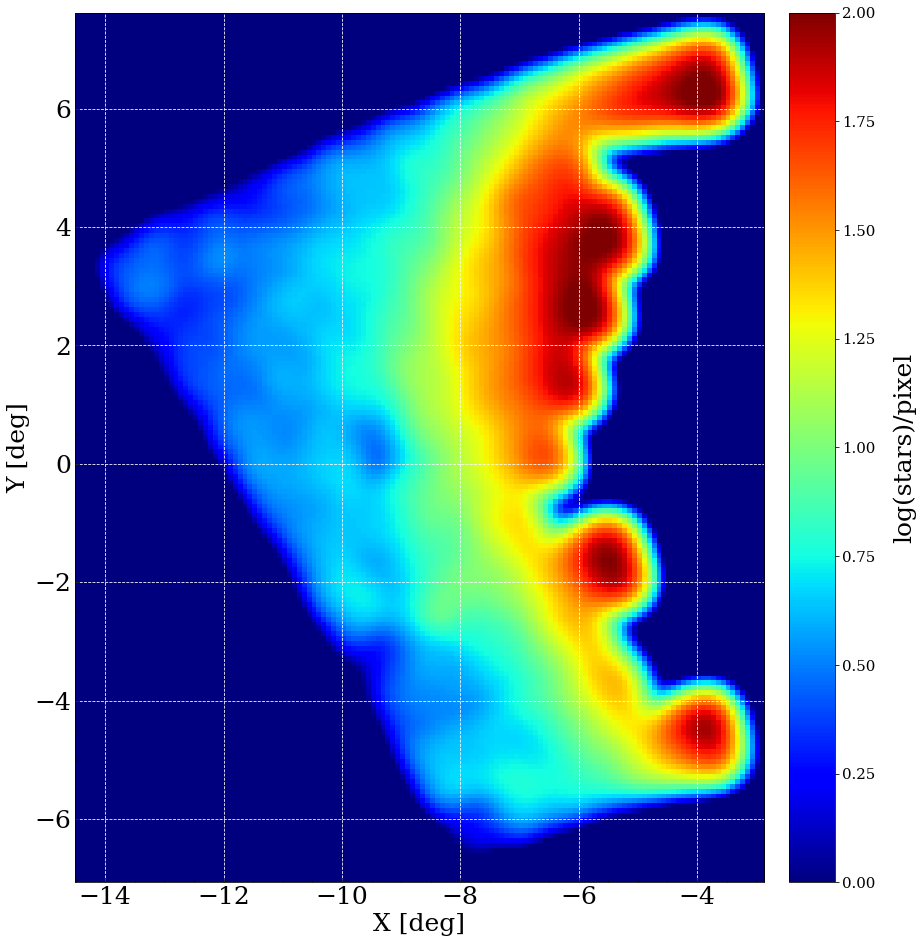}
    \includegraphics[width=.48\textwidth]{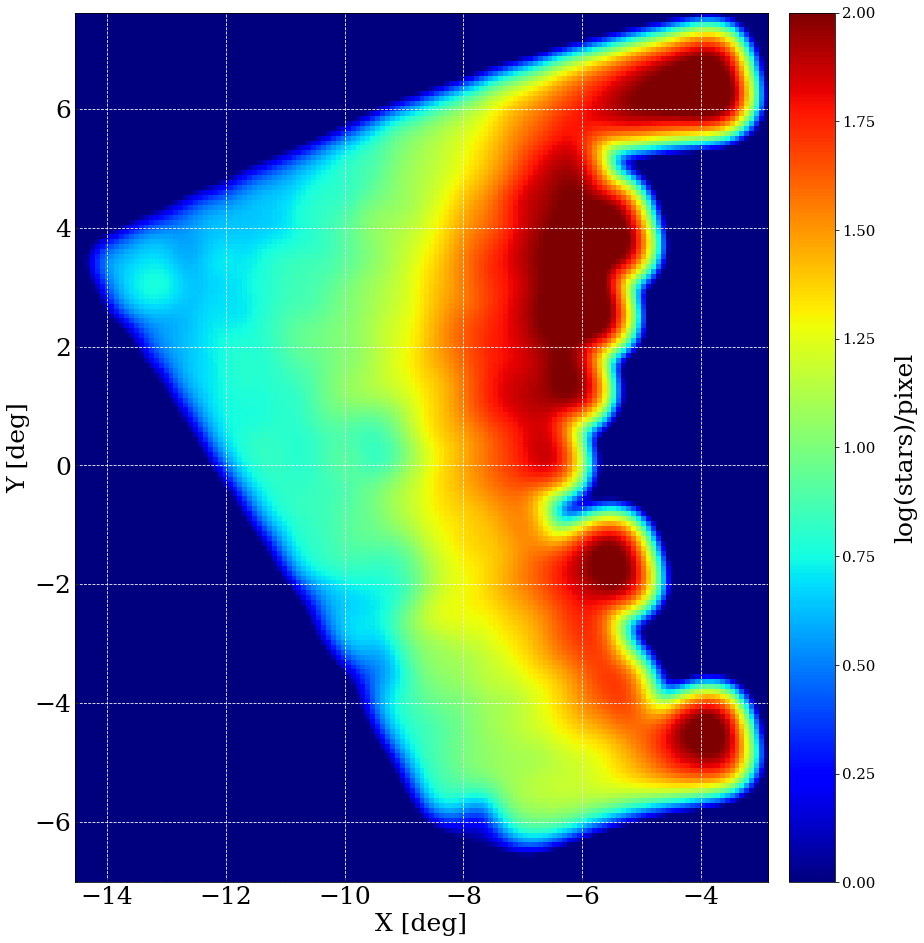} \caption{Density map of the LMC East periphery, constructed by adopting only MSTO stars of an intermediate-age ($7 \leq t \leq 9$~Gyr, left panel) and old ($t \geq 10$~Gyr, right panel) stellar populations. Specifically, we picked up stars within the colour and magnitude range of $0 \leq g - i \leq 0.5$~mag and $21.4 \leq g \leq 21.9$~mag for the younger stellar population, and $0.1 \leq g - i \leq 0.6$~mag and $21.9 \leq g \leq 22.4$~mag, for the older one, respectively. Each pixel has size of $5\arcmin \times 5\arcmin$ and we adopted a Gaussian Kernel Density Estimator with bandwidth = 0.25\degr~to smooth the plot.}
    \label{fig:density_map_LMC_east}
\end{figure*}
The diagram shows that the LMC periphery is made of a mix of stars of different ages, although it is mostly dominated by intermediate-age and old stellar populations, marked by a well-defined red-giant branch (RGB) and red clump (RC).
The blue plume (at $g_0-i_0 \simeq -0.5$~mag and $18 \leq g_0 \leq 19$~mag) suggests the presence of a younger stellar population formed possibly about 500 Myr ago, whereas a significant recent star formation activity seems missing. 
Some regions of the CMD also display a severe contamination by MW foreground or halo stars. In particular, the vertical sequence at $g_0-i_0 \simeq 0.6-0.8$~mag and $g_0 \leq$19 mag, whose faint end intersects the RC evolutionary phase, is likely dominated by stars belonging to the MW thick disk, whereas the blob of stars in the Hess diagram visible between 1.5 $\leq g_0-i_0$ mag $\leq$ 3 and $g_0 >$ 20.5 mag likely represents a mix of halo and disk MW stars \citep[see Figure 13 in][for the expected MW contamination into the CMD]{Ripepi-2014}.\par
To examine the spatial distribution of the different stellar populations, Fig.~\ref{fig:cmd_LMC_east_subplots} displays the same Hess diagram divided into 12 subplots, each one containing stars in circular annuli at increasing distances from the LMC centre\footnote{We set the LMC centre at $(\alpha, \delta) = (81.24\degr,-69.73\degr$) as derived by \citet[][]{ElYoussoufi-2021}.}; built in order to have the same number of stars (i.e. about 350,000) in each galactocentric interval. 
Two isochrones, representative of a young and an old stellar population are superposed onto each Hess diagram for comparison.
As evident from the figure, the LMC eastern tiles cover a range of angular distances between 5.4\degr~and about 15\degr~(i.e. between $\sim 5$ and $\sim 13$ kpc at the LMC distance).\par
A close inspection of the single subplots, starting from the top left panel (i.e. the shell closest to the LMC), reveals the presence of an intermediate-age to old stellar population, as well as a relatively young one, born $\sim 0.5$ Gyr ago, traced by the blue plume. Moving towards larger distances from the LMC, the young population becomes increasingly less evident, and it disappears beyond $\simeq$9\degr.
Therefore, the LMC old stellar population dominates the outer regions, while young stars are more concentrated at smaller distances, as already determined by previous works \citep[e.g.][]{ElYoussoufi-2021} and as usually found in star-forming dwarf and irregular galaxies \citep[e.g.,][and references therein]{Annibali&Tosi2022}.
Within the interval of distances between $\sim$~7\degr~and 10\degr~
the horizontal branch (HB) also stands out, at $-0.6 \leq g_0 - i_0 \leq 0$~mag and $g_0 \simeq 19.0$~mag, representing the signature of a very old population (t > 10 Gyr).
HB stars possibly populate also regions closer to the LMC, where, however, a clear detection is hampered by the presence of the blue plume, partially overlapping the HB. Beyond 10\degr, the HB is no longer detectable.
At larger galactocentric distances, the contamination by MW stars becomes very high, although an old LMC stellar population is still visible also in the last distance bin (bottom right panel). These findings are in agreement with previous results \citep[e.g.][]{Munoz-2006,Saha-2010,Nidever-2019}.\par
Note that the observed increase in background galaxies in the figure is attributed to our method of generating annuli containing an equal number of stars. Specifically, the stellar density decreases from the centre toward the outer regions. Consequently, larger areas are required to gather the same number of stars in these outer regions. Assuming a uniform density of background galaxies within the YMCA footprint (on scales larger than those of galaxy groups and clusters), this methodology leads to the inclusion of significantly more background galaxies as the area enclosed by the annuli increases.\par
Finally, Figure~\ref{fig:density_map_LMC_east} presents a density map of the LMC East periphery, constructed by adopting stars within the MSTO of intermediate-age ($7 \leq t \leq 9$~Gyr, left plot) and an ancient ($t \geq 10$~Gyr, right panel) stellar population. As expected, the density of stars diminishes towards the outer regions, albeit maintaining non-negligible values (especially visible in the right panel), indicating an extension of the LMC beyond the area covered by the YMCA survey in this direction (as also hinted by the bottom rightmost panel of Figure~\ref{fig:cmd_LMC_east_subplots}). Notably, the density exhibits a sudden decline at about 8\degr~from the LMC centre, mirroring the steep drop-off in density reported by \citet{Nidever-2019} in the same direction\footnote{The overdensity of old stars visible at $\simeq$(-13.5,+3) has a suspicious shape of tile 11\_51, therefore we checked if it might be artificial. Specifically, we compared tile 11\_51 with the adjacents tiles 11\_50, 10\_49, and 10\_48. Their seeing is comparable (see Table~\ref{tab:log_observation}), and a visual comparison of their CMDs also did not reveal any remarkable difference in their depth. 
Counting the total number of stars (with any restriction within the CMD), tile 11\_51 contains more stars ($\sim$ 47000 stars) with respect to the other three ($\sim$ 41000 stars in tile 10\_48, $\sim$38000 stars in tile 10\_49, and $\sim$ 33000 stars in tile 11\_50, respectively). This might indicate that the density of the LMC in that region increases again, but the proximity to the border of the survey does not allow us to draw any firm conclusion about it.
}. However, a comprehensive analysis of the radial density profile of the LMC is beyond the scope of this paper.


\subsection{The west side of the LMC out to the periphery of the SMC}
\label{sec:strip_region}

\begin{figure}
    \centering
    \includegraphics[width=.49\textwidth]{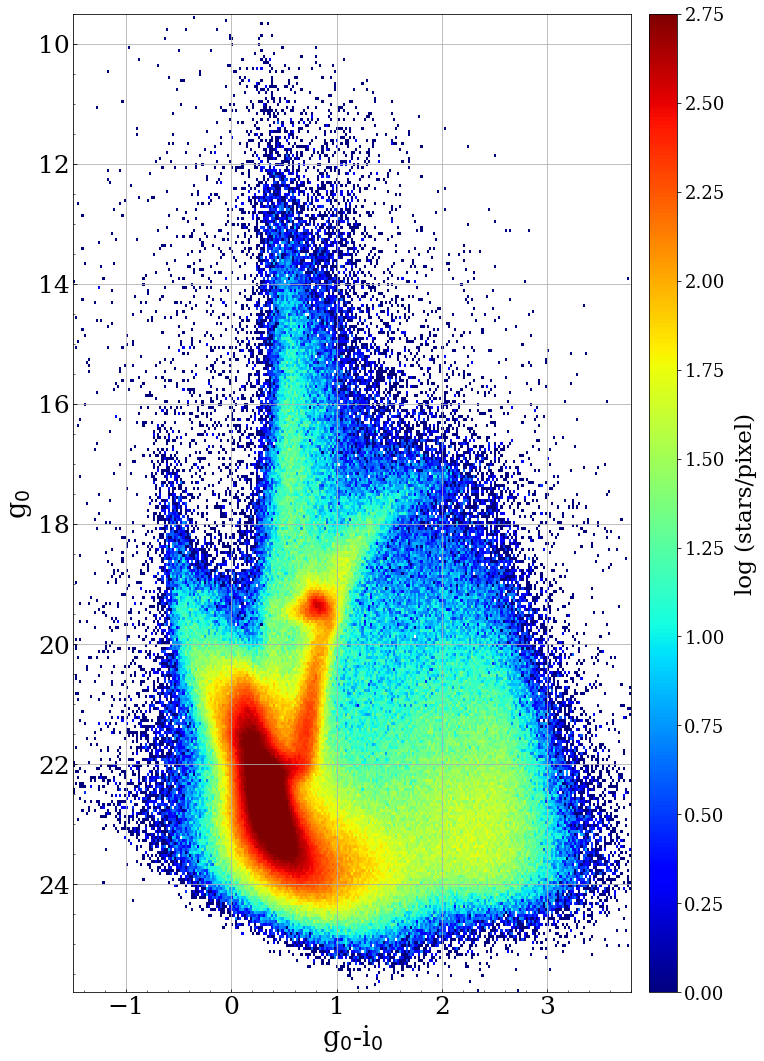}
    \caption{Hess diagram of the tiles in between the LMC and the SMC. Each pixel is $0.02 \times 0.05$ mag in size. Note that to preserve the readability of the figure we do not overlap any isochrone to it as the CMD contains both LMC and SMC stars, which are therefore placed at two different distance moduli and have different metallicities.}
    \label{fig:cmd_strip_total}
\end{figure}

Here we discuss the CMD of the 24 tiles located between the LMC and the SMC, approximately to the north of the classical Bridge (see Fig.~\ref{fig:ymca_footprint}).
The westernmost tiles of this sub-sample lie in the surroundings of the LMC, whereas the easternmost ones cover the northeast periphery of the SMC. 
Therefore, Fig.~\ref{fig:cmd_strip_total}, which displays the overall Hess diagram for the 1.593.456 stars within these 24 tiles, should contain a mix of LMC and SMC stars.
The remarkable RGB and RC visible in this plot show that intermediate-age and old stars are still predominant, although the blue plume is more pronounced with respect to that discussed in Fig.~\ref{fig:cmd_LMC_east_total}, an indication of a more intense star formation activity in recent epochs in these regions.\par
\begin{figure*}[h!]
    \centering
    \includegraphics[scale=0.48]{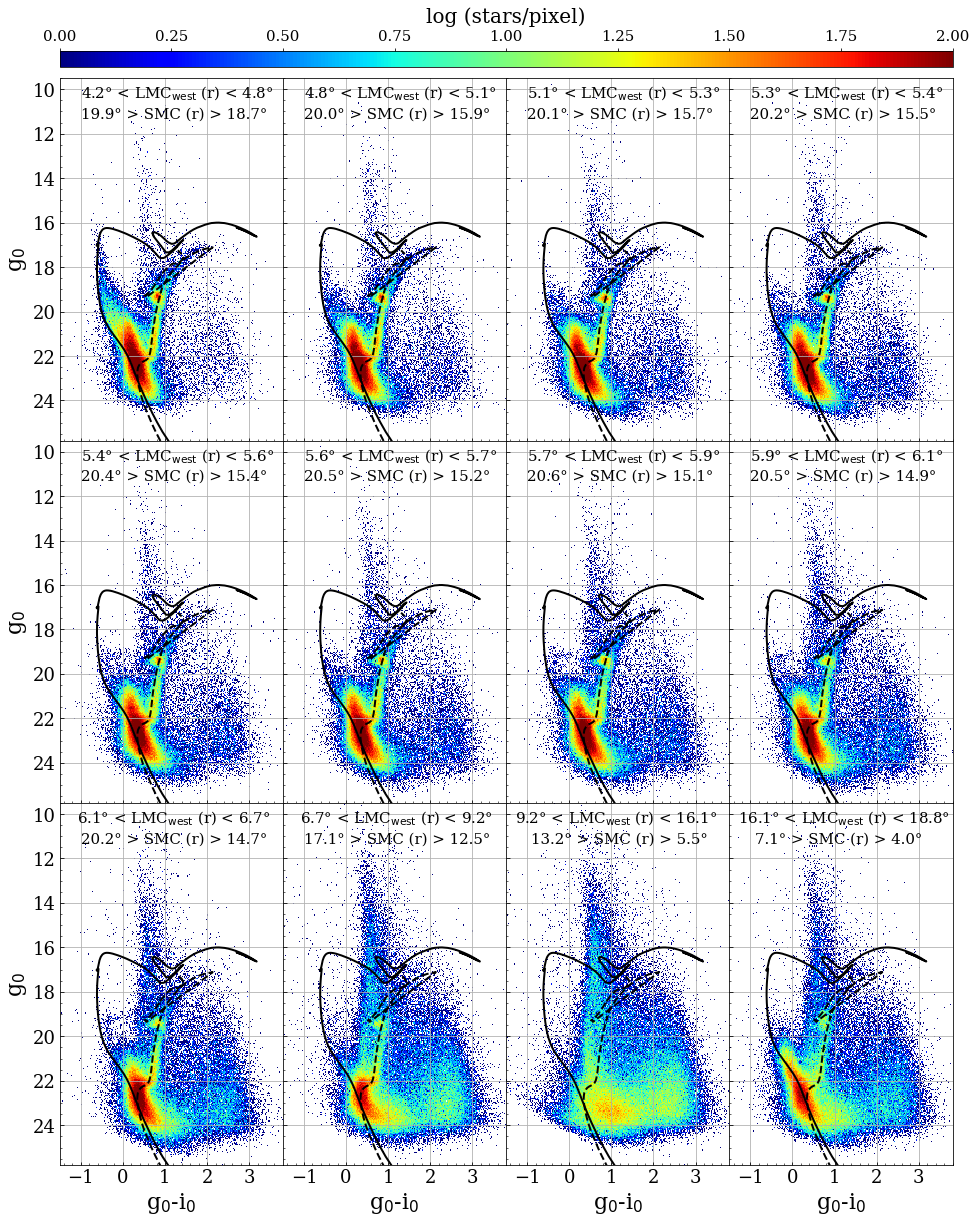}
    \caption{Hess diagrams of the tiles located between the west side of the LMC and the east side of the SMC. They are ordered based on their distance from the LMC centre. Each subpanel shows about 130.000 stars. The isochrones are representative of a young (t = 200 Myr; Z = 0.008 dex, solid line) and old (t $\sim$ 12.5 Gyr, Z = 0.001 dex, dashed line) stellar population, corrected for the distance modulus of the LMC. At the top of each subplot the range of angular distances from the LMC and SMC centres are indicated.}
    \label{fig:cmd_strip_subplots}
\end{figure*}
In Fig.~\ref{fig:cmd_strip_subplots} we show the Hess diagram of the whole region split into 12 different sub-panels, each one containing about 130.000 stars, located between $\sim 4\degr$~up to $\sim 20\degr$~from the LMC centre (the farthest layer is at about 4\degr~from the SMC centre).
Two isochrones representative of a young ($t \sim 200$~Myr; Z = 0.008) and an old ($t \sim 12.5$~Gyr; Z = 0.001) stellar population, and placed at the LMC distance, are also drawn in the figure.
The top left panel, which contains the stars closest to the LMC (less than 5\degr~from the LMC centre), exhibits the presence of stars formed about 200 Myr ago, attesting that recent events of star formation have occurred in these regions. The presence of populations younger than 200 Myr is not supported by a significant population of classical Cepheids in this tile \citep[see][]{Ripepi-2022}.
All these stars belong to the easternmost part of the tile 7\_27, which is spatially very close to the west rim of the LMC bar, where the northern arm of the LMC departs, regions known to host a very young stellar population \cite[e.g.][]{Harris&Zaritsky2009,Indu&Subramaniam2011,Gaia-Luri-2021,Mazzi-2021,Ripepi-2022}.
It is currently accepted that the intense star formation activity observed at $\sim$200 Myr is likely a consequence of the last close passage between the LMC and the SMC occurred about 150--200 Myr ago \citep[][]{Zivick-2018,Zivick-2019,Schmidt-2020,Choi-2022}.
Stars formed about 200 Myr ago are not found beyond 5\degr~from the LMC centre. 
Only intermediate-age and old stellar populations are found at larger galactocentric distances, as already observed towards the east side of the LMC.
Interestingly, the density of LMC stars appears to abruptly drop down beyond $\simeq 9\degr$ ($\simeq$ 8 kpc at the LMC distance), as illustrated in the subpanel located in the third row and third column of the figure.\par
\begin{figure*}[h!]
    \centering
    \includegraphics[width=.48\textwidth]{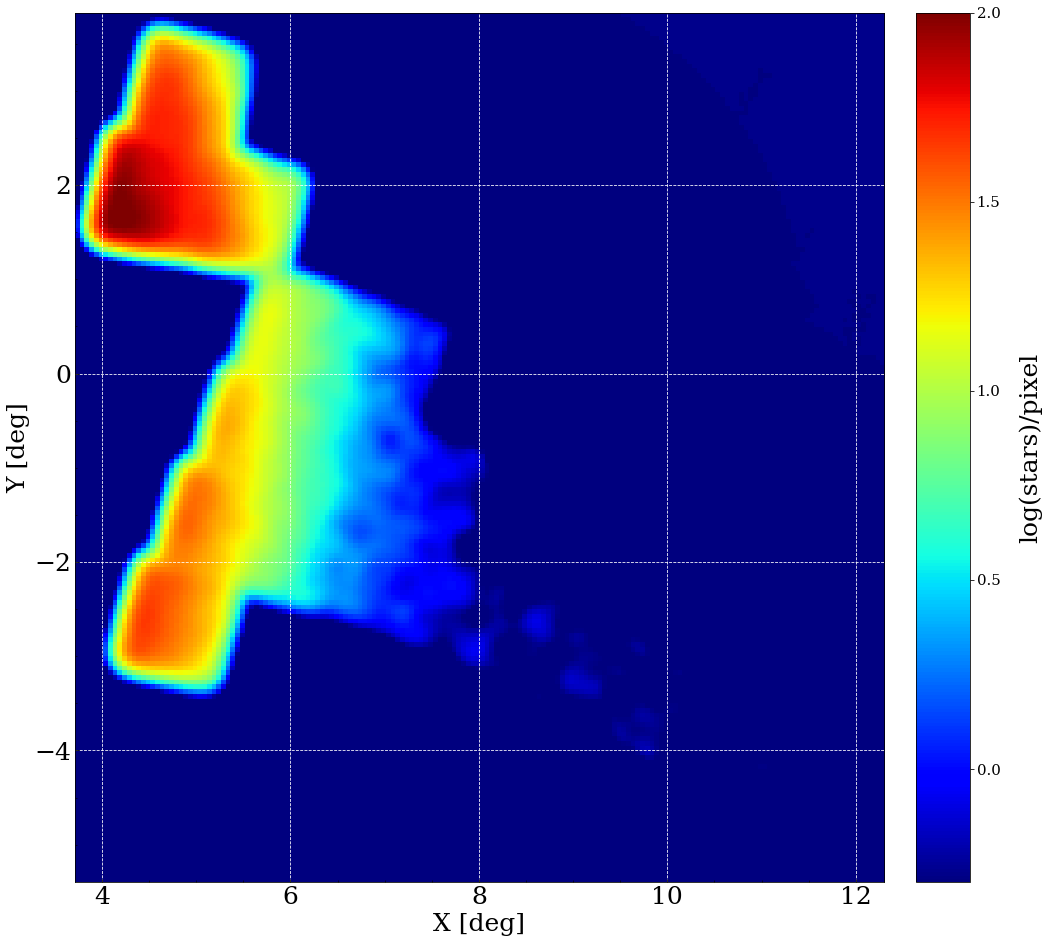}
    \includegraphics[width=.48\textwidth]{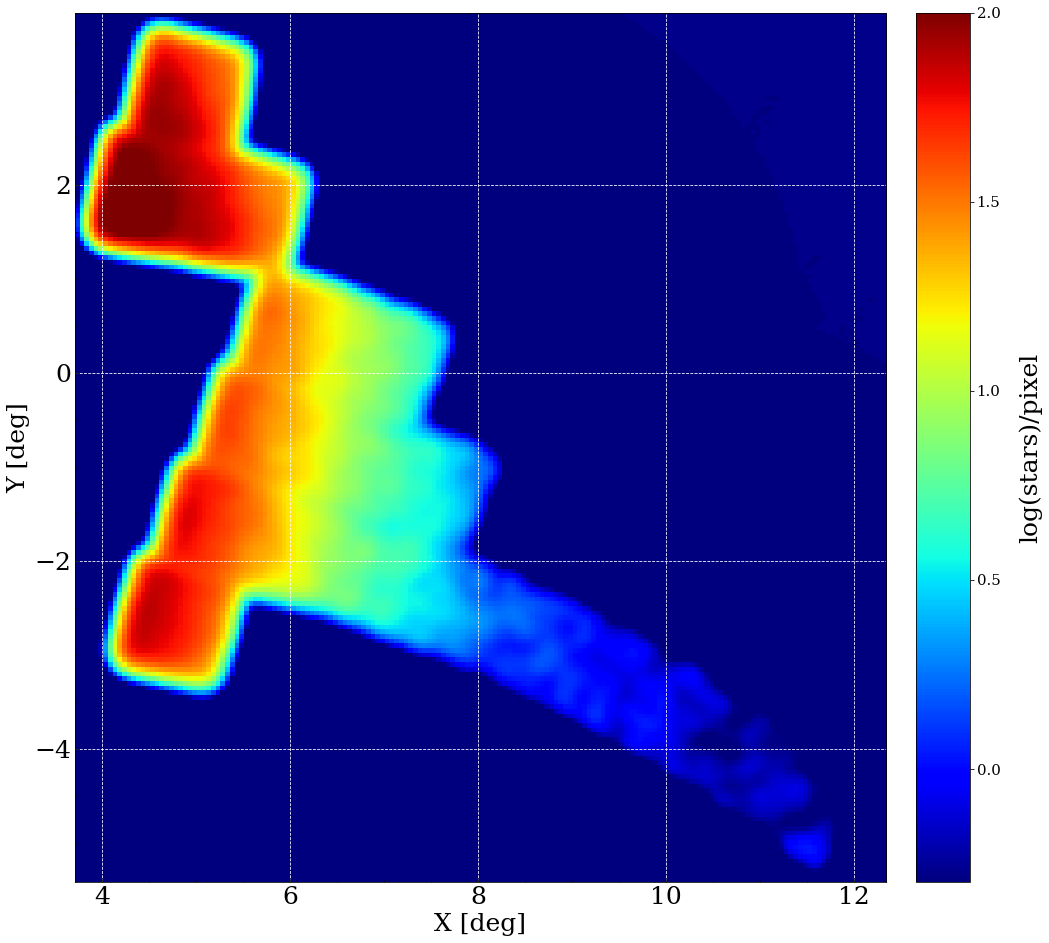}\\
    \includegraphics[width=.48\textwidth]{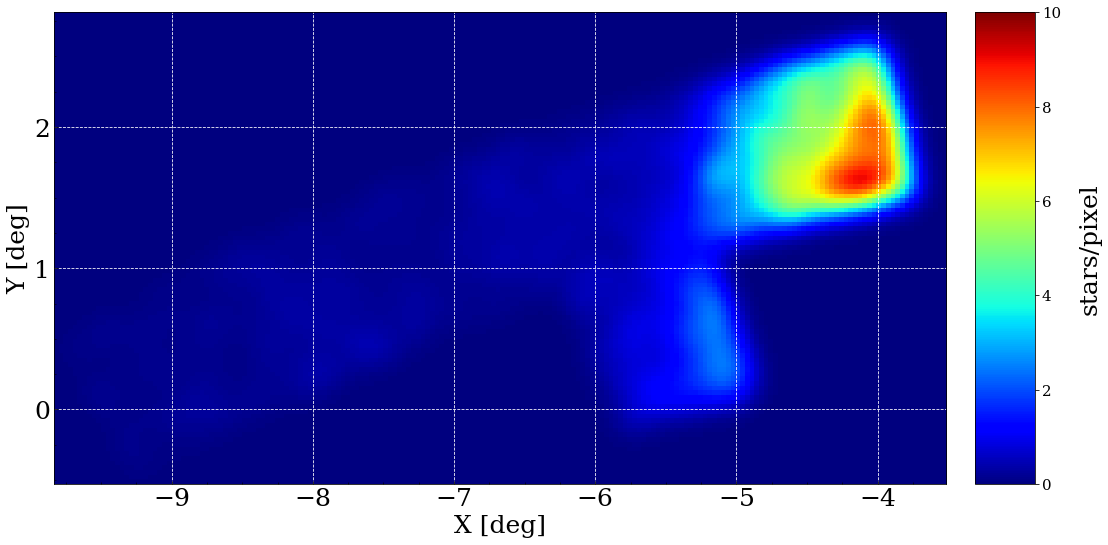}
    \includegraphics[width=.48\textwidth]{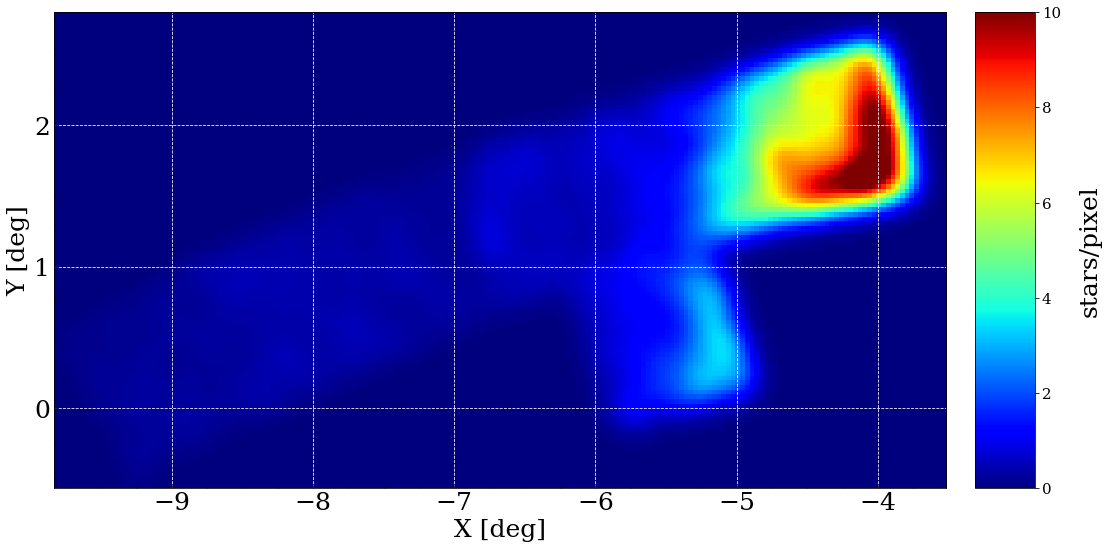}\\
    \caption{\emph{Top:} Density map of the strip region close the LMC, derived by adopting MSTO stars corresponding to intermediate-age ($7 \leq t \leq 9$~Gyr, left panel) and old ($t \geq 10$~Gyr, right panel) stellar populations. We selected stars within colour and magnitude ranges of $0.0 \leq 0.5$~mag and $21.4 \leq g \leq 21.9$~mag for the younger stellar population, and $0.1 \leq g - i \leq 0.6$~mag and $21.9 \leq g \leq 22.4$~mag, for the old one, respectively. Each pixel has size of $3\arcmin \times 3\arcmin$. 
    \emph{Bottom:} Density map of the strip region close the SMC. The region of the CMD adopted to select SMC MSTO stars is $0.0 \leq g - i \leq 0.5$~mag and $21.8 \leq g \leq 22.3$~mag for the younger stellar population, and $0.1 \leq g - i \leq 0.6$~mag and $22.3 \leq g \leq 22.8$~mag for the ancient one, respectively. Each pixel has size of $2\arcmin \times 2\arcmin$. We applied a Gaussian Kernel Density Estimator with bandwidth = 0.1\degr~for smoothing all four subplots.}
    \label{fig:density_map_STRIP}
\end{figure*}

The density map depicted in Figure~\ref{fig:density_map_STRIP}, constructed using MSTO stars in the strip region, provides a more detailed illustration of the sharp decline in density. Specifically, top panels show the density map of the tiles closer to the LMC, indicating that the LMC stellar population abruptly truncates at about 8\degr--8.5\degr~from the LMC centre.
Notably, on the Eastearn side, main sequence stars of the LMC remain observable up to $\simeq 12\degr -15\degr$ (i.e. 10 - 13 kpc), as evidenced in the bottom rightmost panel of Fig.~\ref{fig:cmd_LMC_east_subplots}).
This distinctive feature was previously documented by \citet{Mackey-2018}, who observed a sudden truncation of the LMC disc in the West and Southern regions, facing the SMC, using MSTO stars extracted from the photometric catalogue provided by the Dark Energy Survey. 
Furthermore, \citet{Belokurov&Erkal2019} employed simulations to model the LMC disc's behaviour in presence of both the MW and the SMC. Their findings suggest that a close encounter with the SMC could lead to the truncation observed on the West side of the LMC.\par
In the top rightmost panel, the HB begins to be perceivable and it is clearly visible when the MSTO becomes fainter than $g_0 \simeq 20$~mag, namely beyond 5.7\degr. The HB stage seems to be present up to a distance of $\sim 7\degr$, then it vanishes. Interestingly, in the bottom right-central panel, with stars located between the tiles 4\_12 and 4\_18, that is beyond 9\degr~from the LMC centre, the LMC stellar population is overwhelmed by MW contaminants, and an old main sequence and a feeble sub-giant branch (SGB) are only barely discernible.
The last sub-panel displays the Hess diagram for stars in the North-East periphery of the SMC, in particular between $4\degr \leq r_{\rm SMC} \leq 7\degr$. 
It, therefore, contains SMC stars, and indeed evolutionary phases of intermediate-age and old stars are still clearly visible in the plot.\par
Interestingly, the magnitude of the RC, that can be used as a distance indicator \citep{Girardi2016}, is similar to that of the LMC. 
\begin{figure*}[h!]
    \centering
    \includegraphics[width=.9\textwidth]{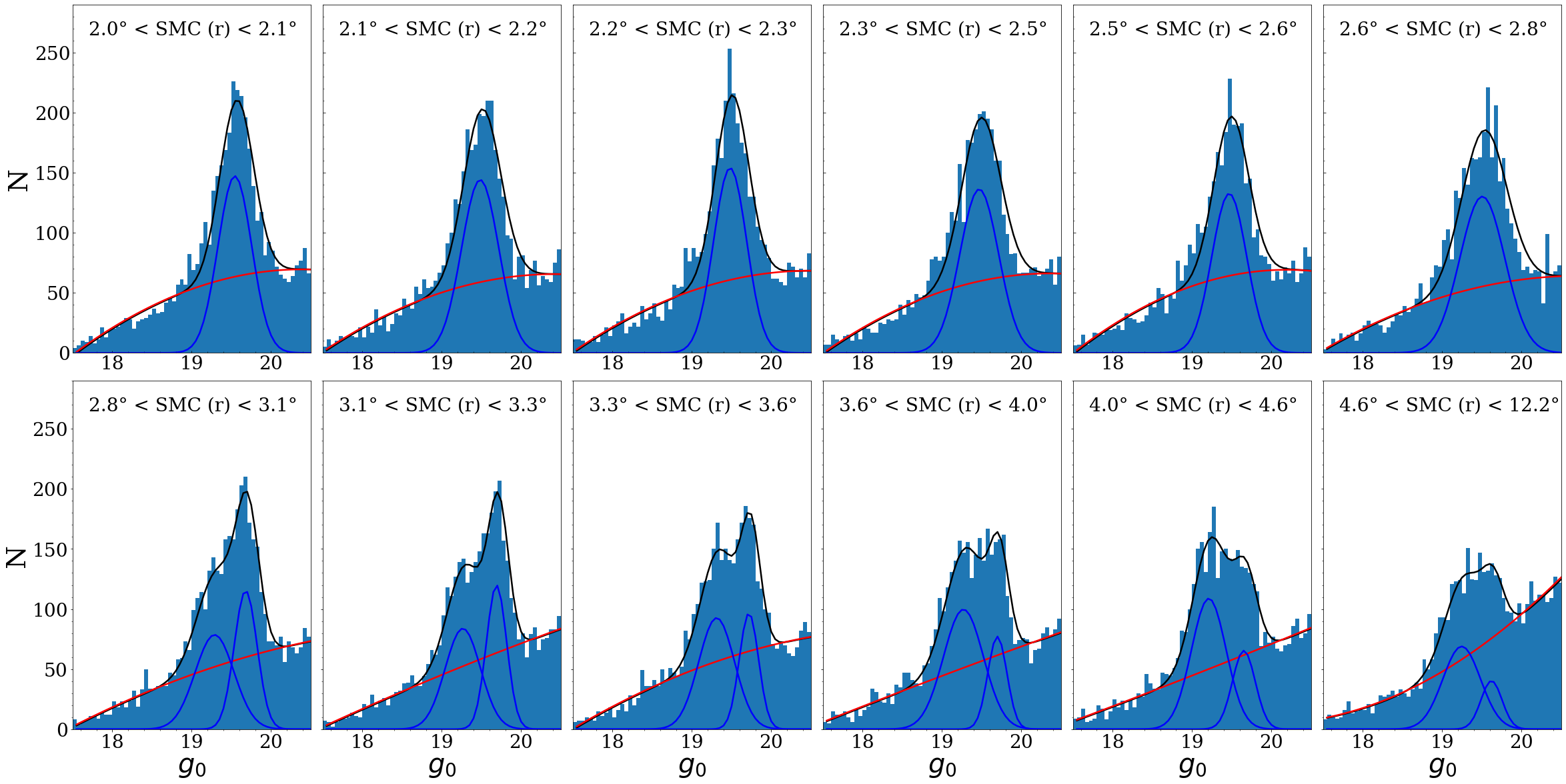}\\
    \includegraphics[width=.9\textwidth]{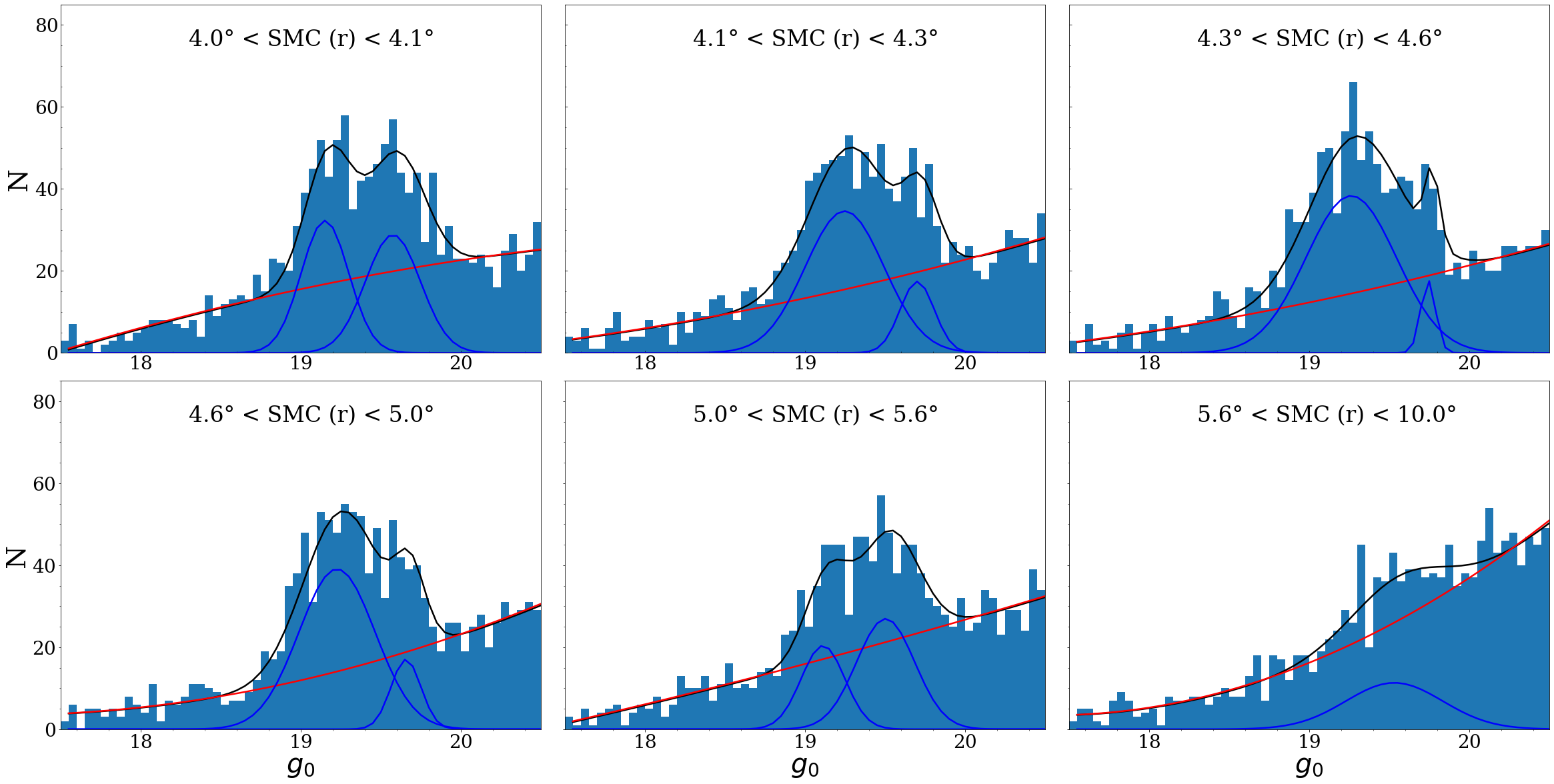}
    \caption{Magnitude distribution in the $g$ band for stars included in both the YMCA and STEP catalogues and having the position angle in the range $55\degr \leq PA \leq 95\degr$. The panels in the first two rows display $\sim$ 4360 stars with magnitudes between $g_0$ = 17.5 mag and $g_0$ = 20.5 mag, while the panels in the last two rows contain $\sim$ 1370 stars within the same range of magnitudes. At the top of each sub-panel the interval of distances from the SMC centre is reported.
    The blue solid lines represent the Gaussian distributions of RC stars, whereas the red solid line is the quadratic polynomial modeling the distribution of stars along the RGB. The black solid line illustrates the best-fit model derived from the combined contributions of both stellar evolutionary phases.}
    \label{fig:rc_analysis}
\end{figure*}
This might indicate that these regions are much closer to us with respect to the average distance of the SMC main body, namely $\rm DM \sim 18.9$ \citep[][]{deGrijs&Bono2015}, and it could be a further consequence of the mutual gravitational interaction between the MCs. 
Recent works already reported the presence of a foreground SMC stellar population within SMC northeastern regions based on the analysis of the RC feature, attributing it to the tidal stripping of the LMC \citep{Nidever-2013,Subramanian-2017,Tatton-2021}. For example, \citet{Omkumar-2021} traced this substructure up to $\sim$5-6\degr~from the SMC centre, thanks to the dataset provided by {\it Gaia} Data Release 2 \citep{Gaia-Brown-2018}, estimating that on average it is $\sim$12 kpc closer to us with respect to the SMC main body.
In addition, \citet{James-2021} observed in the same region a bimodal distribution in the radial velocities of RGB stars, being the two components separated by $\sim 35-45$ km s$^{-1}$. They suggested that RGB stars in the lower radial velocity component belong to the same foreground stellar substructure traced by RC stars.\par
In our pursuit to validate the existence of a closer SMC stellar population, we integrated the YMCA dataset with the STEP\footnote{To correct the STEP catalogue for extinction  we utilized the reddening maps by \citet{Skowron-2021}.} catalogue \citep{Ripepi-2014}, which covers the SMC contiguously from its core to the edge of YMCA tiles. Subsequently, we conducted an analysis of the luminosity function of RC stars in the de-reddened $g$ band, following a methodology akin to that employed by \citet{Omkumar-2021}.
Notably, the RC position in the CMD is highly contaminated by MW halo stars, as depicted in Figs.~\ref{fig:cmd_LMC_east_total}-\ref{fig:cmd_strip_total}. To mitigate this, we cross-matched the YMCA+STEP catalogue with Gaia DR3, adopting a tolerance of 1\arcsec. We exclusively retained stars with reliable astrometric data characterized by {\it RUWE} < 1.4 and {\it astrometric\_excess\_noise\_sig} < 2.
Leveraging the Gaia dataset we excluded stars with distances and tangential velocities incongruent with those expected for the MCs. In particular, as suggested by \citet{Gaia-Luri-2021}, we get rid of foreground MW stars by imposing $\bar{\omega} \leq 3\sigma_{\bar{\omega}}$, where $\bar{\omega}$ is the parallax and $\sigma_{\bar{\omega}}$ is the parallax error. Finally, we selected only stars with PMs within 5$\sigma$ from the average PM of the SMC: $0.7321 - (5 \cdot 0.3728) \leq \mu_{\alpha} \leq 0.7321 + (5 \cdot 0.3728)$ and $-1.2256 - (5 \cdot 0.2992) \leq \mu_{\delta} \leq -1.2256 + (5 \cdot 0.2992)$. These average SMC PM values and their associated uncertainties were derived from \citet{Gaia-Luri-2021}.\par
Following \citet{Girardi&Salaris2001} we modeled the magnitude distribution of $g_0$ adopting the following equation:
\begin{equation}
    N(g_0) = a+b \cdot g_0+ c \cdot g_0^2+ d \cdot {\rm exp[-\frac{(g_0-g_0^{\rm RC})^2}{2\sigma_{g_0^2}}]}
\end{equation}
where the Gaussian function models the RC distribution, while the quadratic polynomial accounts for RGB stars contaminants.
Additionally, we introduced the possibility of including a second Gaussian function to assess the potential presence of a foreground stellar population within the SMC.
We sub-divided the YMCA-STEP catalogue into slices at different angular distances from the SMC centre, each holding an equal number of stars, in the direction of the observed closer RC stellar population.
Specifically, we confined our analysis to stars falling within the position angle range: $55\degr \leq PA \leq 95\degr$.
Employing the {\sc LMFIT} Python package, we conducted individual fits within each slice, considering a model with two Gaussians only when their centres were separated by at least the sum of their widths.\par
Figure~\ref{fig:rc_analysis} illustrates the results obtained from the fit of regions at increasing angular distances, moving from the top left to the bottom right. 
The top two rows correspond to distances ranging between $2.0\degr$~and $12.2\degree$, with approximately 4360 stars in each sub-panel. The last two rows, that aim to better visualize the outer regions that contain less stars, encompass distances between 4\degr~and 10\degr, with approximately 1370 stars in each sub-panel.  
Our analysis reveals a discernible foreground stellar population emerging around $\sim 2.8\degree$ from the SMC centre, becoming more prominent at greater angular distances and becoming the predominant feature beyond $\simeq 3.5\degr$. 
However, beyond 5.0\degr, the faint RC appears again as the dominant stellar population, and beyond $\simeq 5.5\degree$, the foreground stellar population becomes indiscernible.
Table~\ref{tab:RC_gaussian_values} provides the mean and standard deviation of the Gaussians for each slice, indicating that the closer stellar population exhibits an average distance modulus smaller than 0.4 mag, corresponding to about 14.5 kpc at the distance of the SMC. These findings align with the conclusions drawn by \citet[][]{Omkumar-2021}.



\begin{table}
\centering
 
 \caption{Statistics of the Gaussian approximating the RC stellar population within the investigated region.}
 \label{tab:RC_gaussian_values}
 \begin{tabular}{lcccc}
  \hline\hline
  r & $g_0^{\rm RC_{f}}$ & $g_0^{\rm RC}$ \\
  \hline
  2.0\degr - 2.1\degr & - & 19.5 $\pm$ 0.2\\
2.1\degr - 2.2\degr  & - & 19.5 $\pm$ 0.2 \\
2.2\degr - 2.3\degr  & - & 19.5 $\pm$ 0.2 \\
2.3\degr - 2.5\degr  & - & 19.5 $\pm$ 0.2 \\
2.5\degr - 2.6\degr  & - & 19.5 $\pm$ 0.2 \\
2.6\degr - 2.8\degr  & - & 19.5 $\pm$ 0.3 \\
2.8\degr  - 3.1\degr  & 19.3  $\pm$ 0.2  & 19.7  $\pm$ 0.1 \\
3.1\degr  - 3.3\degr  & 19.3  $\pm$ 0.2  & 19.7  $\pm$ 0.1 \\
3.3\degr  - 3.6\degr  & 19.3  $\pm$ 0.2  & 19.7  $\pm$ 0.1 \\
3.6\degr  - 4.0\degr  & 19.3  $\pm$ 0.2  & 19.7  $\pm$ 0.1 \\
4.0\degr  - 4.6\degr  & 19.2  $\pm$ 0.2  & 19.7  $\pm$ 0.1 \\
4.6\degr  - 12.2\degr  & 19.2  $\pm$ 0.2  & 19.6  $\pm$ 0.1 \\
\hline
4.0\degr  - 4.1\degr  & 19.2  $\pm$ 0.1  & 19.6  $\pm$ 0.2 \\
4.1\degr  - 4.3\degr  & 19.2  $\pm$ 0.2  & 19.7  $\pm$ 0.1 \\
4.3\degr  - 4.6\degr  & 19.3  $\pm$ 0.3  & 19.7  $\pm$ 0.0 \\
4.6\degr  - 5.0\degr  & 19.2  $\pm$ 0.2  & 19.7  $\pm$ 0.1 \\
5.0\degr  - 5.6\degr  & 19.1  $\pm$ 0.1  & 19.5  $\pm$ 0.2 \\
5.6\degr - 10.0\degr  & - & 19.5  $\pm$ 0.3\\
    \hline
 \end{tabular}
 \tablefoot{The first column shows the inner and outer radii from the SMC centre of the analyzed annulus. In the second and third columns, we inserted the mean and standard deviation of the Gaussians representative of the foreground population, if detected, and the main SMC population, respectively.}
\end{table}

\subsection{The unexplored north-west periphery of the SMC}

\begin{figure}
    \centering
    \includegraphics[width=.49\textwidth]{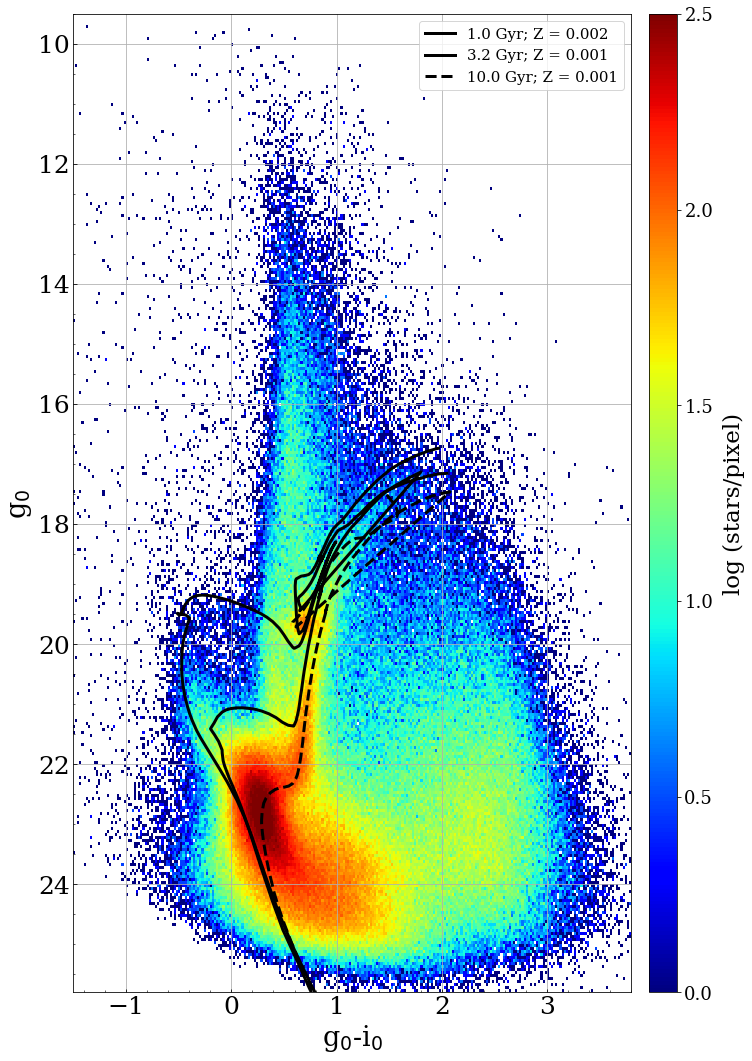}
    \caption{Hess diagram of the 19 tiles centred at the north-western periphery of the SMC. Each pixel is $0.02 \times 0.05$ mag in size. Three different isochrones, with ${\rm DM} = 18.9$~mag and ages and metallicities listed in the top right corner of the plot, are also superposed on the Hess diagram.}
    \label{fig:cmd_smc_total}
\end{figure}

In this section, we discuss the characteristics of the $\sim$700,000 stars placed in the 19 tiles on the northwestern side of the SMC.
As in the previous sections, Fig.~\ref{fig:cmd_smc_total} shows the overall Hess diagram of all stars in these fields, along with three isochrones of different ages, corrected for the distance modulus of the SMC (${\rm DM} = 18.9$~mag).
Unlike the regions previously described, the absence of the blue plume in the plot denotes the absence of any population younger than $\simeq 1.0$~Gyr, while the RC and RGB evolutionary phases are evident, indicating that the periphery of the SMC is made only by intermediate-age and old stars.\par
\begin{figure*}[h!]
    \centering
    \includegraphics[scale=0.48]{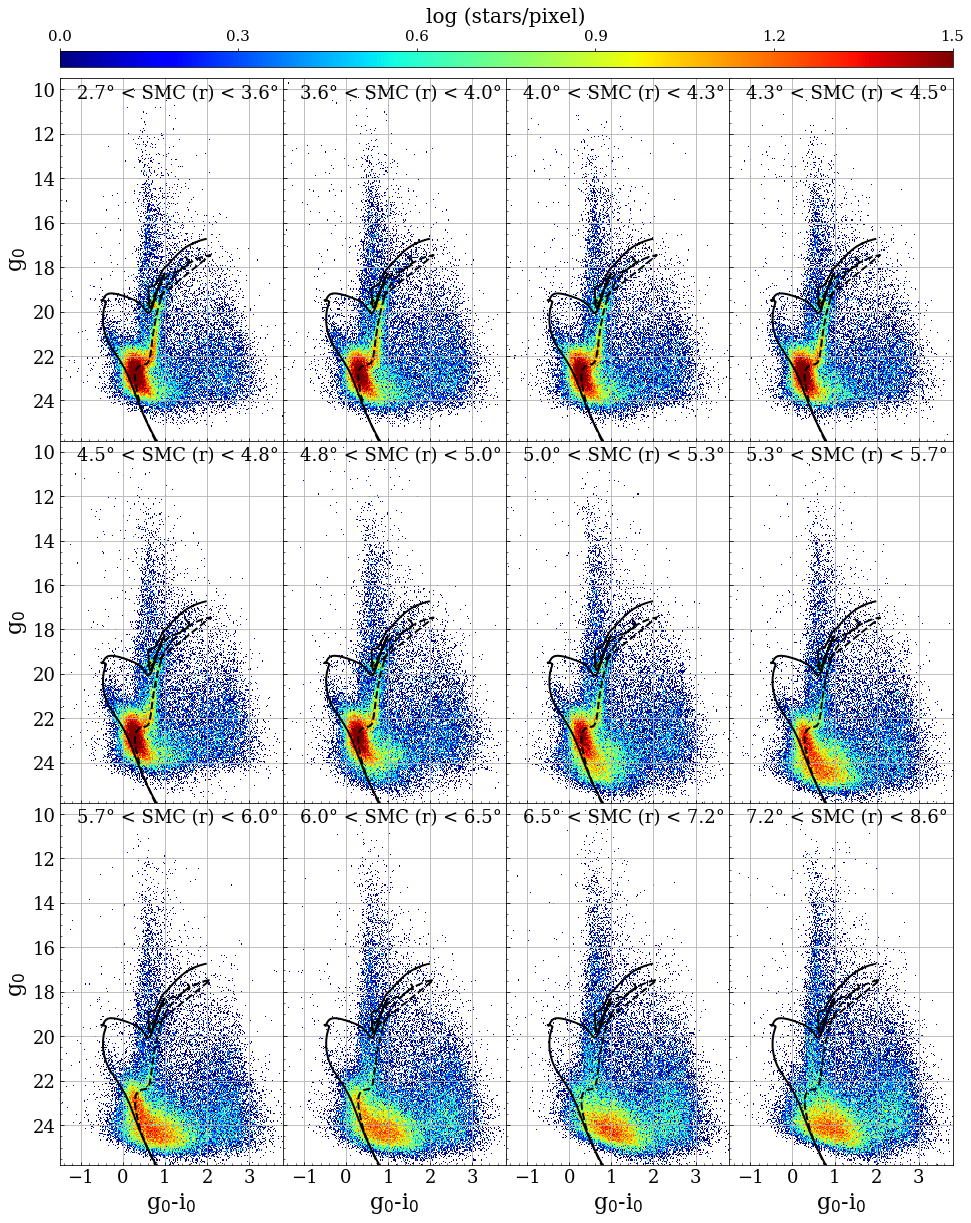}
    \caption{Hess diagrams of the fields on the northwest side of the SMC. Each subpanel contains about 60.000 stars, and they are ordered according to their distance from the SMC centre. Two isochrones (t = 1 Gyr; Z = 0.001 dex, solid line and t = 10 Gyr, Z = 0.001 dex, dashed line), corrected for the distance modulus of the SMC, are superposed as a comparison. At the top of each subplot the range of distances from the SMC centre is indicated.}
    \label{fig:cmd_smc_subplots}
\end{figure*}
\begin{figure*}[h!]
    \centering
    \includegraphics[width=.48\textwidth]{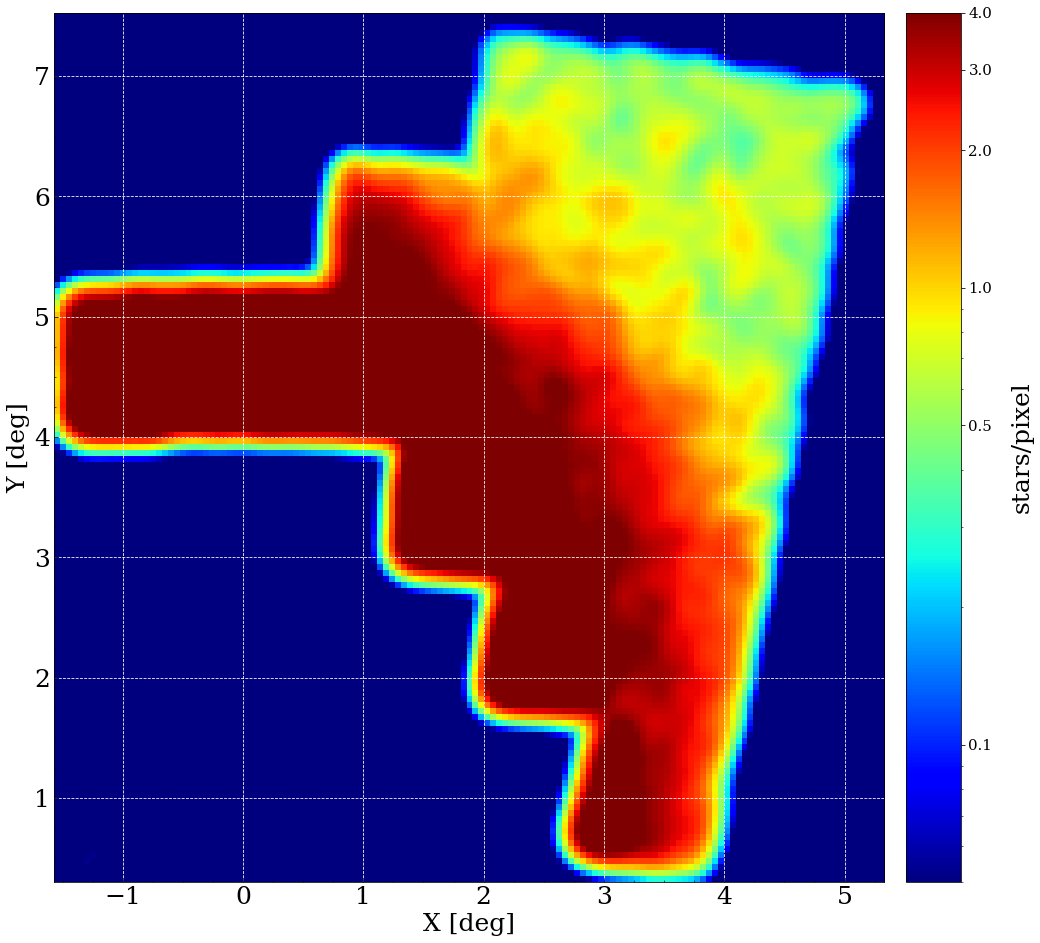}
    \includegraphics[width=.48\textwidth]{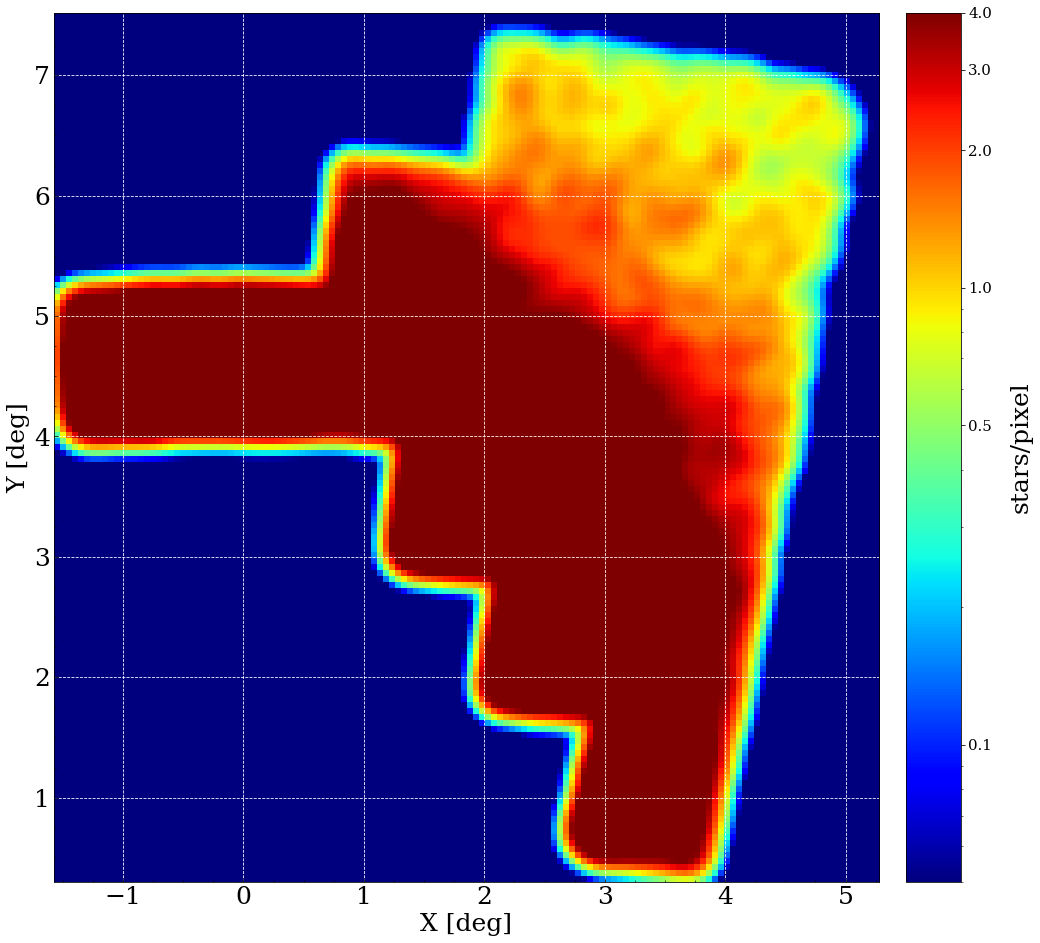}
    \caption{Density map of the SMC north-western periphery, derived by adopting MSTO stars corresponding to intermediate-age ($7 \leq t \leq 9$~Gyr, left panel) and old ($t \geq 10$~Gyr, right panel) stellar populations. Specifically, we selected stars within colour and magnitude ranges of $0.0 \leq g - i \leq 0.5$~mag and $21.8 \leq g \leq 22.3$~mag for the left panel, and $0.1 \leq g - i \leq 0.6$~mag and $22.3 \leq g \leq 22.8$~mag for the right panel, respectively. Each pixel has size of $3\arcmin \times 3\arcmin$ and we applied a Gaussian Kernel Density Estimator with bandwidth = 0.1\degr~for smoothing the plot.}
    \label{fig:density_map_SMC}
\end{figure*}
To analyze the trend of the stellar population with the distance from the SMC centre, in Fig.~\ref{fig:cmd_smc_subplots}, we display 12 Hess diagrams for stars at different galactocentric radii. Each subplot contains about 60,000 stars and we also draw two isochrones of 1 Gyr and 10 Gyr to guide the eye. These tiles cover an interval of galactocentric distances between $\sim$ 3\degr~and 8.5\degr~, corresponding to $\sim$ 3 and $\sim$ 9 kpc at the mean distance of the SMC.
The shell closest to the SMC reveals that the youngest stellar population in this area is possibly older than 1 Gyr, suggesting a total absence of recent star formation in the northwest side of the SMC. On the contrary, it is known that the SMC main body and its eastern direction (i.e. the Wing) underwent a very intense recent period of star formation \citep[][]{Cignoni-2013,Rubele-2018}. The intermediate-age and old stellar population coexist up to $\sim$5.7\degr, beyond which the RGB and RC begin to fade. In the farthest two subpanels the densest areas of the Hess diagram are at $g_0 - i_0 \sim 1$~mag and $g_0 \sim 24$~mag, which likely represent MW halo stars \citep[see also Fig. 13 in][]{Ripepi-2014}. 
In all these plots the HB is less evident with respect to what was observed in the outskirts of the LMC (see Figures~\ref{fig:cmd_LMC_east_subplots} and \ref{fig:cmd_strip_subplots}), a clue that the star formation activity at look-back times greater than 10 Gyr was not so energetic as in the LMC \citep{Cignoni-2013}. Many authors reported a substantial difference in the SFHs of the MCs for the period between 10 Gyr and 12-13 Gyr, casting doubt on them having formed as a pair \citep[see e.g.][and references therein]{Rezaeikh-2014,Massana-2022}.\par
Figure~\ref{fig:density_map_SMC} displays the density map of the SMC north-western periphery, as traced by MSTO stars belonging to intermediate-age ($7 \leq t \leq 9$~Gyr, left plot) and ancient ($t \geq 10$~Gyr, right panel) stellar populations.
Unlike the LMC, at a radial distance of about 6\degr, corresponding to 6.3~kpc, the density of the younger population sharply approaches zero, suggesting the border of the SMC in this direction. This feature can also be hinted at by looking in subplots of Figure~\ref{fig:cmd_smc_subplots} displaying radial distances greater than $\sim$6.5\degr. 

\section{Discovery of unknown star clusters in the MC periphery} 
    
In \citet{Gatto-2020}, we presented a list of 78 new candidate SCs, which were discovered in the outskirts of the LMC using an algorithm designed to identify over-densities characteristic of SCs in the sky. 
That work was based on 21 YMCA + 2 STEP tiles in the vicinity of the LMC available at that time. Here we report the results of the cluster-finding algorithm applied to the remaining 79 YMCA tiles, including those located in the periphery of the SMC and in the region between the LMC and SMC.

\subsection{Main steps of the cluster finding algorithm}

The working flow of the algorithm is thoroughly described in Sect. 3 of \citet{Gatto-2020}. Here, we provide a brief summary of the process.
The algorithm takes the coordinates of stars as its only input parameter, and processes each tile individually to identify regions of the sky where the number of stars is higher than the local background. 
To achieve this, it performs a Kernel Density Estimation (KDE) on each tile using both the \emph{tophat} and the \emph{gaussian} functions as kernels, with a bandwidth of 0.2\arcmin\footnote{To maximize the chances of discovering new SCs, \citet{Piatti-2018} suggested that the bandwidth of the kernel function should match the size of the smallest stellar systems being researched.}.
The algorithm then automatically determines the centres and radii of the detected over-densities \citep[see Sect. 3.2 of][]{Gatto-2020}.
The final outcome is a list of over-densities for each tile, including some that are not actual physical systems (usually called asterisms) and must be removed, as described in the next Section. 

\subsection{Cleaning the colour-magnitude diagram from field star contamination}

To distinguish between real SCs and asterisms we first have to decontaminate the CMD of the SC candidates from field stars. Following \citet{Gatto-2020}, we adopt the method described by \citet{Piatti&Bica2012} to clean the CMDs, which is well-suited for sparse SCs, such as those found in this work. In this paper, we apply a method slightly different from that adopted in \citet{Gatto-2020}; therefore, we summarize the method below.
In this procedure, the CMD of each SC candidate (over-density) is compared with those of six surrounding fields with the aim to identify probable field stars and remove them from the SC CMD. Each field is defined as a circular region with a radius three times that of the SC candidate, thereby ensuring a sufficient sample size of the local stellar background even in low-density environments. For each of the six fields, boxes centred on each star are constructed, with sides that vary based on the local density of the CMD stars. In particular, small boxes are constructed in denser areas \citep[see Fig. 12 in][for an example of the procedure]{Piatti&Bica2012}. 
Note that in the following steps we leave out stars within the CMD region defined by $g - i > 1.5$~mag and $g > 20$~mag, as these are likely MW dwarf disk stars \citep[see Fig.~13 in][]{Ripepi-2014} that could alter the outcome of the cleaning procedure.\par
The median number of stars in the six fields is estimated and normalized to the SC area, providing an estimate of the expected number of contaminant stars ($N_{\rm bkg}$) in the SC area. 
Subsequently, in order to determine the likelihood of a star belonging to the candidate SC, we randomly extracted $N_{\rm bkg}$ stars for each field, and then identified and subtracted the $N_{\rm bkg}$ closest stars in the CMD of the candidate SC for each of the six surrounding fields.
This process was repeated for all six fields and the probability of a star belonging to the cluster was calculated as $P = 1 - (N_{\rm subt})/6$, where $N_{\rm subt}$ refers to the number of times a star was subtracted.
The position of the stars with a high membership probability on the CMD was then utilized to distinguish between real physical systems and asterisms, and to estimate the age of the SCs through an isochrone fitting procedure (see Sect.~\ref{sec:isochrone_fitting}).\par
As done in \citet{Gatto-2020}, to quantitatively gauge the goodness of each SC we also estimate the number of  stars hosted by  the SC divided by the standard deviation above the local average mean. Hence, we define
\begin{equation}
    G = (N_{\rm cl} - N_{\rm bkg})/ \sqrt{N_{\rm bkg}}
    \label{eq:G}
\end{equation}
where $N_{\rm cl}$ is the number of stars within the SC radius. 
With respect to \citet{Gatto-2020},
we preferred a more conservative approach by taking into consideration only candidate SCs with $G > 5.0$.
The CMDs of the SC passing this selection were subsequently analysed in more detail as explained in the following section.


\subsection{Determine SC age through isochrone fitting}
\label{sec:isochrone_fitting}

To determine the age of each SC candidate, we adopted a visual isochrone matching procedure. The PARSEC isochrones by \citet{Bressan-2012} were used and the distance modulus was fixed to either ${\rm DM} = 18.49$~mag or ${\rm DM} = 18.9$~mag for SCs located in the LMC and SMC peripheries, respectively. The metal content was fixed based on the age-metallicity relation derived by \citet{Piatti&Geisler2013} and \citet{Parisi-2015} for the LMC and SMC, respectively. 
We varied the age of the isochrones to seek the best one that matches the positions of the stars with high membership values (i.e. $P \geq 50\%$) on the de-reddened CMD.
In particular, we focus on the following evolutionary phases to aid the matching procedure: MS, RC, RGB, and sub-giant branch (SGB). 
To be conservative, we considered an age uncertainty of $\sigma_{\log t} = 0.2$~dex,  which includes both the statistical error from the visual procedure and the systematic error due to the elongation of the LMC/SMC along the line of sight.
To assess the accuracy of the SC ages estimated via this procedure, we provide a comprehensive comparison of our SC ages with those in common with the existing literature in Appendix~\ref{app:age_comparison}.

\subsection{Candidate SCs in the MC periphery}
\label{sec:candidate_SCs}

\begin{table*}
\caption{Main parameters of the SCs investigated in this work.}  

    \label{tab:sc_params}

\footnotesize\setlength{\tabcolsep}{1.9pt}
    \begin{tabular}{l|c|c|c|c|c|c|c|c|c|c|c|l}
    \hline\hline
    ID* & RA & Dec & R & $\log t$ & ${\rm N_{\rm stars}}$ & G & $n_0$ & $r_c$ & $r_c$ & Tile & Flag & Simbad Name \\
     & (J2000) & (J2000) & (') &  & &  & (arcmin$^{-2}$) & (') & (pc) & & & \\
     \hline
YMCA-0049 & 11.8554 & -68.9199 & 1.10 & 9.60 & 422 & 47.7 & $676 \pm 7$ & $0.32 \pm 0.13$ & $5.6 \pm 2.2$ & 6\_5 & A & L~32; OGLE-CL SMC 310\\
YMCA-0050 & 14.7444 & -68.9153 & 0.85 & 9.70 & 210 & 31.6 & $382 \pm 5$ & $0.28 \pm 0.03$ & $4.8 \pm 0.5$ & 6\_6 & A & ESO 51-9; OGLE-CL SMC 309\\
YMCA-0051 & 25.4233 & -71.1624 & 0.70 & 9.20 & 165 & 22.9 & $249 \pm 8$ & $0.34 \pm 0.19$ & $5.0 \pm 2.7$ & 4\_9 & A & HW~84; OGLE-CL SMC 305\\
YMCA-0052 & 28.2998 & -70.7526 & 0.95 & 9.85 & 79 & 14.0 & $187 \pm 9$ & $0.14 \pm 0.02$ & $2.0 \pm 0.4$ & 4\_10 & A & OGLE-CL MBR 6\\
YMCA-0053 & 66.7160 & -67.2567 & 0.50 & 9.75 & 54 & 6.6 & $83 \pm 9$ & $0.24 \pm 0.12$ & $3.5 \pm 1.7$ & 7\_26 & B & \\
YMCA-0054 & 67.3571 & -67.4461 & 0.40 & 9.40 & 62 & 6.9 & $337 \pm 10$ & $0.14 \pm 0.05$ & $2.0 \pm 0.7$ & 7\_26 & A & OGLE-CL LMC 0841\\
YMCA-0055 & 67.6149 & -67.7673 & 0.25 & 9.45 & 27 & 5.2 & $136 \pm 17$ & $0.16 \pm 0.12$ & $2.4 \pm 1.7$ & 7\_26 & C & \\
YMCA-0056 & 67.6596 & -66.9542 & 0.35 & 9.25 & 77 & 15.7 & $486 \pm 10$ & $0.16 \pm 0.05$ & $2.3 \pm 0.7$ & 7\_26 & A & HS~8; KMHK~5\\
YMCA-0057 & 67.8233 & -67.7799 & 0.20 & 9.40 & 22 & 5.5 & $228 \pm 16$ & $0.12 \pm 0.10$ & $1.8 \pm 1.4$ & 7\_26 & C & OGLE-CL LMC 0834\\
YMCA-0058 & 68.8670 & -67.7115 & 0.50 & 9.30 & 112 & 9.2 & $284 \pm 10$ & $0.17 \pm 0.02$ & $2.5 \pm 0.3$ & 7\_27 & A & HS~13; KMHK~11\\
YMCA-0059 & 69.4153 & -66.1983 & 1.30 & 9.25 & 836 & 33.8 & $692 \pm 25$ & $0.38 \pm 0.06$ & $5.5 \pm 0.9$ & 8\_28 & A & ESO 84-30; LW~11; NGC~1644; SL~9 \\
YMCA-0060 & 69.7900 & -67.4985 & 0.25 & 9.40 & 38 & 5.9 & $164 \pm 14$ & $0.19 \pm 0.06$ & $2.8 \pm 0.9$ & 7\_27 & C & \\
YMCA-0061 & 69.9930 & -67.3831 & 0.35 & 9.80 & 62 & 6.0 & $176 \pm 21$ & $0.21 \pm 0.13$ & $3.1 \pm 1.9$ & 7\_27 & C & \\
YMCA-0062 & 93.9911 & -64.9733 & 0.60 & 9.25 & 178 & 17.0 & $615 \pm 19$ & $0.19 \pm 0.06$ & $2.7 \pm 0.8$ & 9\_40 & A & KMHK~1710; LW~444 \\
YMCA-0063 & 94.4250 & -65.5469 & 0.40 & 9.45 & 86 & 11.2 & $456 \pm 9$ & $0.15 \pm 0.02$ & $2.1 \pm 0.3$ & 9\_40 & A & OGLE-CL LMC 0758\\
YMCA-0064 & 94.4670 & -65.3440 & 0.30 & 9.45 & 33 & 5.0 & $85 \pm 8$ & $0.21 \pm 0.16$ & $3.1 \pm 2.4$ & 9\_40 & C & \\
YMCA-0065 & 96.1209 & -66.5061 & 0.45 & 9.20 & 109 & 13.7 & $448 \pm 18$ & $0.18 \pm 0.05$ & $2.7 \pm 0.7$ & 8\_39 & A & OHSC~34; KMHK~1751\\
YMCA-0066 & 96.2493 & -66.5821 & 0.40 & 9.40 & 58 & 6.6 & $235 \pm 14$ & $0.12 \pm 0.02$ & $1.8 \pm 0.4$ & 8\_39 & B & \\
YMCA-0067 & 97.0305 & -67.2605 & 0.35 & 9.35 & 35 & 6.2 & $139 \pm 6$ & $0.18 \pm 0.07$ & $2.7 \pm 1.1$ & 7\_38 & B & \\
YMCA-0068 & 97.4257 & -70.5889 & 0.80 & 9.25 & 226 & 13.4 & $309 \pm 12$ & $0.25 \pm 0.04$ & $3.6 \pm 0.6$ & 4\_33 & A & OHSC~36; KMHK~1757 \\
YMCA-0069 & 97.4821 & -70.4781 & 0.35 & 9.45 & 39 & 6.1 & $124 \pm 6$ & $0.19 \pm 0.15$ & $2.8 \pm 2.1$ & 4\_33 & C & \\
YMCA-0070 & 97.5571 & -64.3277 & 2.50 & 10.10 & 3291 & 227.8 & $1013 \pm 10$ & $0.74 \pm 0.05$ & $10.8 \pm 0.8$ & 10\_43 & A & ESO 87-24; KMHK~1756; LW~481;\\  
 &  &  &  &  &  &  &  &  &  &  &  & NGC~2257 ; SL~895\\
YMCA-0071 & 98.0190 & -70.5724 & 0.40 & 9.35 & 42 & 5.0 & $176 \pm 12$ & $0.11 \pm 0.02$ & $1.7 \pm 0.4$ & 4\_33 & B & \\
YMCA-0072 & 98.2564 & -71.1269 & 1.00 & 9.00 & 417 & 32.8 & $377 \pm 7$ & $0.43 \pm 0.11$ & $6.3 \pm 1.5$ & 4\_33 & A & KMHK~1760; LW~483; SL~897\\
YMCA-0073 & 106.9160 & -69.9839 & 0.50 & 9.75 & 51 & 23.9 & $355 \pm 8$ & $0.16 \pm 0.06$ & $2.3 \pm 0.9$ & 5\_38 & A & KMHK~1762; OHSC~37\\
\hline
    \end{tabular}
    \tablefoot{
The meaning of the different columns is ID, RA, Dec: YMCA identifier and the coordinates of the SCs; R: estimated radius; $\log t$: logarithm of the age; N$_{\rm stars}$: number of stars within the SC radius; G: value of the parameter defined in Eq.~\ref{eq:G}; $n_0$ central number surface density as derived from the fit of the radial density function (Eq.~\ref{Eq.RDP}); $r_c$ estimated core radius expressed in arcmin and parsec (Eq.~\ref{Eq.rc}); Tile: tile in which each SC resides; Flag: robustness ranking of SC: \emph{A} and \emph{B} represent the highest and the intermediate confidence level; \emph{C} reports SCs which necessitate deeper follow-up images to be confirmed.
The last column reports additional IDs of already catalogued SCs as reported in {\tt SIMBAD Astronomical Database}.\\
*: ID numbers begin from 49 since SCs from YMCA-0001 up to YMCA-0048 were presented in \citet{Gatto-2020}.\\
To properly match the position of the stars in the CMD of NGC~2257 (YMCA-0070), we adopted Z = 0.0006 and a distance modulus of $\rm{DM} = 18.37$~mag, as reported in \citet{Milone-2023}.}
    \end{table*}

Table~\ref{tab:sc_params} lists the main parameters of the 25 SCs identified within the 79 remaining YMCA tiles\footnote{The cluster finder algorithm also led to the discovery of YMCA-1, an ancient SC likely associated with the LMC. However, main properties of YMCA-1 were deeply discussed in \citet{Gatto-2021} and \citet{Gatto-2022a}, and therefore YMCA-1 is not included in Table~\ref{tab:sc_params}.}. Within these regions, sixteen of them were already reported in the literature, while the remaining nine are new discoveries. 
As a result, we have increased the number of previously known SCs in these fields by $\simeq50\%$. 

It is important to note that we did not find any new SC within the YMCA tiles located in the periphery of the SMC, and therefore all the 9 newly detected candidate SCs reside in the outskirts of the LMC.
Among the main SC parameters, the table provides the G value\footnote{The already known SCs have $G \geq 5.5$, and the new SC candidates have $G$ values in the range $5.0 \leq G \leq 6.6$.} defined in the Eq.~\ref{eq:G} and a flag assigned subjectively to each SC to classify them as likely real physical systems versus possible asterisms. Our assessment of each SC is based on its statistical significance, such as the ${\rm G}$ parameter, or the discernible presence in the CMD of distinct evolutionary phases, such as the RC or a well-defined RGB.
Additionally, Figures~\ref{fig:known_clusters_cmd} and \ref{fig:new_clusters_cmd} illustrate the CMDs of all the SCs studied in this work.\par
\begin{figure}
    \centering
    \includegraphics[width=0.45\textwidth]{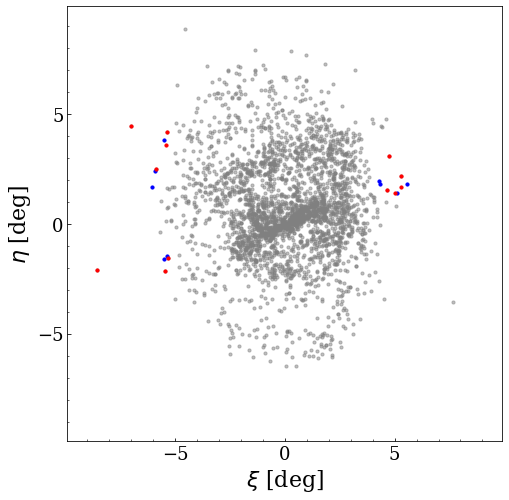}
    \caption{Relative positions of the 21 SCs located in the surroundings of the LMC, with respect to the LMC centre. Coloured points indicate the clusters identified in this work, in red those already known in the literature, in blue the newly discovered ones. Grey points represent SCs reported in the catalogue by \citet[][]{Bica-2008}.}
    \label{fig:scs_position}
\end{figure}

Figure~\ref{fig:scs_position} illustrates the relative spatial position of the 21 SCs located in the LMC periphery, with respect to the centre of the LMC itself. The majority of the SCs detected in this work are located in close proximity to those previously known. In particular, all newly discovered SCs reside within 7\degree~from the LMC centre, consistent with previous findings \citep[][]{Pieres-2016, Gatto-2020}, in spite of the fact that the LMC disc extension is significantly larger \citep[e.g.][]{Saha-2010}. Note that this result is not due to an observational bias, as the YMCA tiles used in this work cover regions of the LMC up to $\sim$15\degree~from its centre.
The same Fig.~\ref{fig:scs_position} suggests the presence of a possible SC clump at ($\xi$, $\eta$) $\simeq$ (5\degree,-2\degree), corresponding to the tiles $7\_26 - 7\_27 - 8\_28$.
These tiles are amongst the ones with the highest number of stars and the closest to the LMC centre. Consequently, in principle, it may be hypothesized that the SC over-density in this region is a result of the highest number of stars located in these tiles. However, this hypothesis is challenged by the observation of a low number of SCs in tiles with a similar number of stars (e.g. tile $8\_39 - 9\_40$) or by the complete absence of SCs in tiles at similar radial distances (e.g. tile $6\_24$).
As the SC agglomeration is located at the edge of the LMC bar, a more plausible explanation may be a higher level of star formation occurred in this region in the last Gyrs \citep{Mazzi-2021}.\par

\subsection{The age distribution of the SCs in the LMC outskirts}

\begin{figure}
    \centering
    \includegraphics[width=0.45\textwidth]{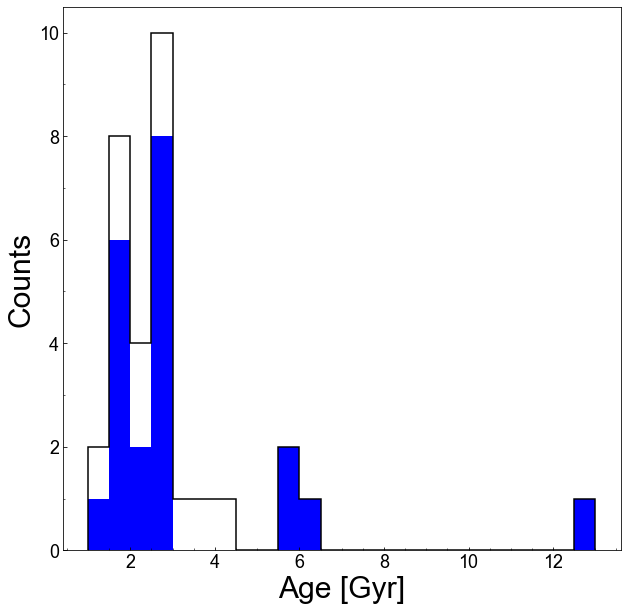}
    \caption{Age distribution of the 21 SCs identified in the LMC periphery, as examined in this work (depicted by the blue filled histogram). Additionally, we integrate data from 10 SCs previously analyzed in \citet{Gatto-2020} with $G \geq$ 5 (represented by the black histogram).}
    \label{fig:sc_age}
\end{figure}

Figure~\ref{fig:sc_age} shows the age distribution of the 21 candidate SCs in the LMC outskirts, as well as of 10 SCs reported in \citet{Gatto-2020} with $G \geq$ 5. The distribution exhibits a prominent peak at approximately 2.0-3.0~Gyr, which is consistent with the findings of \citet{Gatto-2020}. The number of SCs steadily decreases towards younger ages ($t \leq 2$ Gyr), while on the opposite side, after 3~Gyr, namely slightly before the beginning of the period dubbed as the age gap, it sharply drops down.
The sudden SC formation enhancement at $\sim 3.0$ Gyr may be associated with a past close encounter between the MCs, which triggered a new episode of massive SC formation.
This is independently supported by different works that tried to reconstruct the history of the orbital path of the MCs, and most of them agree that a past pericentric passage between the LMC and SMC happened between 1 and 3 Gyr ago \citep[see for example,][and references therein]{Bekki&Chiba2005,Patel-2020}.
Interestingly, \citet[][]{Pieres-2016} reported a similar peak at $\sim$ 2.7 Gyr by analysing the CMD of 109 SCs to the Northern side of the LMC. However, they also evidenced a major peak at $\sim$ 1.2 Gyr that is not present in our SC sample. 
Indeed, we detected only two SCs younger than 1.5~Gyrs and none of them is younger than 1 Gyr, suggesting that no recent SC formation event occurred in the LMC outskirts, at least in our analyzed tiles. The lack of young SCs (younger than 1 Gyr) in our sample is consistent with the general consensus that the periphery of the LMC is inhabited by an old population \citep{Saha-2010,Piatti&Geisler2013,Mazzi-2021}.\par
Alternatively, these results may indicate that an episodic increase in the SC formation $\sim 1-2$~Gyrs ago occurred in the Northern region of the LMC, investigated by \citet[][]{Pieres-2016}, leaving quite unaltered the fields investigated in this work.
To explore this possibility, we took advantage of \citet{Mazzi-2021}'s work that derived the spatial-resolved SFH of the LMC in 96 deg$^2$, down to sub-spatial regions of 0.125 deg$^2$, using near-infrared observations from the VMC \citep{Cioni-2011}.  
Examining their fig. 5, we noticed that during the 1-1.58 Gyr interval (left-lower panel in their figure), the LMC star formation was more intense within the central bar, and in a morphological feature resembling a spiral arm that originates from the West edge of the bar and wraps to the North, encompassing the regions investigated by \citet{Pieres-2016}, in a counter-clockwise direction. 
At the greater look-back time, the SF is more evenly distributed around the LMC, with only the bar displaying a significantly more intense SF rate.
Based on these observations, it seems plausible to infer that the region explored by \citet{Pieres-2016} was undergoing an enhanced SC formation at about 1~Gyrs ago, while the LMC outskirts in other directions, including the fields analyzed in this work\footnote{Note that our fields are slightly beyond the VMC tiles used by \citet{Mazzi-2021} to retrieve the LMC SFH.}, were in a more quiescent phase, after a more intense SF activity occurred 2-3~Gyrs ago throughout the LMC, as also indicated by the secondary peak at 2.7~Gyrs identified by \citet{Pieres-2016}.\par
Even more importantly, as illustrated in Fig.~\ref{fig:sc_age}, we confirm the existence of three reliable candidate SCs within the age gap, in particular having ages of 5.5-6.0~Gyr. 
Amongst them, we have already demonstrated \citep{Gatto-2022c} that KMHK~1762 (YMCA-0073) is the third ever confirmed age gap SC, besides ESO121-03 and KMHK~1592 \citep{Piatti2022b}. The other two candidate SCs residing in the age-gap are YMCA-0053 and YMCA-0061.
On these bases, and considering also the detailed discussion in \citet{Gatto-2020}, we shall conclude that the age gap in the SC population of the LMC, may either be significantly narrower than previously believed, lasting no more than 3 Gyr (e.g. between 7 - 10 Gyr), or it may be an artefact of the photometric and spatial limitations attained by the previous surveys. 
If confirmed, our results would fix the long-lasting debated issue about a different evolutionary path between SC and stellar field components in the LMC.
We cannot exclude that low-mass SCs formed in the (now narrower) age gap were tidally disrupted (shocked) during the LMC-SMC encounter around 3 Gyr ago which thereby removed those clusters, whilst triggering the formation of new SCs around that time. This would allievate the cluster-field tension also.\par 
In order to confirm the reality of the newly discovered SC candidates, especially those within the age gap, further photometric follow-up is necessary. Indeed, given that the new SC candidates are generally old and faint, very deep photometry, up to four magnitudes below the MSTO (i.e. $V\sim$ 26.5 mag) obtained in very good seeing conditions (to reduce confusion) would be valuable to confirm them as real physical systems. 
The advent of the Rubin-LSST telescope is expected to greatly aid in this endeavour, as it promises to deliver deep photometry across the entire southern sky, and has a dedicated observational program for the MCs \citep[][]{Olsen-2018}.

\subsection{Density profiles}
\label{sec:fit_rdp}

To estimate the structural parameters of the SCs under investigation we derived
their radial density profiles (RDP) and we fitted them with the Elson, Fall, and Freeman \citep[EFF;][]{Elson&Fall&Freeman1987} profile:
\begin{equation}
    n(r) = n_0 \times \left(1 + \frac{r^2}{\alpha^2}\right)^\frac{-\gamma}{2} + \phi
    \label{Eq.RDP}
\end{equation}
where $n(r)$ represents the number of stars per squared arcminute as a function of the distance from the centre of the cluster, $n_0$ is the central surface density, $\alpha$ is the core parameter, $\gamma$ is the slope parameter, and $\phi$ is the background value, which is a free parameter of the fit. 
We obtained the best fitting parameters through the module {\it curve fit} of the {\sc Python} library {\it scipy}. These parameters can be used to calculate the core radius of the SCs using the equation:
\begin{equation}
    r_c = \alpha \times \sqrt{2^{2/\gamma} - 1}
    \label{Eq.rc}
\end{equation}
We inserted the values of $n_0$ and the core radius for each SC in both arcminute and parsec in Table~\ref{tab:sc_params} (columns eight to ten). To obtain the physical size of the core radius, we applied a correction for the adopted distance modulus of the LMC or the SMC, depending on the SC membership.
Figures~\ref{fig:known_clusters_rdp} and \ref{fig:new_clusters_rdp} display the RDPs and their best fit profile for the known SCs analyzed in this work, and for the new candidate SCs, respectively.

\section{Summary}

We have presented and discussed the data acquired with the YMCA survey, a wide-field optical imaging project conducted with the VLT Survey Telescope using $g$ and $i$ filters. The primary objective of YMCA is to reconstruct the evolution of the MCs and unravel their complex interaction history. 
It covers 110 square degrees in the periphery of both the MCs. Specifically, this survey covered the East and West side of the LMC, the North-West side of the SMC, and a strip of 1\degr~in size above the Magellanic Bridge that links the two Clouds.
These targeted regions complement the optical STEP survey, which probed the main body of the SMC and the Bridge, utilizing the same instrument and with similar observing constraints. Together, YMCA and STEP provide a deep catalogue covering more than 160 square degrees within the Magellanic System. The YMCA catalog of stars adopted in this work will be published at the CDS.\par
The depth and accuracy achieved by YMCA allows us to resolve individual stars down to $\simeq$ 1.5-2.0 mag below the MSTO of the oldest stellar population of the MCs, opening a view into the early stages of their evolution.
While preliminary studies based on subsamples of the YMCA catalogue already demonstrated the potential of this survey in addressing critical questions about the past MCs' evolutionary path \citep{Gatto-2020,Gatto-2021,Gatto-2022c}, here we presented an overall analysis of the stellar field population within the periphery of the MCs, including their SC system, by leveraging the full YMCA dataset.\par
The analysis of the deep CMD constructed with the YMCA photometric catalogue suggests:
\begin{itemize}
    \item The peripheries of the LMC and the SMC are dominated by intermediate-age and old stellar populations.
    This finding is common amongst star-forming dwarf galaxies in the Local Group. \citep[see the review by][and references therein]{Annibali&Tosi2022}
    \item While the North-West side of the SMC does not exhibit the presence of young stars, a moderate blue plume, indicating a younger (t $\sim$ 300 - 500 Myr) stellar population is evident within the YMCA tiles closest to the LMC, on both its West and East sides. Notably, the tile 7\_27, situated near the West edge of the LMC bar, shows a more pronounced blue plume, suggesting that a more recent event (t $\simeq$ 200 Myr) of star formation may have occurred.
    \item Consistently with observations by \citet{Mackey-2018}, the LMC disc appears to be truncated on its Western side at about $9\degr$, namely 8 kpc from the LMC centre. This feature might be the consequence of a close encounter with the SMC \citep[][]{Belokurov&Erkal2019}.
    \item We confirmed the presence of a foreground stellar population, hinted by the presence of a double RC in the CMD, towards the North-Eastern direction of the SMC \citep[see also][]{Omkumar-2021, James-2021}. The closer substructure emerges at $\simeq 2.8\degree$ from the SMC centre, and is clearly visible up to $\simeq 5.5\degree$, in line with previous studies \citep{Omkumar-2021}. 
    \end{itemize}
In addition, we investigated SCs properties within the YMCA footprint to get additional insights into the MCs' past evolution. This pursuit involved an in-depth analysis of SCs properties already documented in the literature by means of the deeper YMCA survey, as well as an extensive research to detect unknown SCs adopting a customized algorithm applied to the YMCA dataset. In particular:
\begin{itemize}
    \item We detected 9 new candidate SCs in the outskirts of the LMC.
    \item In agreement with previous studies, the age distribution of the SCs showcases a pronounced peak at $\simeq$ 2--3 Gyr. This finding potentially sheds light on the timeline of a past close encounter between the LMC and the SMC, the likely main actor of such a sudden increase in the SC formation activity.
    \item The age distribution of the SCs examined in this work challenges the actual existence of the so-called age gap in the LMC, proposed more than three decades ago \citep{Jensen-1988,DaCosta1991}, and already questioned in \citet{Gatto-2020} and in \citet{Gatto-2022c}.
\end{itemize}
Deeper photometric follow-ups are crucial to confirm the existence of a population of SCs born during the age-gap period, and thus to finally resolve the long-standing debated discrepancy between the star field and SC formation histories observed in this broad period of the LMC lifetime.\par

\begin{acknowledgements}

This research was made possible through the use of the AAVSO Photometric All-Sky Survey (APASS), funded by the Robert Martin Ayers Sciences Fund and NSF AST-1412587.\\
This work has made use of data from the European Space Agency (ESA) mission
{\it Gaia} (\url{https://www.cosmos.esa.int/gaia}), processed by the {\it Gaia} Data Processing and Analysis Consortium (DPAC, \url{https://www.cosmos.esa.int/web/gaia/dpac/consortium}). Funding for the DPAC has been provided by national institutions, in particular the institutions participating in the {\it Gaia} Multilateral Agreement.\\
M.G. acknowledges the INAF AstroFIt grant 1.05.11. M.B. acknowledges the financial support by the Italian MUR through the grant PRIN 2022LLP8TK\_001 assigned to the project LEGO – Reconstructing the building blocks of the Galaxy by chemical tagging (P.I. A. Mucciarelli), funded by the European Union – NextGenerationEU.
Project PRIN MUR 2022 (code 2022ARWP9C) “Early Formation and Evolution of Bulge and HalO (EFEBHO)", PI: Marconi, M.,  funded by European Union – Next Generation EU.
Large grant INAF 2023 MOVIE (PI: M. Marconi).
\\
In this work we used the following softwares: TOPCAT \citep{Topcat}{}{}, STILTS \citep{STILTS}{}{}, DAOPHOT IV/ALLSTAR \citep{Stetson1987,Stetson1992}, SExtractor \citep{SExtractor}, NumPy \citep{Numpy}{}{}, pandas \citep[][]{Pandas-2010,Pandas-2020}{}{}, Matplotlib \citep{Matplotlib}{}{}, Scikit-learn \citep{scikit-learn}{}{}, LMFIT \citep[][]{LMFIT}{}{}.

\end{acknowledgements}

%
%

\bibliographystyle{aa}
\bibliography{unatesi}

\begin{appendix}



\section{Age comparison with literature}
\label{app:age_comparison}

As a crucial validation step of the procedure described in Sect.~\ref{sec:isochrone_fitting}, we undertook a comparative analysis of the estimated ages for SCs derived using our visual fitting procedure with those available in literature. This was done to ensure that fixing two parameters, namely distance modulus and metallicity, did not introduce any bias into the age assessment process. 
Specifically, we identified 16 SCs (comprising 3 in the vicinity of SMC, 1 within the studied Strip, and the remaining 12 in the outskirts of the LMC), all of which had been previously documented in various works (refer to Section~\ref{sec:candidate_SCs} for details). Among these 16 SCs, age values were successfully retrieved for 10 of them. Therefore, for 6 SCs that were already known we provide for the first time their age estimations.
It is noteworthy that KMHK1762 was excluded from the analysis of the aforementioned 10 SCs, as in a dedicated study \citep{Gatto-2022c}, we demonstrated that, considering the fainter limiting magnitude of the YMCA dataset, KMHK~1762 is significantly older than previously thought.\par
\begin{figure}[h!]
    \includegraphics[width=.45\textwidth]{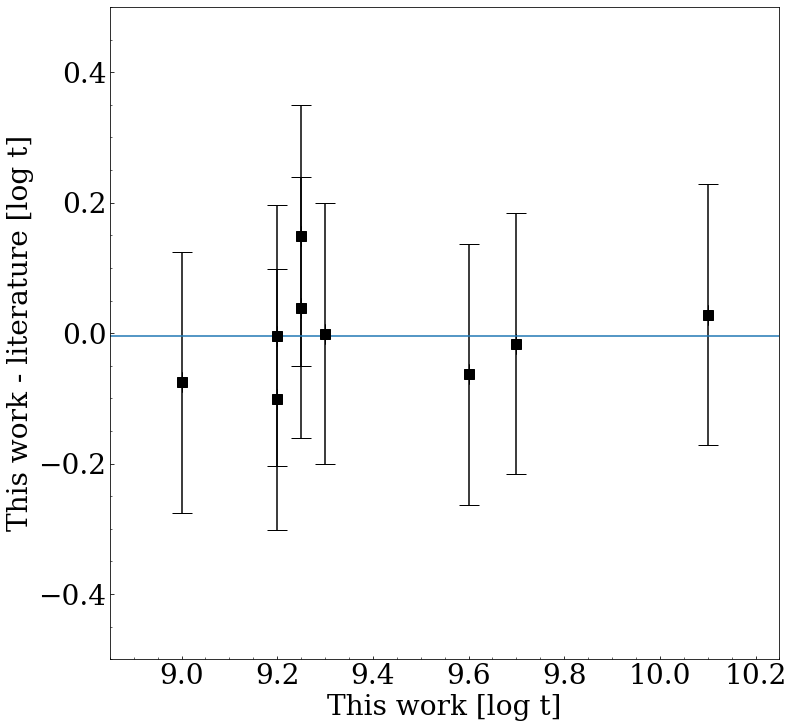}
    \caption{Age difference between this work and literature values as a function of the age estimated in this work. The horizontal blue line indicates the mean of the difference, which is $\log (t_{\rm our}) - \log (t_{\rm lit})$ = -0.005 dex. Errorbars on the y-axis represent our estimated uncertainties, namely $\sigma_{\log t} = 0.2$~dex. Literature values have been taken from \citet{Parisi-2014}, \citet{Piatti&Bastian2016}, \citet{Piatti-2018}, 
    \citet{Maia-2019}, and \citet{Milone-2023}.}
    \label{fig:age_lit_comparison}
\end{figure}

Table~\ref{tab:sc_age_literature} displays the list of SCs with ages reported in literature and their references.
Figure~\ref{fig:age_lit_comparison} illustrates the comparison between the ages determined through our visual isochrone matching and those obtained from prior studies for 9 SCs (6 located in the outskirts of the LMC and 3 in the periphery of the SMC). Importantly, the age values for these 9 SCs were sourced from five distinct works\footnote{In cases where a SC had been studied by multiple works, we retrieved the age value provided by the most recent work.} (see caption of the figure). Moreover the age estimations were derived using diverse methodologies, resulting in a highly heterogeneous dataset.
Nevertheless, the plot reveals a good agreement with respect to the literature's age estimations, with a mean difference of $\log (t_{\rm our}) - \log (t_{\rm lit})$ = -0.005 dex, as indicated by the blue line in the figure. All SC age estimations are consistent within their errors when compared to values provided by other works. Notably, this agreement does not exhibit any discernible trend with age. This finding indicates that YMCA photometry is sufficiently accurate even at faint magnitudes to allow us to visually identify the actual magnitude of the MSTO of these SCs. 
In light of these considerations, we assert our confidence that the steps taken in our isochrone matching procedure did not introduce any significant effect on the determination of SC ages.

\begin{table}
\caption{Comparison between the SC ages estimated in this work and literature values.}
    \label{tab:sc_age_literature}

\footnotesize\setlength{\tabcolsep}{3pt}
    \begin{tabular}{l|c|c|c|c}
    \hline\hline
    ID & $\log t$ (Our) & $\log t$ (Lit.) & Reference & Simbad Name\\
     \hline
YMCA-0049 & 9.60 & $9.66 \pm 0.02$ & 1 & L~32\\
YMCA-0050 & 9.70 & $9.72 \pm 0.04$ & 1 & ESO 51-9\\
YMCA-0051 & 9.20 & $9.20 \pm 0.05$ & 1 & HW~84\\
YMCA-0056 & 9.25 & $9.10 \pm 0.10$ & 2 & KMHK~5\\
YMCA-0058 & 9.30 & $9.30 \pm 0.10$ & 2 & KMHK~11\\
YMCA-0059 & 9.25 & $9.21 \pm 0.05$ & 3 & NGC~1644\\
YMCA-0065 & 9.20 & $9.30 \pm 0.02$ & 4 & KMHK~1751\\
YMCA-0070 & 10.10 & $10.07 \pm 0.02$ & 3 & NGC~2257\\ 
YMCA-0072 & 9.00 & $9.08 \pm 0.05$ & 5 & KMHK~1760\\
\hline
    \end{tabular}
    \tablefoot{The different columns display: YMCA SC IDs, ages estimated in this study, ages from previous studies with references, literature IDs from {\tt SIMBAD Astronomical Database}. Our estimated uncertainty for SC ages is $\sigma_{\log t} = 0.2$~dex, as described in Sect.~\ref{sec:isochrone_fitting}.\\
    References are: (1) \citet{Parisi-2014}; (2)  \citet{Piatti-2018}; (3) \citet{Milone-2023}; (4) \citet{Piatti&Bastian2016}; (5) \citet{Maia-2019}.}
    \end{table}

\section{CMD of SCs in the MC periphery}
\label{app:CMD_all_clusters}

In this section we present the CMDs of all SCs analyzed in this work, a comparison with the CMD of the local stellar field, and the relative spatial distribution of their star with respect to their estimated centre. In particular, in Fig.~\ref{fig:known_clusters_cmd} we show these plots for the SCs already reported in the literature, whereas Fig.~\ref{fig:new_clusters_cmd} displays the same subplots for the candidate SCs discussed in this work.

\begin{figure*}

    \includegraphics[width=0.35\textwidth]{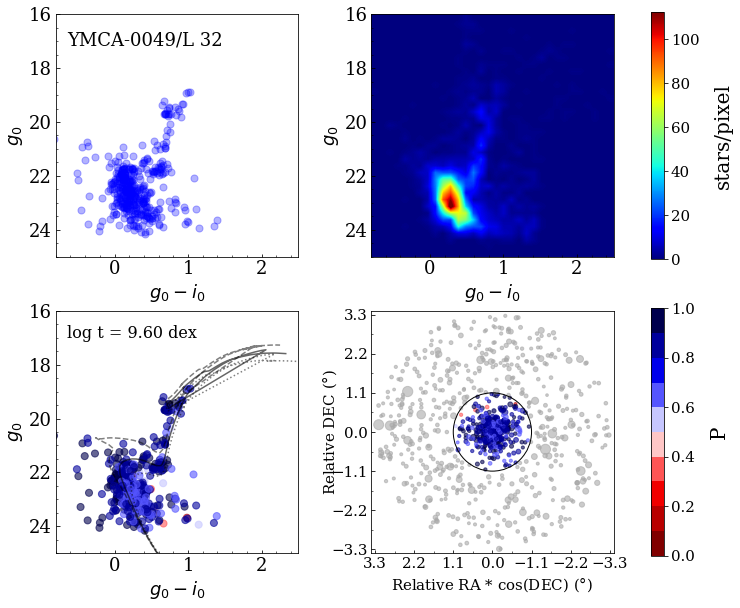}
    \includegraphics[width=0.35\textwidth]{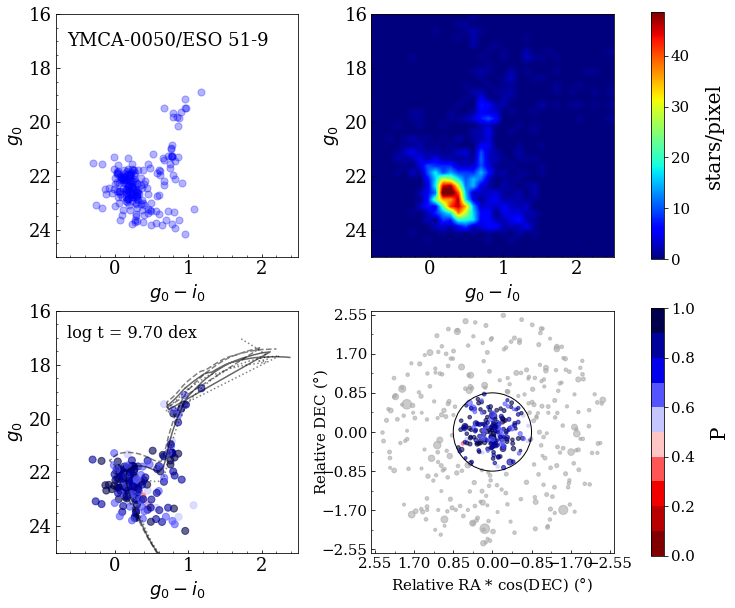}
    \includegraphics[width=0.35\textwidth]{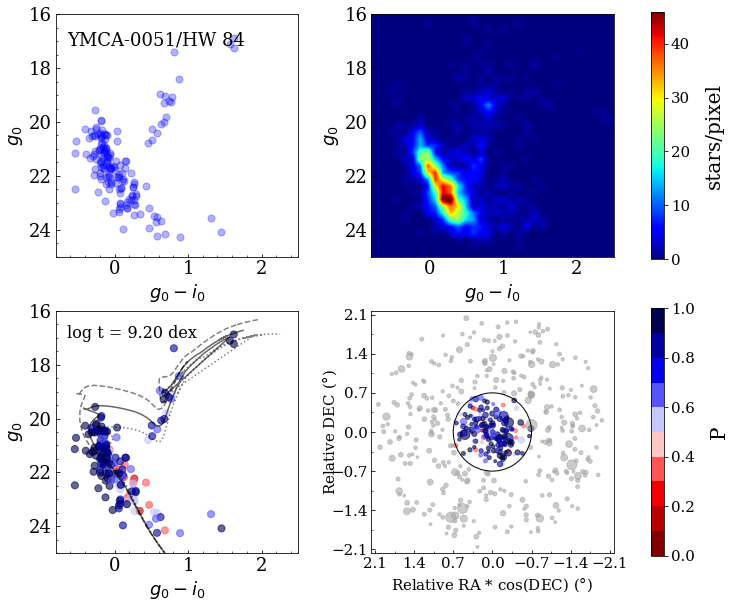}\\
    \includegraphics[width=0.35\textwidth]{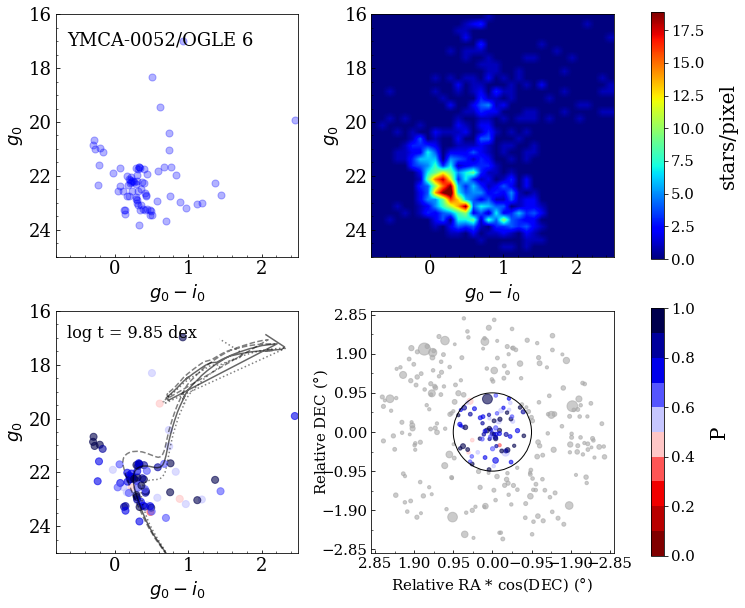}
    \includegraphics[width=0.35\textwidth]{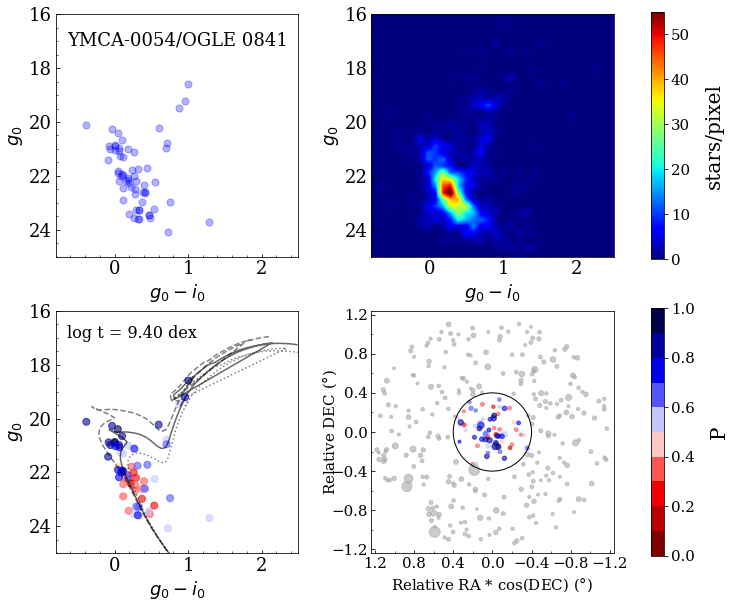}
    \includegraphics[width=0.35\textwidth]{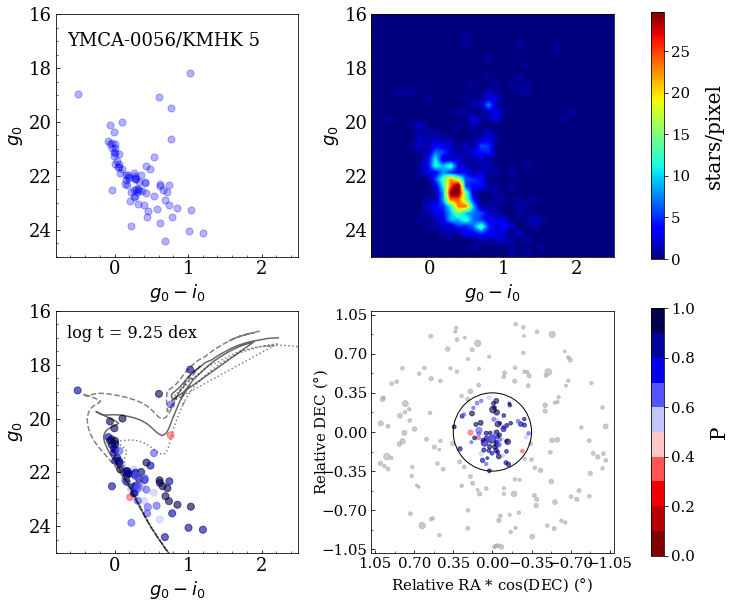}\\
    \includegraphics[width=0.35\textwidth]{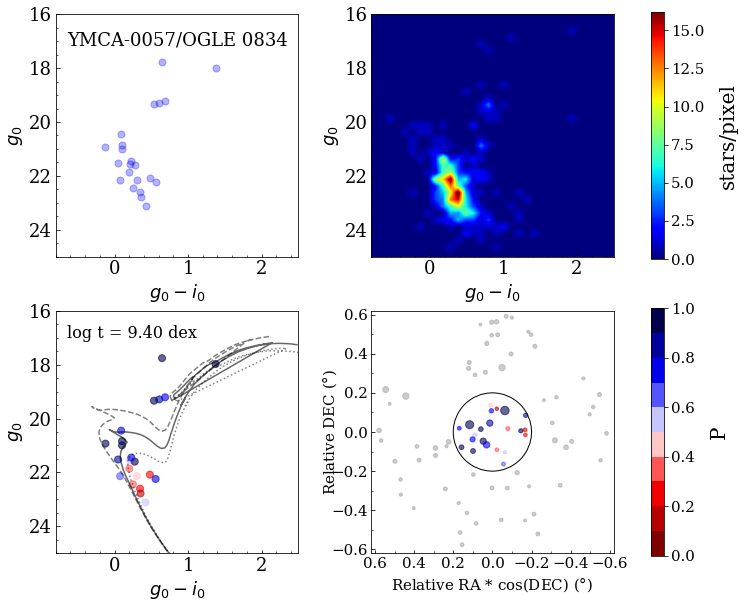}
    \includegraphics[width=0.35\textwidth]{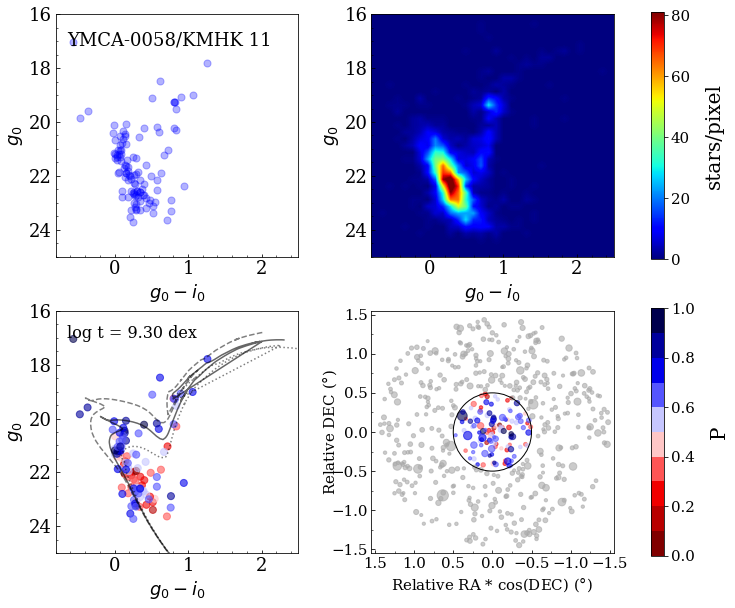}
    \includegraphics[width=0.35\textwidth]{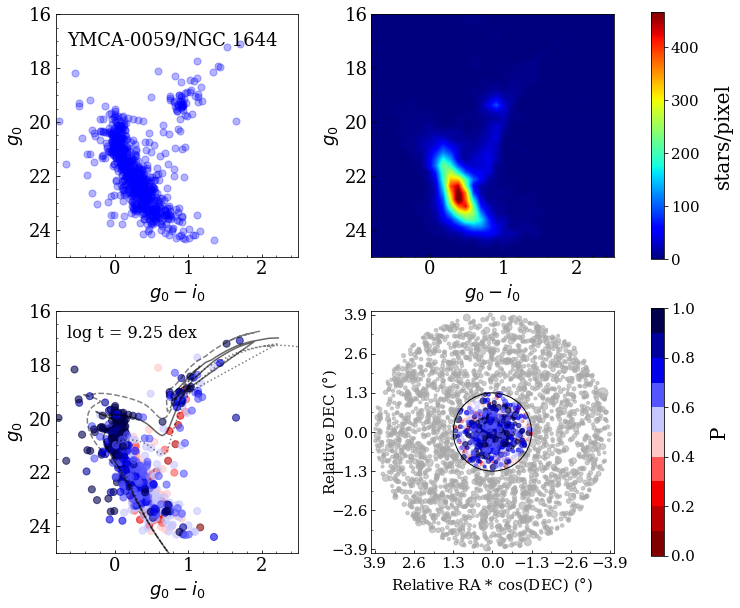}\\
    \includegraphics[width=0.35\textwidth]{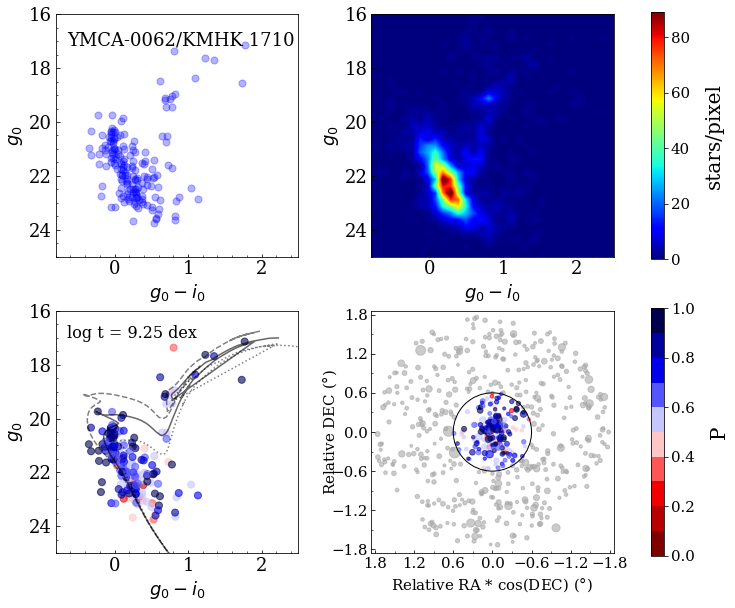}
    \includegraphics[width=0.35\textwidth]{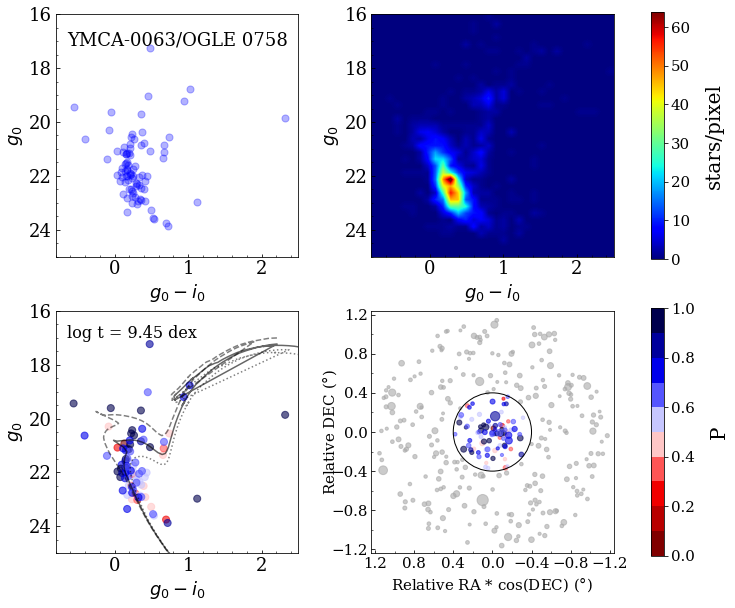}
    \includegraphics[width=0.35\textwidth]{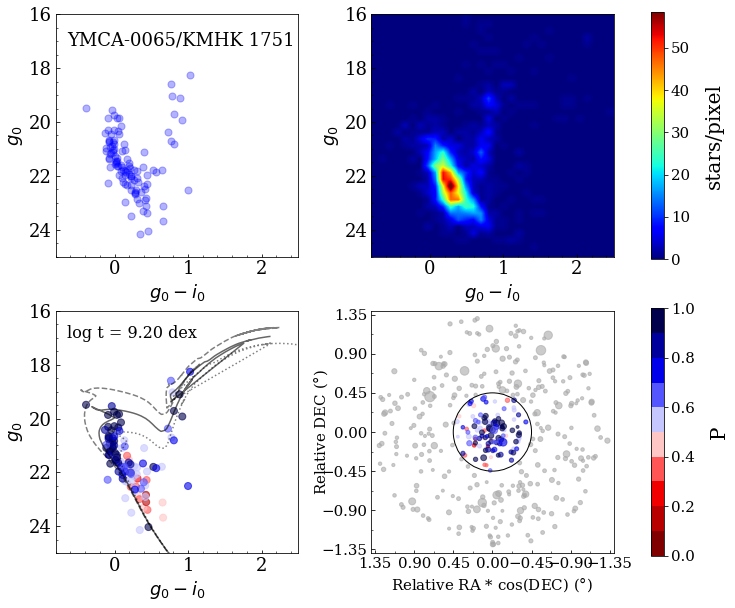}\\
    \caption{SCs already known analyzed in this work. For each of them this figure illustrates: \emph{Top Left}: CMD of the candidate SC; \emph{Top Right}: Hess diagram of the six fields used as representative of the local stellar field contamination; \emph{Bottom Left}: CMD of the SC with stars colour coded according to their membership probability of being a cluster member along with the best fitting isochrone (solid line) and isochrones at log(t) = $\pm$0.2 dex from the best one; \emph{Bottom Right}: Spatial distribution of stars with respect to the SC centre. The solid circle marks the estimated SC radius.} 
    \label{fig:known_clusters_cmd}
\end{figure*}

\begin{figure*}\ContinuedFloat
    \includegraphics[width=0.35\textwidth]{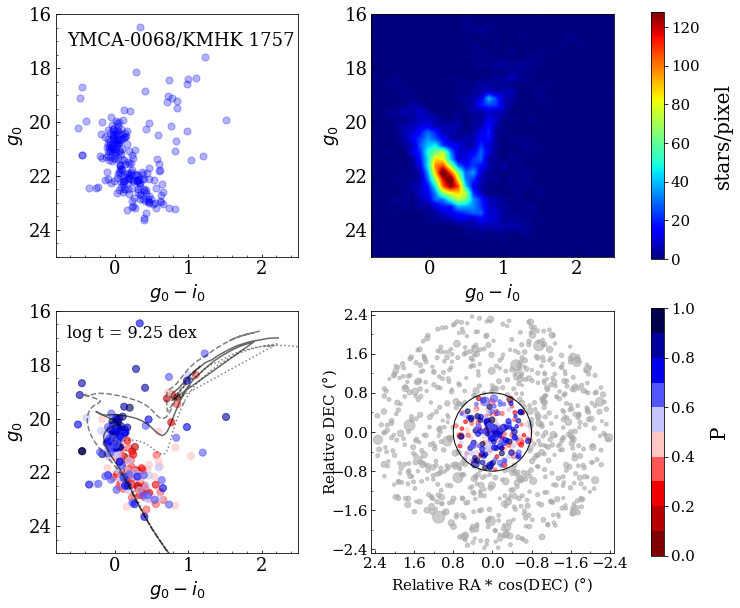}
    \includegraphics[width=0.35\textwidth]{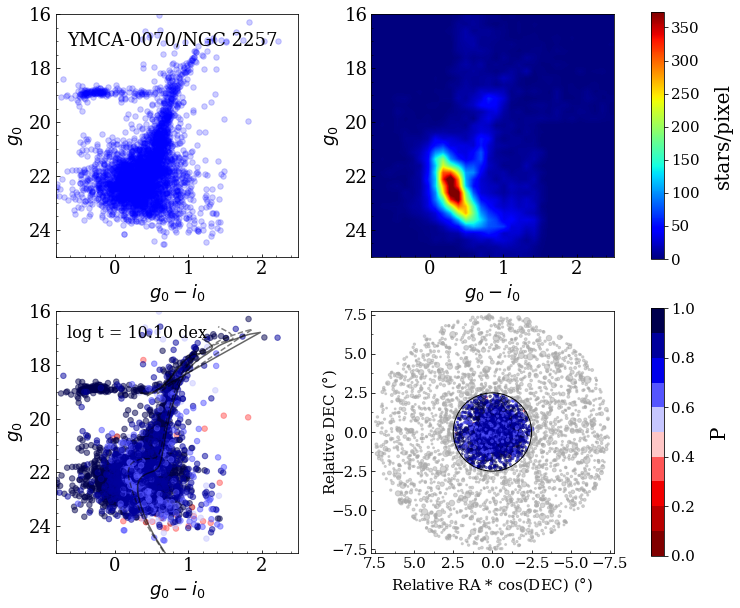}
    \includegraphics[width=0.35\textwidth]{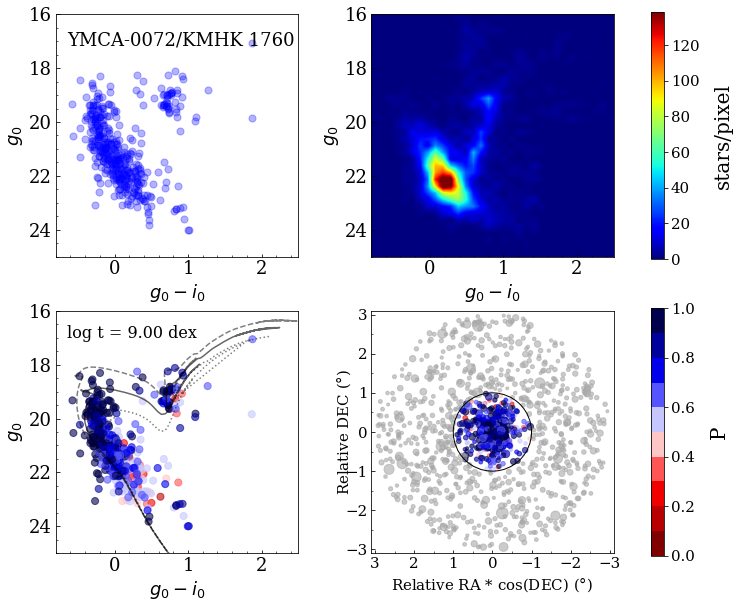}\\
    \includegraphics[width=0.35\textwidth]{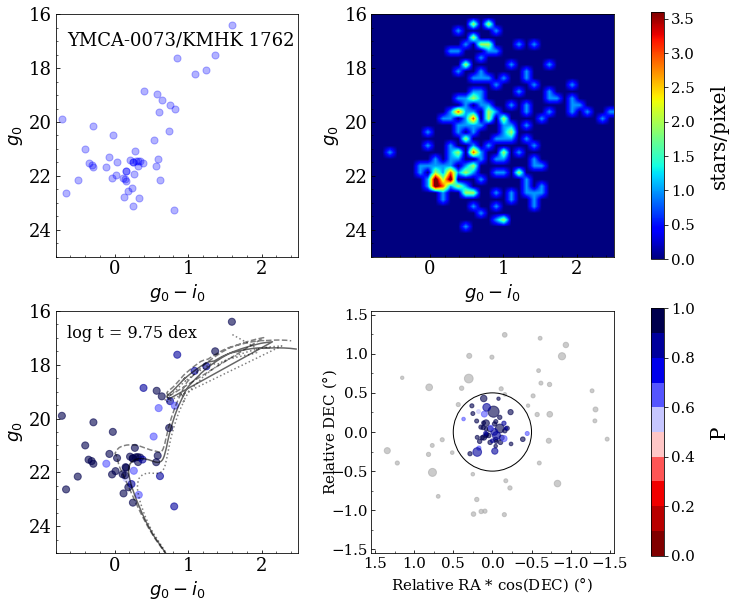}
    \caption{Continued}
\end{figure*}

\begin{figure*}
    \includegraphics[width=0.35\textwidth]{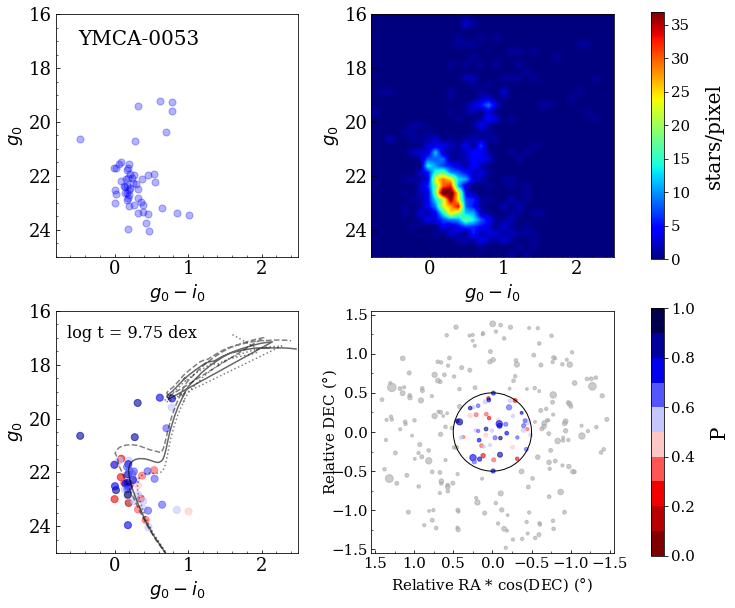}
    \includegraphics[width=0.35\textwidth]{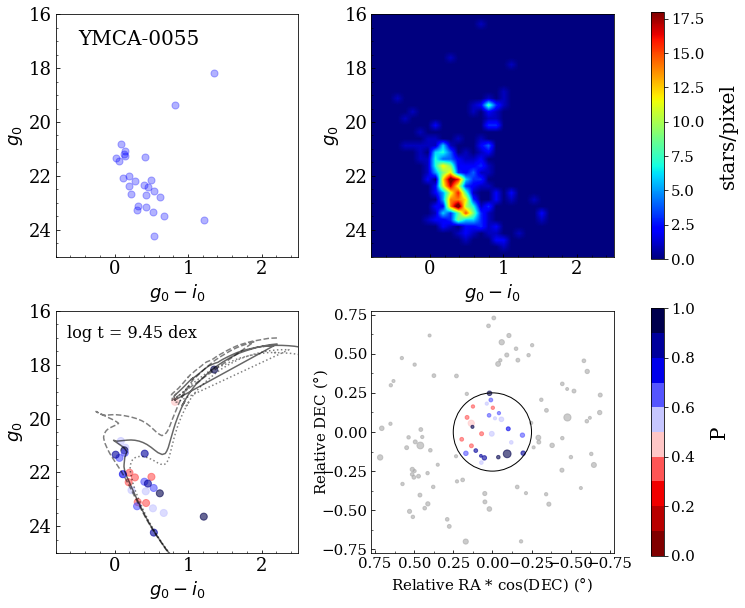}
    \includegraphics[width=0.35\textwidth]{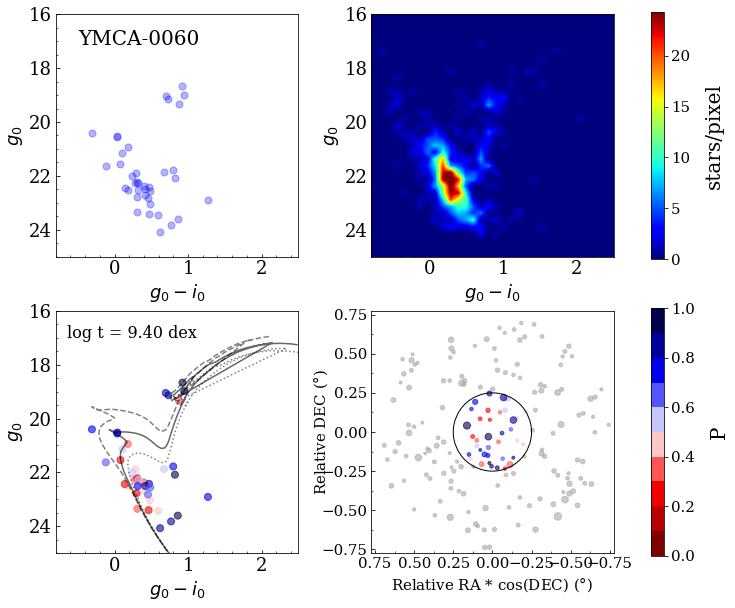}\\
    \includegraphics[width=0.35\textwidth]{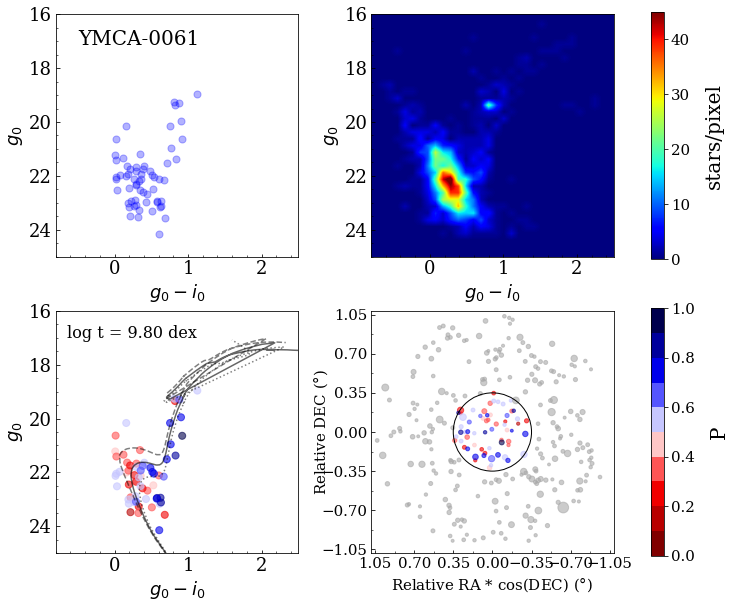}
    \includegraphics[width=0.35\textwidth]{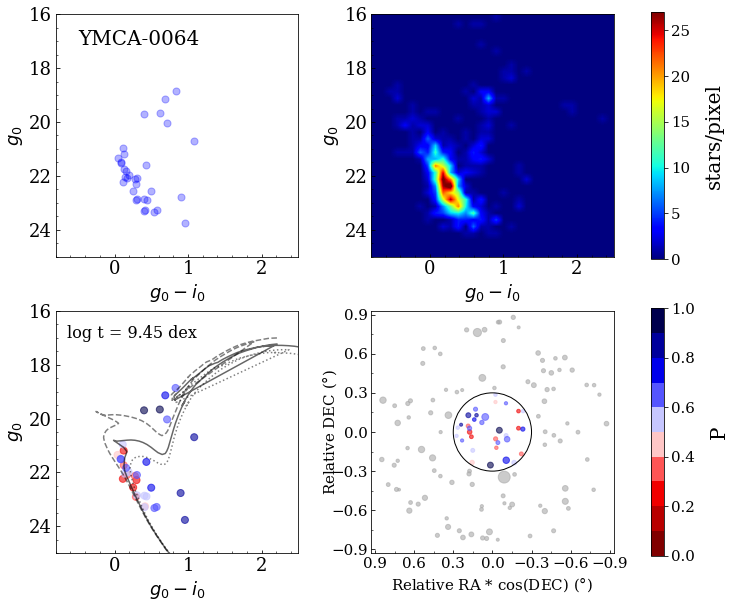}
    \includegraphics[width=0.35\textwidth]{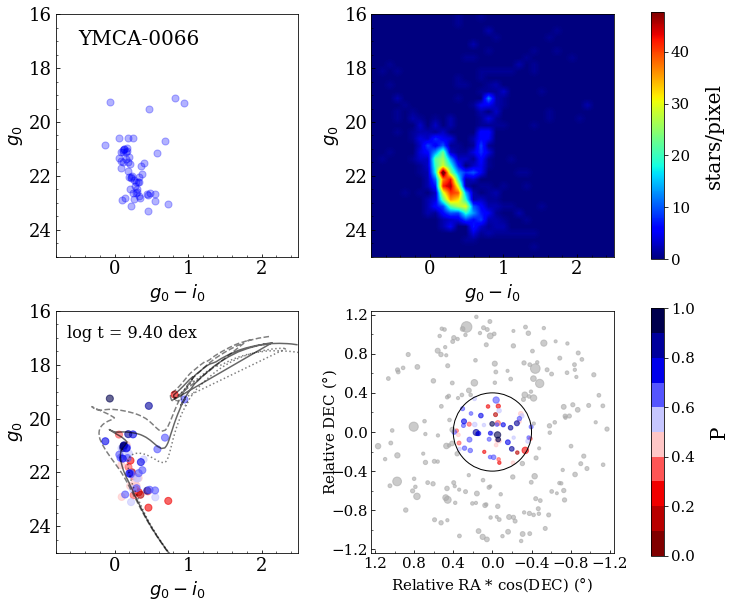}\\
    \includegraphics[width=0.35\textwidth]{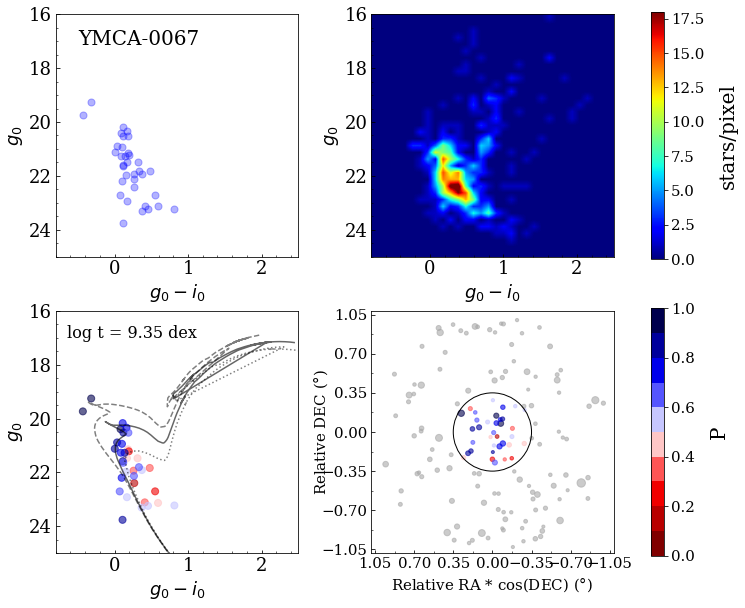}
    \includegraphics[width=0.35\textwidth]{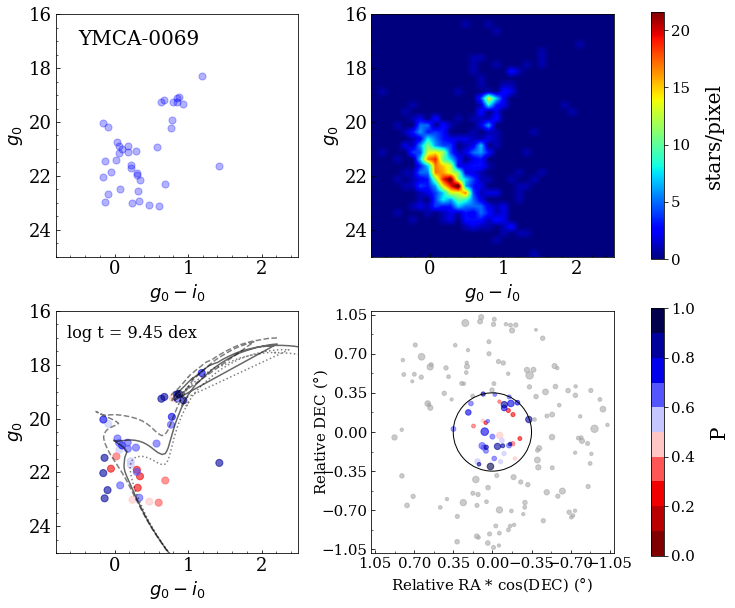}
    \includegraphics[width=0.35\textwidth]{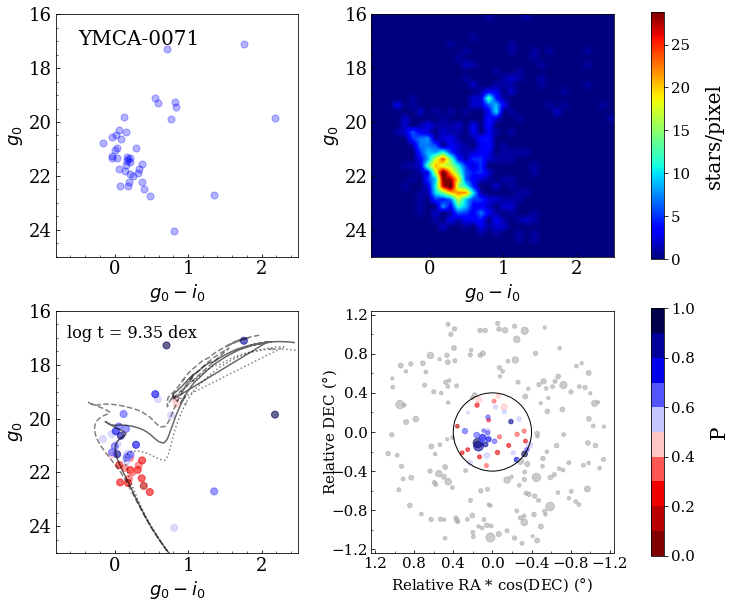}\\
    
    \caption{Same as Figure~\ref{fig:known_clusters_cmd}, but for the new candidate SCs.}
    \label{fig:new_clusters_cmd}
\end{figure*}

\section{Radial density profile of SCs in the MC periphery}
\label{app:RDP_all_clusters}

In this section we present the RDPs and their best fit EFF profile for all SCs analyzed in this work. In Fig.~\ref{fig:known_clusters_rdp} we show the RDPs for the SCs already reported in literature, whereas Fig.~\ref{fig:new_clusters_rdp} displays the RDPs for the candidate SCs discussed in this work.

\begin{figure*}
    \includegraphics[width=0.35\textwidth]{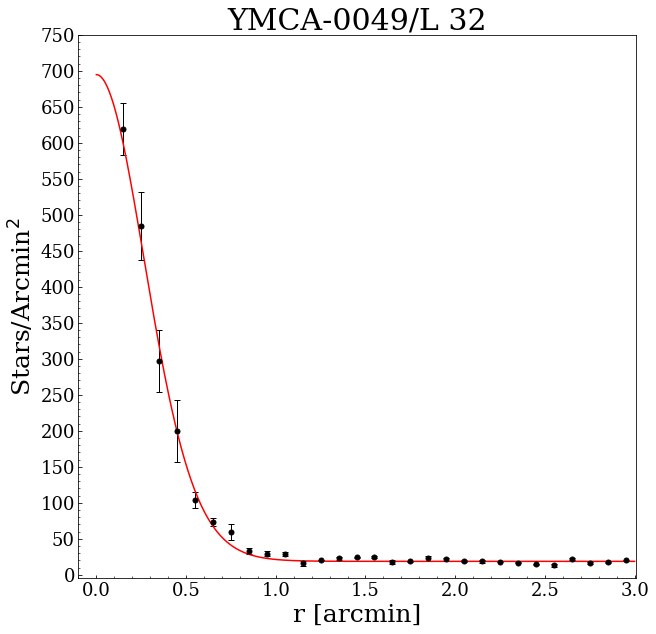}
    \includegraphics[width=0.35\textwidth]{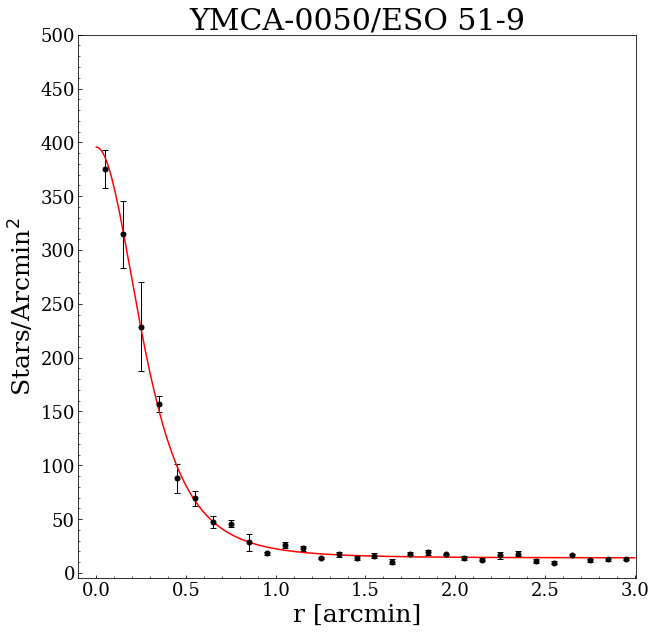}
    \includegraphics[width=0.35\textwidth]{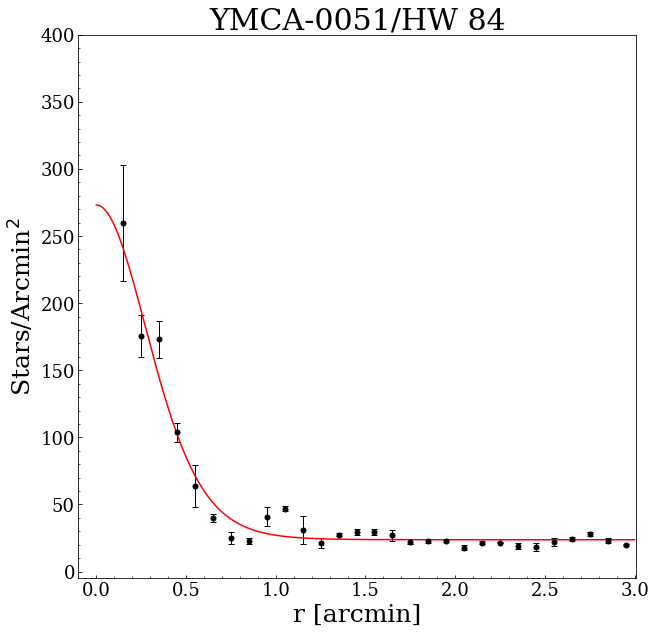}\\
    \includegraphics[width=0.35\textwidth]{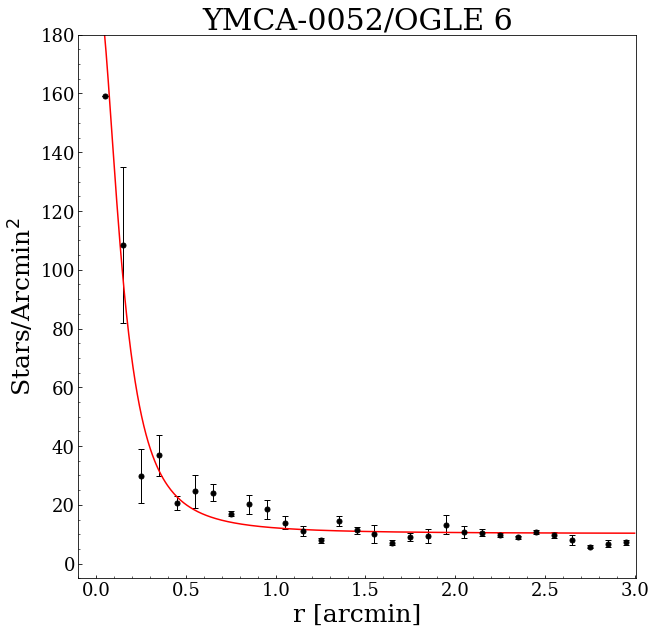}
    \includegraphics[width=0.35\textwidth]{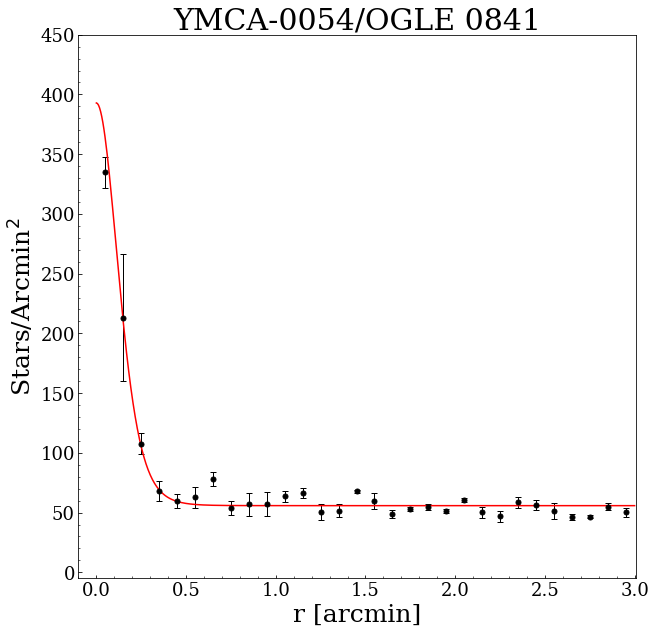}
    \includegraphics[width=0.35\textwidth]{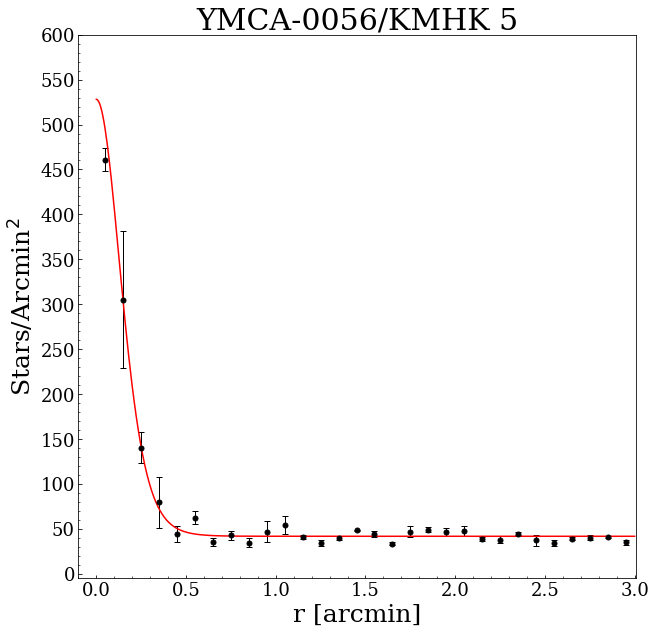}\\
    \includegraphics[width=0.35\textwidth]{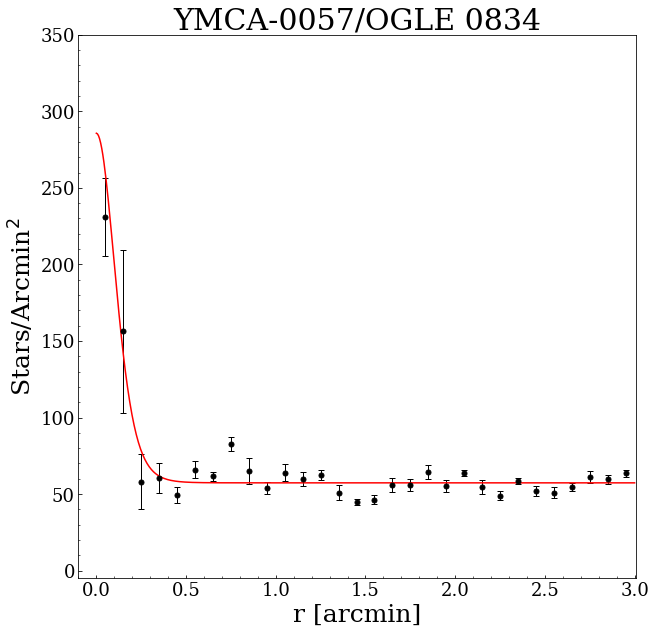}
    \includegraphics[width=0.35\textwidth]{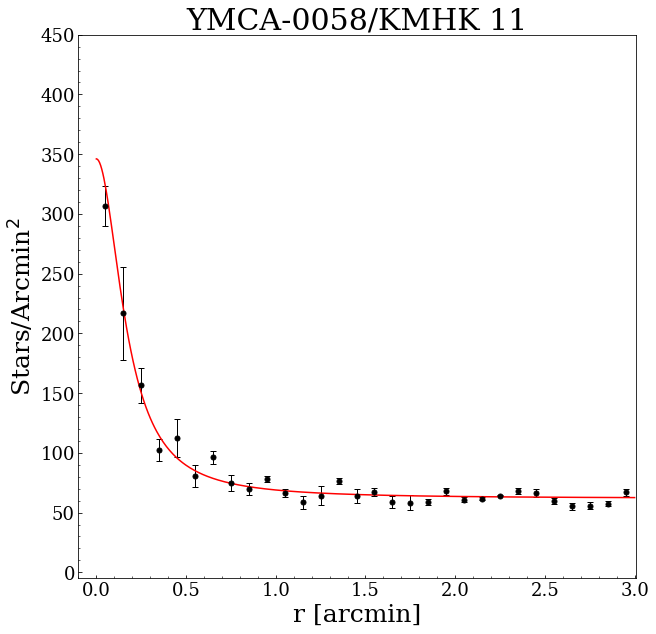}
    \includegraphics[width=0.35\textwidth]{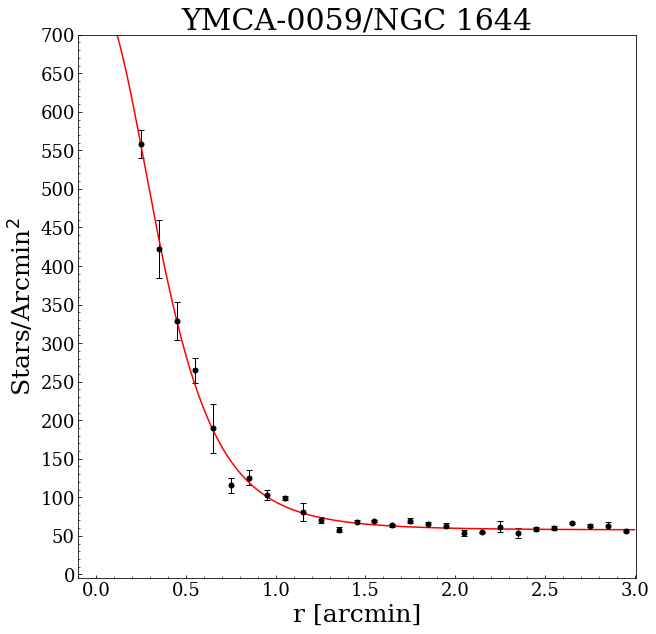}\\
    \includegraphics[width=0.35\textwidth]{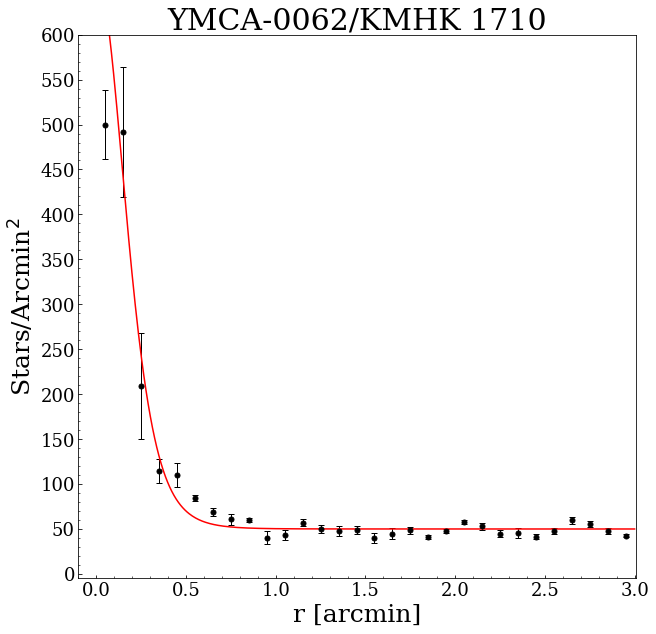}
    \includegraphics[width=0.35\textwidth]{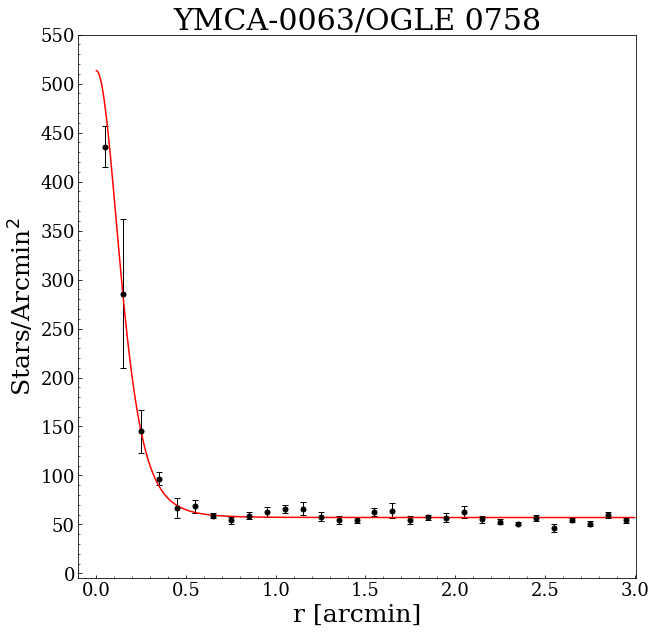}
    \includegraphics[width=0.35\textwidth]{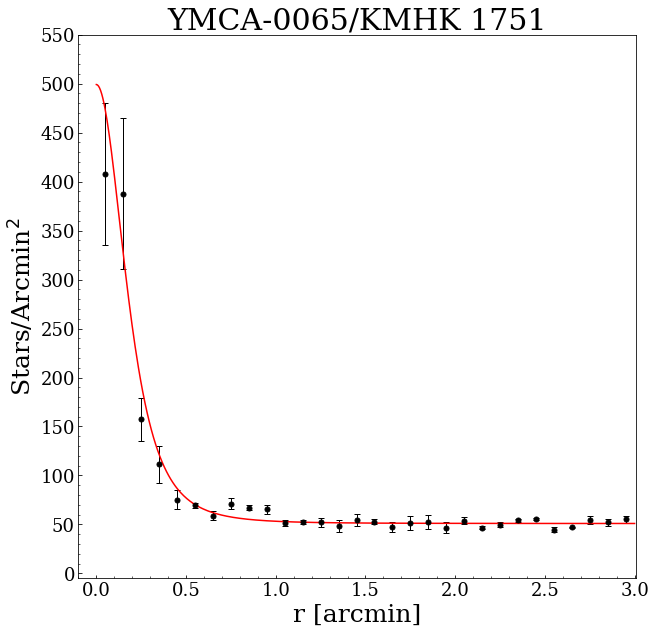}\\
    \caption{RDPs for all SCs already known. The red solid line indicates the best EFF profile (see Sect.~\ref{sec:fit_rdp} for the description of the fit procedure).} 
    \label{fig:known_clusters_rdp}
\end{figure*}

\begin{figure*}\ContinuedFloat
    \includegraphics[width=0.35\textwidth]{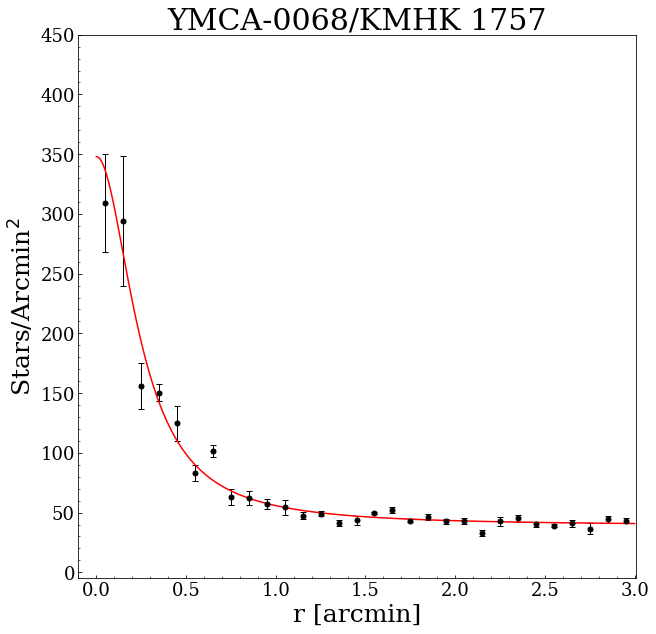}
    \includegraphics[width=0.35\textwidth]{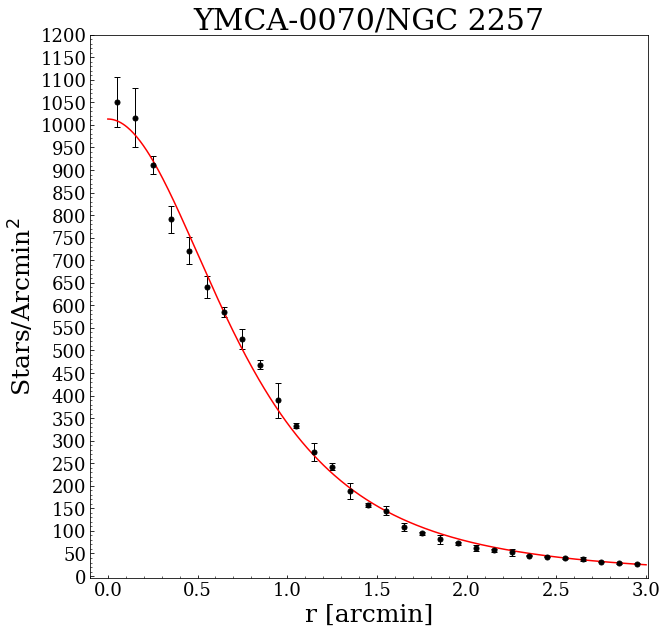}
    \includegraphics[width=0.35\textwidth]{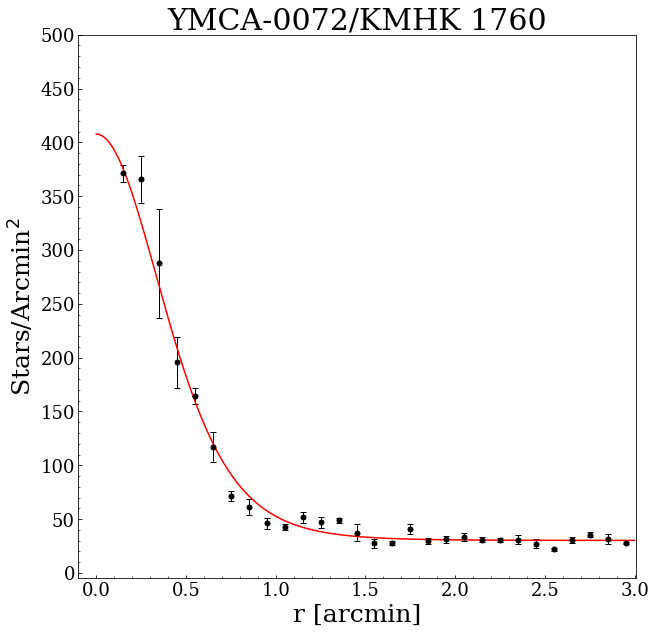}\\
    \includegraphics[width=0.35\textwidth]{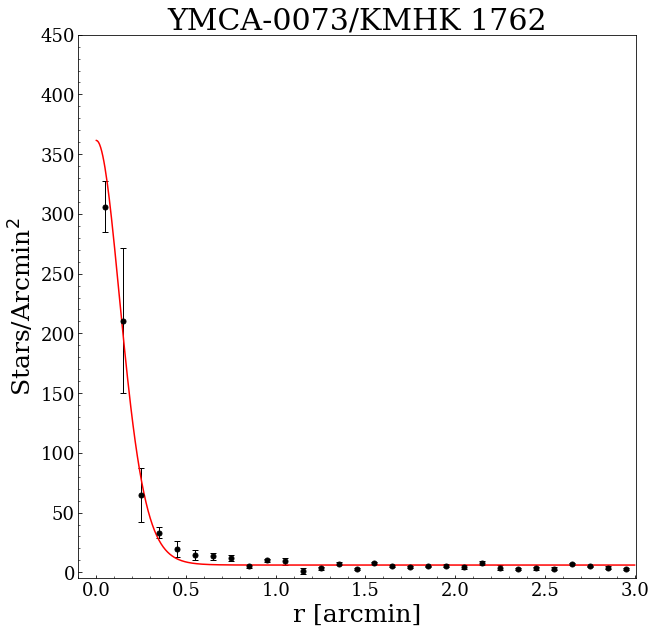}
    \caption{Continued}
\end{figure*}

\begin{figure*}
    \includegraphics[width=0.35\textwidth]{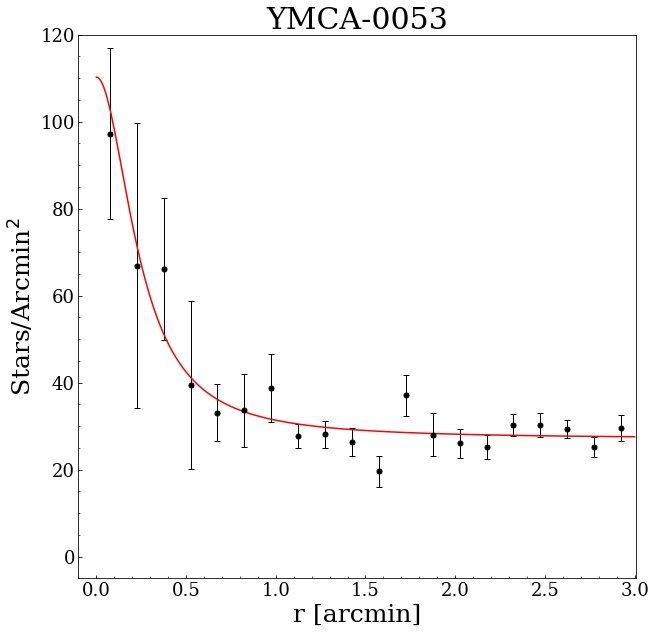}
    \includegraphics[width=0.35\textwidth]{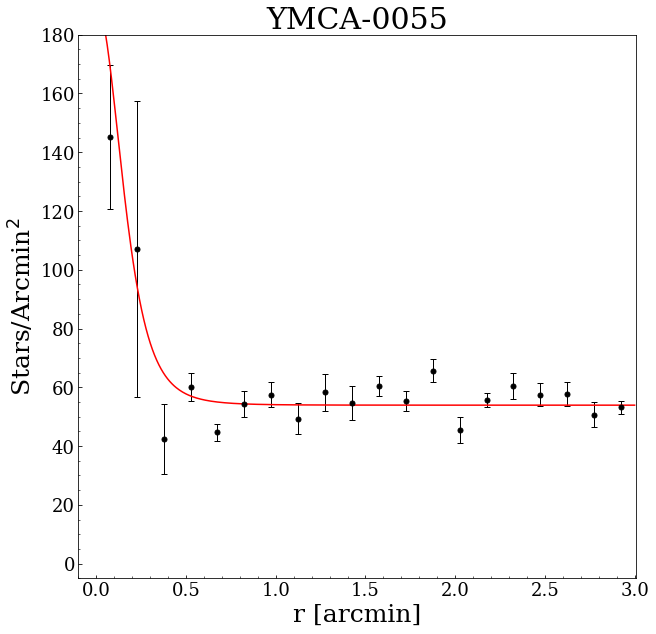}
    \includegraphics[width=0.35\textwidth]{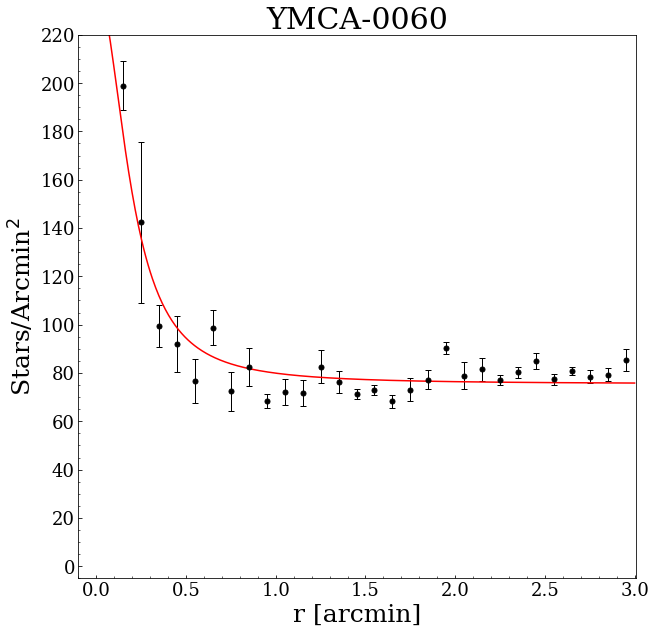}\\
    \includegraphics[width=0.35\textwidth]{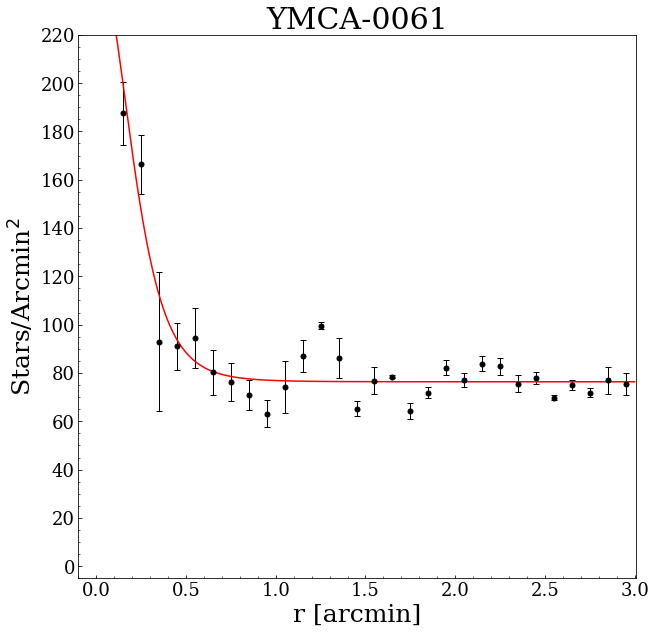}
    \includegraphics[width=0.35\textwidth]{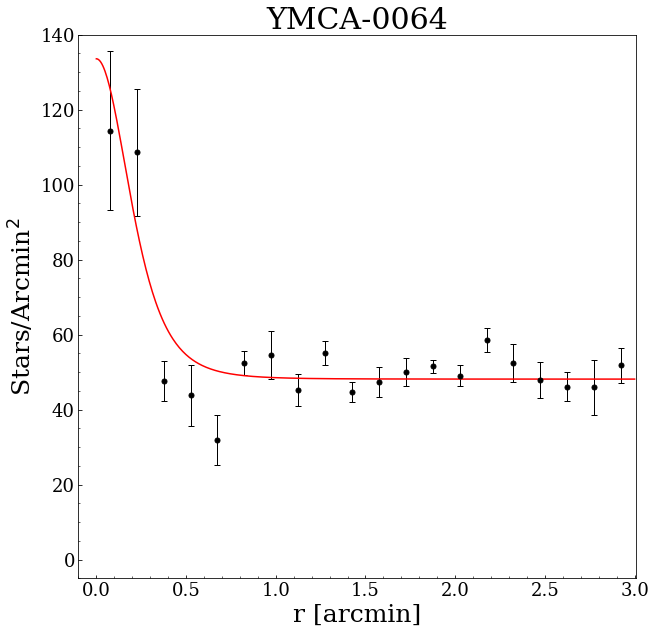}
    \includegraphics[width=0.35\textwidth]{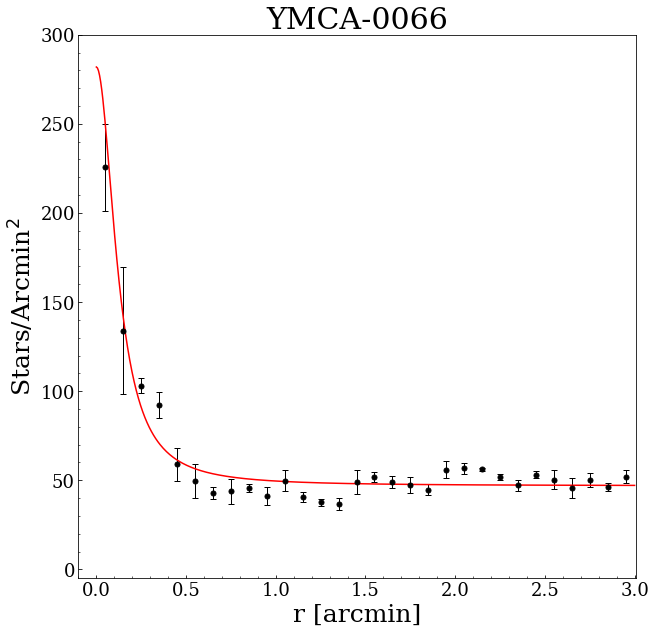}\\
    \includegraphics[width=0.35\textwidth]{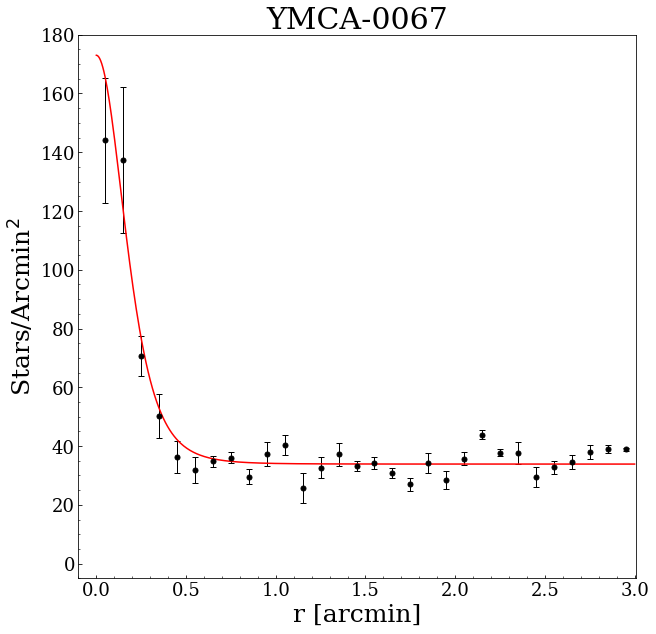}
    \includegraphics[width=0.35\textwidth]{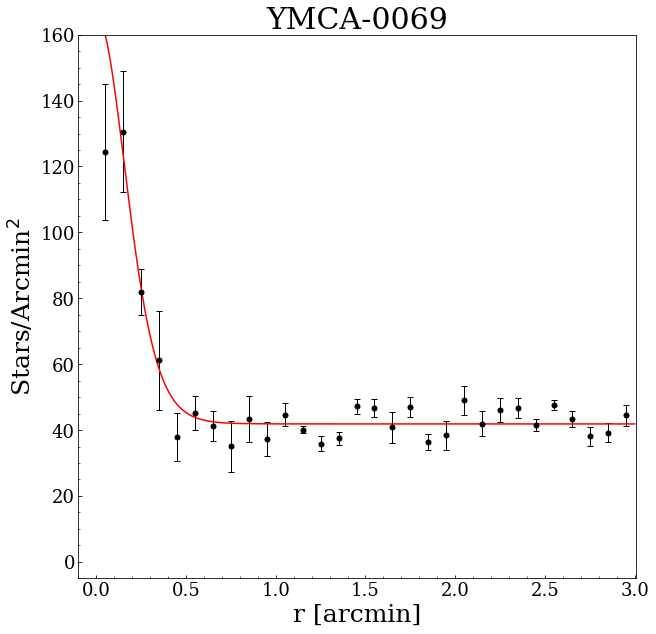}
    \includegraphics[width=0.35\textwidth]{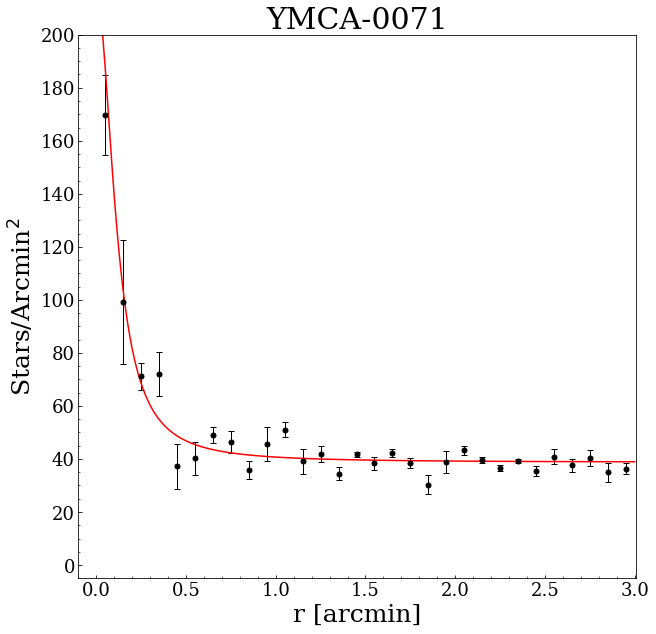}\\
    
    \caption{Same as Figure~\ref{fig:known_clusters_rdp}, but for the new candidate SCs.}
    \label{fig:new_clusters_rdp}
\end{figure*}
\end{appendix}
\end{document}